\documentclass[twocolumn,showpacs,preprintnumbers,amsmath,amssymb,prb]{revtex4-1}
\usepackage{graphicx}% Include figure files
\usepackage{dcolumn}% Align table columns on decimal point
\usepackage{bm}% bold math
\usepackage{color}
\usepackage{physics}
\usepackage{multirow}
\usepackage[normalem]{ulem}
 
\newcommand{\R}{{\bm{R}}}

\renewcommand{\d}{{\bm{\delta}}}
\newcommand{\e}{{\bm e}}

\newcommand{\G}{{\bf G}}
\newcommand{\K}{{\bf K}}
\renewcommand{\a}{{\bf a}}

\renewcommand{\k}{{\bm{k}}}
\newcommand{\q}{{\bm{q}}}
\renewcommand{\b}{{\bf b}}
\newcommand{\rr}{{\bm{r}}}

\def\lsim{\lower.35em\hbox{$\stackrel{\textstyle<}{\textstyle\sim}$}}
\def\gsim{\lower.35em\hbox{$\stackrel{\textstyle>}{\textstyle\sim}$}}

\begin{document}

\title{Nematic versus Kekul\'e phases in twisted bilayer graphene under hydrostatic pressure}

\author{Miguel S\'anchez S\'anchez$^1$, Israel D\'iaz$^1$, Jos\'e Gonz\'alez$^2$, and Tobias Stauber$^1$ }

\affiliation{
$^{1}$ Instituto de Ciencia de Materiales de Madrid, CSIC, E-28049 Madrid, Spain\\
$^{2}$ Instituto de Estructura de la Materia, CSIC, E-28006 Madrid, Spain
}
\date{\today}

\begin{abstract}
We address the precise determination of the phase diagram of magic angle twisted bilayer graphene under hydrostatic pressure within a self-consistent Hartree-Fock method in real space, including all the remote bands of the system. We further present a novel algorithm that maps the full real-space density matrix to a $4\times4$ density matrix based on a $SU(4)$ symmetry of sublattice and valley degrees of freedom. We find a quantum critical point between a nematic and a Kekul\'e phase, and show also that our microscopic approach displays a strong particle-hole asymmetry in the weak coupling regime. We arrive then at the prediction that the superconductivity should be Ising-like in the hole-doped nematic regime, with spin-valley locking, and spin-triplet in the electron-doped regime.
\end{abstract}

%\maketitle
\maketitle
{\it Introduction.}
Twisted bilayer graphene (TBG) forms when two graphene layers are rotated with respect to each other with a relative twist angle $\theta $. Under a set of commensurable angles $\theta_i$,\cite{Lopes_dos_Santos_2007} the system constitutes a perfect crystalline structure ({\it moir\'e lattice}) where Bloch's theorem applies. Moreover, for so-called magic angles a vanishing Fermi velocity resulting in flat bands near the charge neutrality point (CNP) has been predicted.\cite{Li10,suarezmorell} The first {\it magic angle} is found to be 
$\theta \sim 1.05^\circ$.\cite{bistritzer}

In 2018, TBG tuned around the first magic angle was shown to host insulating phases \cite{Cao_2018} near half-filling of the hole-like moir\'e minibands next to a superconducting dome phases,\cite{Cao_2018unconv} similar to what happens in cuprates.\cite{cuprates} What is more, correlated phases such as anomalous Hall ferromagnetism \cite{Sharpe19,Sharpe21} and quantum Hall effect \cite{koshinohall,PhysRevX.8.031089} have been predicted and observed, and are moreover linked to non-trivial Chern numbers.\cite{ledwith20,Xie2021,Pierce2021} 

The observed superconductivity (SC) is often attributed to the presence of electron pairing mechanisms that yield broken-symmetry states \cite{Oh_2021,bitanroy,goodwin,Lake22} and strange-metal behavior,\cite{Cao20,Gonzalez20,Jaoui22,Stauber22} but also electron-phonon pairing has been discussed.\cite{Lian_2019,mcdonaldphonon} Similar correlation effects and robust SC were further observed in twisted $N$-layer graphene for $ 2 \leq N \leq 5 $.\cite{Park22} Notably, in the case $N>2$, a Pauli limit violation up to a factor of $\sim3$ was seen,\cite{Park22,Park21,Zeyu21,Cao21} reinforcing the idea that the SC in these layered systems is indeed unconventional.\cite{Lake21,Christos22,Christos23,Gonzalez23}

The competition between different symmetry breaking patterns is difficult to address due to the emergent $U(4)$ symmetry realized by the electron system in the strong coupling limit.\cite{Kang19,seo19,Stepanov2020,Bultinck20,bernevig221,bernevig321,bernevig421,dimitru22,Wagner22} 
Even though these moir\'e systems seem to be well-controlled compared to e.g. cuprates as they can be electrically doped, there is still no consensus on the precise phase diagram that should depend sensitively on the surrounding dielectric environment.\cite{Liu21,Jaoui22} This could explain that STM measurements show Kekul\'e patterns in the electron density,\cite{Nuckolls23,Kim23} while former experiments in a different setup with two metallic gates have given evidence of a nematic phase.\cite{Cao21Nematic}

We will undertake the precise determination of the phase diagram of TBG by applying a self-consistent Hartree-Fock method in real space, including all the remote bands of the system.\cite{Vafek20} This microscopic real space implementation represents a key advantage, as the different phases turn out to be very sensitive to the on-site Hubbard repulsion, which otherwise cannot be disentangled from the long-range Coulomb interaction in momentum space. 

Taking into account all the remote bands is also a must to discern the competition among states whose energy difference is sometimes below 0.1 meV. For this purpose, we will resort to a slight simplification and consider magic angle TBG under hydrostatic pressure at a larger twist angle.\cite{jarillopressure} This is in the spirit of the continuum model,\cite{bistritzer} where there is a single dimensionless coupling with a critical value at the flat-band regime, therefore making possible to trade larger moir\'e unit cells by smaller ones, at the cost of decreasing the interlayer distance. This is confirmed by the fact that SC can be tuned experimentally by applying hydrostatic pressure (thereby reducing the interlayer distance).\cite{Yankowitz_2019}

\begin{figure}[h]
\includegraphics[width=0.99\columnwidth]{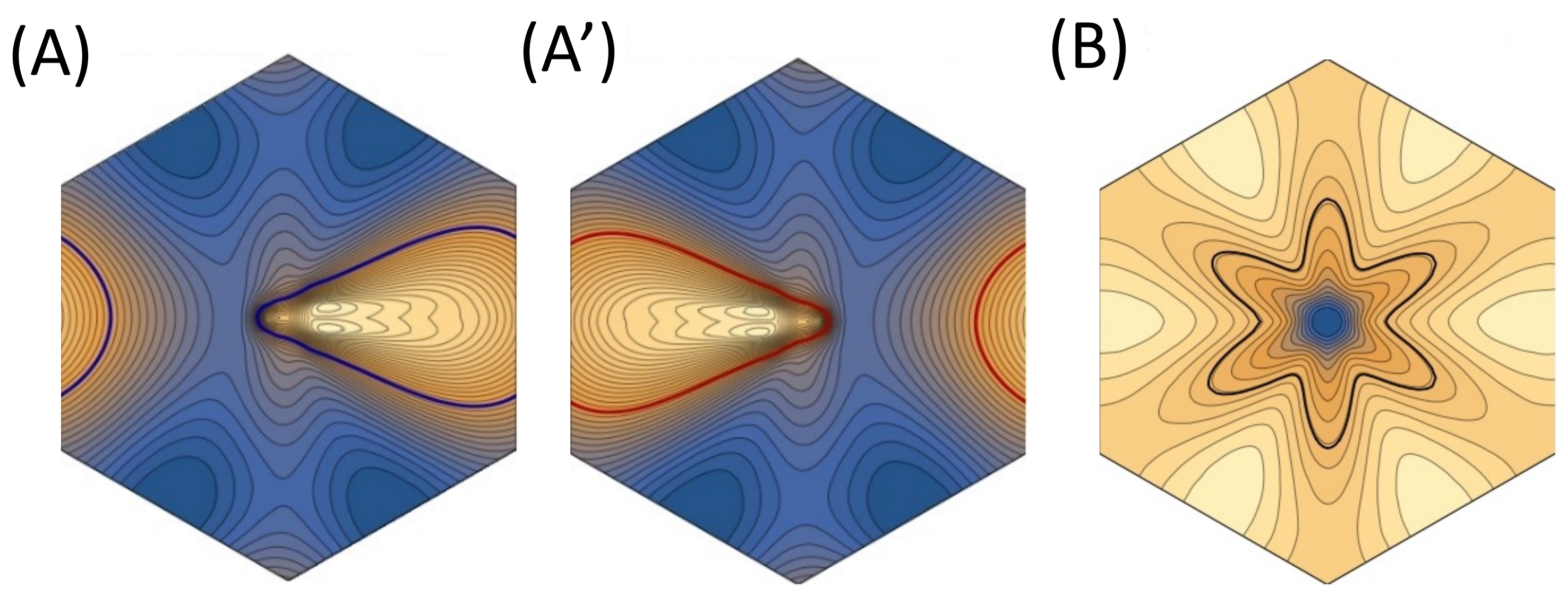}
\caption{Energy contour maps showing the Fermi lines in the second valence band for (A) spin-up and (A') spin-down electrons in the moir\'e Brillouin zone of TBG with twist angle $\theta \approx 3.5^\circ$ under hydrostatic pressure, for interaction strength $\alpha=0.1$ eV$\times a$ ($a/\sqrt{3}$ being the C-C distance) and filling fraction of 2.4 holes per moir\'e unit cell.  (B) Energy contour map showing the Fermi line in the second conduction band for the same interaction strength and filling fraction of 2.3 electrons per moir\'e unit cell. Contiguous contour lines differ by a constant step of 0.5 meV, from lower energies in blue to higher energies in light color.}
\label{one}
\end{figure}

We will show that in the hole-doped regime, there is a quantum critical point separating the strong-coupling regime of the Coulomb interaction, from an intervalley-coherent ground state, and the intermediate coupling regime where the ground state is instead valley polarized. Most significantly, this quantum phase transition implies a change in the symmetry of the bands, which are $C_6$-invariant in the strong-coupling side, but only invariant under reflection by a mirror plane for larger dielectric screening, as shown in Fig. \ref{one}. 

Our microscopic approach also displays a strong particle-hole asymmetry in the weak coupling regime. We then arrive at the prediction that in the hole-doped nematic regime there is Ising SC with spin-valley locking whereas in the electron-doped regime there is spin-triplet SC arising from the strong anisotropy of the $e$-$e$ interaction. 

{\it Tight-binding Hamiltonian.}
We consider twisted bilayer graphene under hydrostatic pressure for the magic twist angle condition at $\theta=3.5^\circ$ whose band structure is similar to the one at $\theta=1.16^\circ$ without pressure, see the Supplemental Material (SM).\cite{SM} The non-interacting tight-binding Hamiltonian reads
\begin{align}
H_0=-\sum_{n,m}\sum_{i,j}\sum_{\sigma}t_{n,m}^{i,j}\psi^\dagger_{n,i,\sigma}\psi_{m,j,\sigma}\;,
\end{align}
where the hopping matrix element $t_{n,m}^{i,j}$ only depends on the distance between lattice sites, i.e., $t_{n,m}^{i,j}=t(|\R_n-\R_m+\d_i-\d_j|)$ with $\R_n$ denoting the lattice vector of unit cell $n$ and $\d_i$ the position of lattice site $i$ with respect to the unit cell. The spin is labeled by $\sigma$. We use the Slater-Koster parametrization
\begin{align}
\nonumber
t(\rr)= V_{pp\pi}e^{\frac{a_{0}-r}{r_0}} \Big(1- \left(\frac{\rr\cdot\e_z}{r}\right)^2\Big) + V_{pp\sigma}e^{\frac{d_{\perp}^0-r}{r_0}} \left(\frac{\rr\cdot\e_z}{r}\right)^2,
\end{align}
where $a_0=1.42$\r{A} is the C-C distance, $r_0=0.319a_0$ the decay parameter, and $d_\perp^0=3.34$\r{A} denotes the equilibrium interlayer distance. Due to the layered structure, the projection $\rr\cdot\e_z$ is either zero or $d_\perp=\frac{1}{1.134}d_{\perp}^0$, the interlayer distance of the compressed lattice for $\theta=3.48^\circ$. Further, we set $V_{pp\pi}=2.7$eV and $V_{pp\sigma}=-0.48$eV.\cite{Moon13}

{\it Coulomb interaction.} The total Hamiltonian shall be written as $H=H_0 + H_{\rm int}$ where the interaction term is split into a long-ranged Coulomb interaction and a short-ranged on-site Hubbard term, $H_{int} =H_V+H_U$:
\begin{align}  
H_V&=\frac{1}{2} \sum_{n,m} \sum_{i,j}\sum_{\sigma,\sigma'} V_{n,m}^{i,j}\psi_{n,i,\sigma}^{\dagger}\psi_{m,j,\sigma'}^{\dagger}\psi_{m,j,\sigma'}\psi_{n,i,\sigma}\;,\\
H_U&=\frac{U}{2} \sum_{n,i}\sum_{\sigma}\psi_{n,i,\sigma}^{\dagger}\psi_{n,i,\bar\sigma}^{\dagger}\psi_{n,i,\bar\sigma}\psi_{n,i,\sigma}\;,
\end{align}
where $\bar\sigma$ denotes the opposite spin-projection. Again, the Coulomb potential shall only depend on the distance between lattice sites, $V_{n,m}^{i,j}= v(|\R_n - \R_m +\d_i-\d_j|) $, and is implemented by the double-gated potential\cite{Throckmorton12}
\begin{align}
v(|\rr|)=\frac{e^2}{4\pi \epsilon_0 \epsilon}\sum_{n} \frac{(-1)^n}{|\rr+ n\xi \boldsymbol{{z}}|} \xrightarrow[r \gg \xi]{} \frac{e^2}{4\pi\epsilon_0\epsilon} \frac{2\sqrt{2}e^{{-\pi r/\xi}}}{ \xi\sqrt{r/\xi}}\;,
    \label{potential}
 \end{align}
where $\epsilon$ stands for the intrinsic dielectric constant of the system, see SM.\cite{SM} Here, we will use $\epsilon$  or alternatively $\alpha=e^2/4\pi\epsilon_0\epsilon=\tilde\alpha$ eV$\times a$ with $a=\sqrt{3}a_0$ as a variable to change the strength of the interaction. We will further choose $\xi=10$nm.\cite{Saito_2020,Stepanov_2020}
    
{\it Hartree-Fock solution.} The interacting system shall be treated within the restricted Hartree-Fock (HF) approach, i.e., we will only consider spin-symmetric solutions and the spin-quantum number $\sigma$ shall be suppressed from now on. Concretely, we will study the ground-state mainly at integer filling factor $\nu=0,\pm2$ for on-site Hubbard interactions $U=0.5,4$eV and $\epsilon=12,15,20,30,60$, corresponding to $\tilde\alpha=0.5,0.4,0.3,0.2,0.1$. For some parameters, we find two different solutions/phases; then, the solution with lowest energy is chosen which gives rise to a phase transition at ($\nu=-2$, $U=4$eV, $\epsilon\approx 20 $) and ($\nu=2$, $U=0.5$eV, $\epsilon\approx 20$), as shown in Fig. 2. For $\epsilon\approx100$, there is a phase transition to a gapless phase which will not be addressed here. More details are given in the SM.\cite{SM}

\begin{figure}[!t]
    \centering
    \includegraphics[width=0.99\columnwidth]{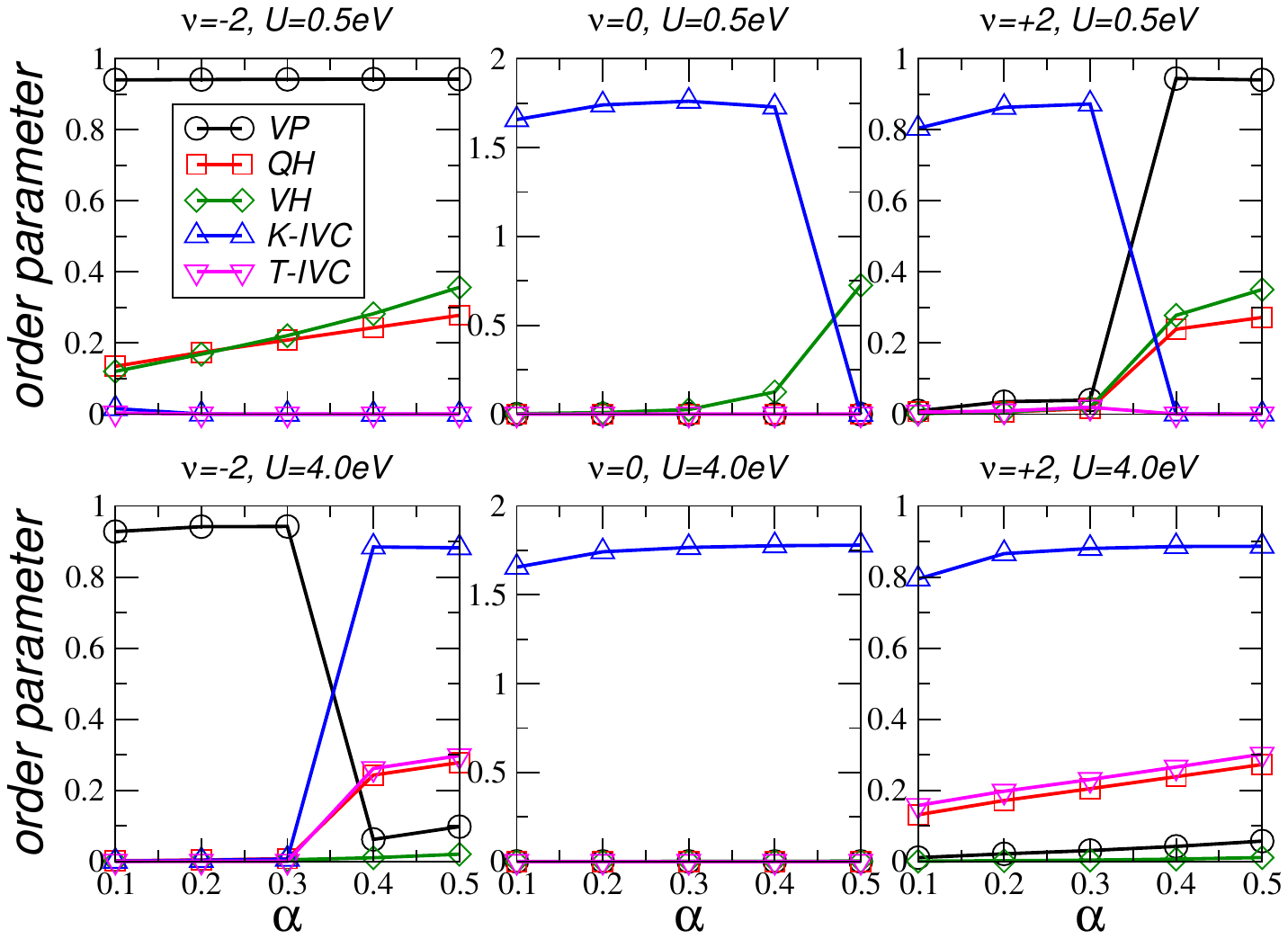}
    \caption{Phase diagrams of magic angle twisted bilayer graphene for integer filling factors $\nu=-2$ (left), $\nu=0$ (center), and $\nu=2$ (right) and Hubbard on-site interaction $U=0.5$eV (upper panels) and $U=4$eV (lower panels) as function of the coupling strength $\alpha=\frac{e^2}{4\pi\epsilon_0\epsilon}$ in units of eV$\times a$. The phases are characterized by the order parameters for valley polarization $\langle\tau_z\rangle$ (VP), quantum Hall $\langle\sigma_z\tau_z\rangle$ (QH), valley Hall $\langle\sigma_z\rangle$ (VH), Kramers intervalley coherence $|\langle\sigma_y\tau_{x/y}\rangle|$ (K-IVC), and $T$-symmetric intervalley coherence $|\langle\sigma_x\tau_{x/y}\rangle|$ (T-IVC), see SM.}
    \label{orderparameters}
\end{figure}

For most parameters, the energy bands have already converged after $\sim\!\!20$ iterations. For the convergence of the order parameters, though, many more iterations are needed. This is due to the emergent symmetry of the ground-state whose band-structure is invariant under a $U(4)$-rotation.\cite{Kang19,seo19,Bultinck20,bernevig321} The self-consistent Hartree-Fock equations thus quickly find the (almost) degenerate manifold of states that spontaneously break the $U(4)$ symmetry, but many additional iterations are necessary to reach the true ground-state. We believe that only within an atomistic tight-binding model, reliable results regarding this symmetry broken state can be obtained.

{\it $SU(4)$-reduced HF density matrix.} The Hartree-Fock ground state is characterized by the real-space HF (one-body) density matrix 
\begin{align}
\label{DensityMatrix}
\rho_{ij}^{HF}=\frac{1}{N_c}\sum_{\varepsilon_\k \le \varepsilon_F}\langle \psi_{\k,i}^\dagger\psi_{\k,j}\rangle 
 %e^{-i \k \cdot (\boldsymbol{R}_i + \d_i - \boldsymbol{R}_j - \d_j)} 
\;,
\end{align}
where the subscripts $i, j$ refers to the atom at position $\d_i, \d_j$ in the unit cell, and $\psi_{n,i}=\frac{1}{\sqrt{N_c}}\sum_{\k\in1.BZ}\psi_{\k,i}e^{i\k\cdot \R_n}$ with $N_c$ the number of moir\'e unit cells and $\varepsilon_{F}$ the Fermi energy.

%\textcolor{red}{(((((}Due to the large moir\'e unit cell, we can relate the many-body density matrix to a single-particle density, as outlined in the SM. We thus have $\rho_{ij}^{HF}=\psi_i^*\psi_j$ with $|\psi\rangle\propto\sum_i\psi_i|\rr_i\rangle$ where $|\rr_i\rangle$ stands for the localized state at lattice site $\rr_i$.\textcolor{red}{)))))}

In the continuum limit without spin and for energies close to the Fermi level, the wave function is characterized by four envelope functions for each layer related to sublattice and valley degree of freedom. With $\K_\ell$ and $\K_\ell'$ denoting the $K$-points for each layer, we write the single-particle wave function $|\psi\rangle=\sum_i\psi_i|\rr_i\rangle$ as
\begin{align}
\notag
|\psi\rangle&=\sum_{\ell,\gamma}\sum_{\rr_i\in \gamma, \ell}\!\!\!\left[e^{i\K_\ell\cdot\rr_i}f_{\gamma,K,\ell}(\rr_i)+e^{i\K_\ell'\cdot\rr_i}f_{\gamma,K',\ell}(\rr_i)\right]|\rr_i\rangle\;,
\end{align}
with $\gamma=A,B$ denoting the sublattice and $\ell$ the layer. This decomposition naturally leads to the single-particle density matrix $\rho_{\gamma,\delta,\ell;\gamma',\delta',\ell'}(\rr, \rr')=f_{\gamma,\delta,\ell}^*(\rr)f_{\gamma',\delta',\ell'}^{}(\rr')$ that contains all information of the long-wavelength theory, with $\delta=K,K'$. Here, we will mainly discuss quantities related to the $4\times4$ density matrix by tracing over the lattice sites and the two layers,
\begin{align}
\label{ReducedDM}
\rho_{\gamma,\delta;\gamma',\delta'}=\int_{A_M}d\rr\sum_\ell f_{\gamma,\delta,\ell}^*(\rr)f_{\gamma',\delta',\ell}^{}(\rr)\;.
\end{align}
As shown in the SM, $f_{\gamma,\delta,\ell}^*(\rr)f_{\gamma',\delta',\ell}^{}(\rr)$ with a coarse-grained position vector $\rr$ can be obtained by performing closed loops $\sum_{m=1\dots n}e^{i\varphi_m}\psi_{i_m}^*\psi_{j_m}^{}$ with $j_1=i_n$ due to constructive/destructive interference. With the identification $\psi_{i}^*\psi_{j}^{}\to\rho_{ij}^{HF}$, we can relate the HF density matrix of Eq. (\ref{DensityMatrix}) to the {\it reduced} HF density matrix based on Eq. (\ref{ReducedDM}). Finally, we can decompose $\rho$ into Hermitian components, $\rho_{\gamma,\delta;\gamma',\delta'} = \sum_{\alpha \beta} \rho_{\alpha,\beta} [\sigma_\alpha\otimes\tau_\beta]_{\gamma, \delta; \gamma' \delta'} = \sum_{\alpha \beta} \langle\sigma_\alpha\otimes\tau_\beta \rangle [\sigma_\alpha\otimes\tau_\beta]_{\gamma, \delta; \gamma' \delta'} $ where $\sigma_\alpha, \tau_\beta$ denote the Pauli matrices for sublattice and valley degree of freedom including the unity matrix with $\alpha,\beta=0,1,2,3$. The tensor-product shall be suppressed from now on. By adding a constant term proportional to unity and proper scaling,\cite{SM} we can finally fix the trace to Tr$\rho=2+\nu/2$.\cite{Bultinck20}

The algorithm can be used for any hexagonal system that is subjected to a structure much larger than the atomic scale. Also extensions to include layer and spin-degrees of freedom are possible.

{\it Phase diagram.} The phase diagram can be characterized by the valley polarized (diagonal order) and intervalley coherent (off-diagonal order) phase whose order parameters assume almost quantized values in their respective phases. The $SU(4)$-reduced HF density matrix based on Eq. (\ref{ReducedDM}) can then be simplified to
\begin{align}
\rho=\left(
\begin{array}{cccc}
\rho_{A,K;A,K}&0&0&\rho_{A,K;B,K'}\\
0&\rho_{B,K;B,K}&\rho_{B,K;A,K'}&0\\
0&\rho_{A,K';B,K}&\rho_{A,K';A,K'}&0\\
\rho_{B,K';A,K}&0&0&\rho_{B,K';B,K'}
\end{array}\right)\;,
\end{align} 
where approximate expressions for $\rho_{\gamma,\delta;\gamma',\delta'}$ as function of $\alpha$ and $U$ are given in the SM.\cite{SM} The general state is thus described by two Bloch-spheres in the two basis sets 
 $\{|A,K\rangle,|B,K'\rangle\}$ and $\{|A,K'\rangle,|B,K\rangle\}$.\cite{Bultinck20} For $\nu=0$, we obtain a state with $\rho^2=\rho$ after proper normalization, see SM;\cite{SM} for $\nu=\pm2$, the $SU(4)$-reduced HF density matrix 
 defines a mixed state with $\rho^2\neq\rho$. 

Valley polarization (VP) only occurs for $\nu=\pm2$ and the bands are almost completely polarized up to 95\%. For smaller twist angles realizing the magic angle condition, this polarization is even increasing and can reach up to 98\% for the real magic angle at $\theta\approx 1.05^\circ$. For $\nu=\pm3$, the lowest conduction band per spin-channel is half-filled and the occupied states are again almost fully valley polarized with up to 85\%. Only for $\nu=\pm1$, this (almost) complete polarization is lost. 

Intervalley coherence (IVC) occurs predominantly for $\nu=0$ and the bands are IVC polarized up to 92\%. For $\nu=\pm2$, the order parameter is half the value of charge neutrality, however, we could have defined our order parameter including a factor 2 that would take into account the proper normalization of the initial wave function. Then, the valence band at $\nu=-2$, both valence bands at $\nu=0$, and the conduction band at $\nu=2$ are predominantely intervalley coherent (per spin-channel). 

Fig. \ref{orderparameters} presents a summary of our main results and shows selective order parameters for the different interaction strengths and filling factors. We observe several phase transitions and strong particle-hole asymmetry in the weak coupling regime; in the strong coupling regime, however, particle-hole symmetry is almost completely restored. 

\begin{figure}[!h]
    \centering
    \includegraphics[width=0.99\columnwidth]{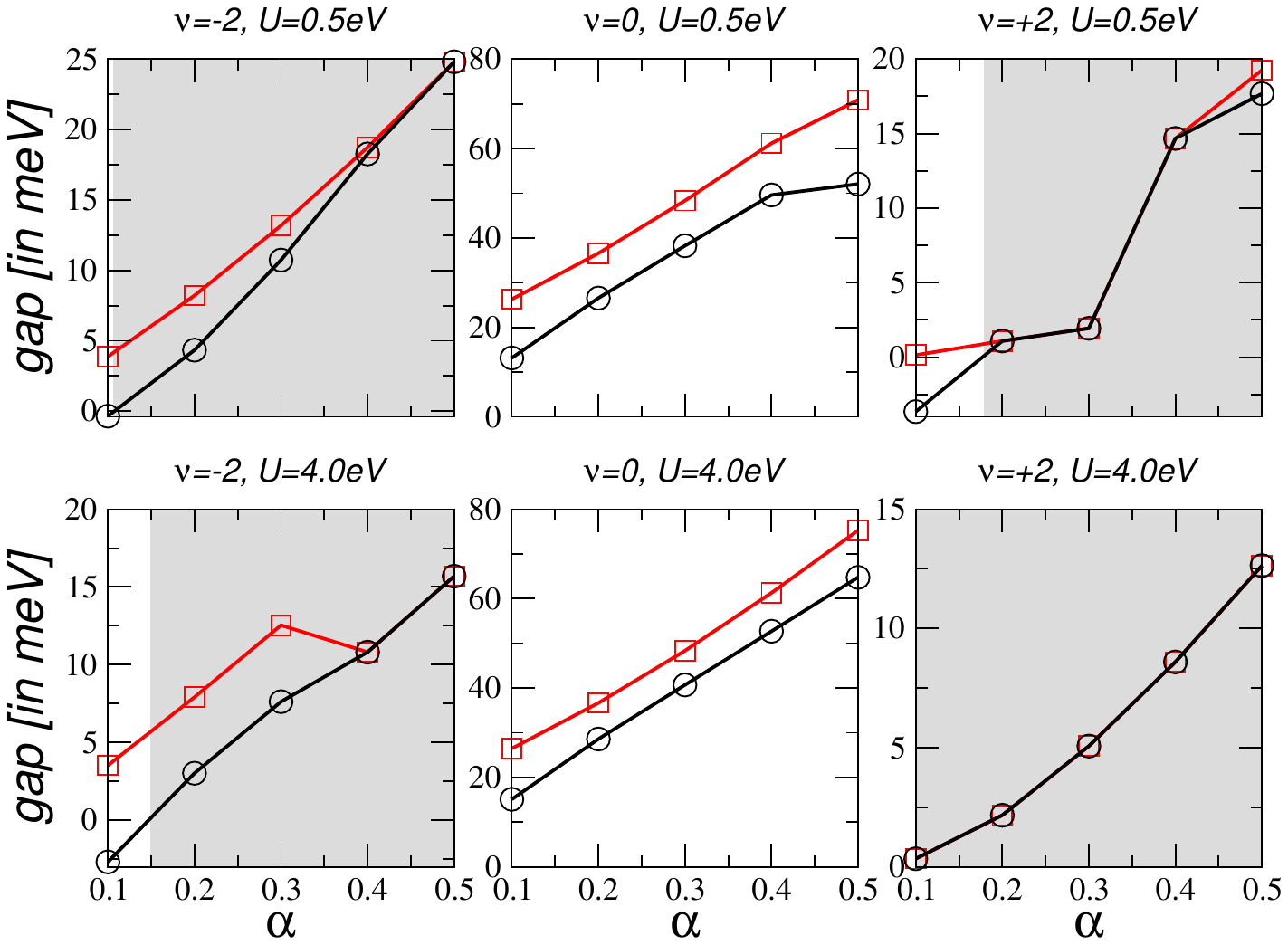}
    \caption{The total (possibly indirect) gap (black circles) and the (averaged) gap at the $K$-points (red squares) of the correlated insulator phase for integer filling factors $\nu=-2$ (left), $\nu=0$ (center), and $\nu=2$ (right) and Hubbard on-site interaction $U=0.5$eV (upper panels) and $U=4$eV (lower panels) as function of the coupling strength $\alpha=\frac{e^2}{4\pi\epsilon_0\epsilon}$ in units of $eV\times a$. The shaded region indicates a topological gap with Chern number $|C|=1$ (for one spin-channel).}
    \label{gap}
\end{figure}

In Fig. \ref{gap}, we show the total (possibly indirect) gap and the (averaged) gap at the $K$-points as function of the same parameters. We also calculate the Chern number employing the method of Ref. \onlinecite{Fukui05}. For $\nu=\pm2$, the gap is always topological with Chern number $|C|=1$ per spin channel. This needs to be contrasted with the gap at charge neutrality that is considerable larger and always trivial with $C=0$. 

{\it Mesoscopic wave function.} After proper normalization of the $SU(4)$-reduced HF density matrix,\cite{SM} we find valley coherence with $\rho^2=\rho$ for $\nu=0$. This density matrix can be well approximated by $\rho_{\rm{KIVC}}=|\varphi\rangle\langle\varphi|$ where $|\varphi\rangle$ denotes the K-IVC state.\cite{Bultinck20,SM} This allows us to make the following separation ansatz for the envelope wave functions, valid on the moir\'e unit cell:
\begin{align}
\label{WaveFunctionAnsatz}
|f(\rr)\rangle= \phi(\rr)|\varphi\rangle\;,
\end{align}
with
$|f(\rr)\rangle=(f_{A,K}(\rr),f_{B,K}(\rr),f_{A,K'}(\rr),f_{B,K'}(\rr))^T$ and summation over $\ell$ is implied. Furthermore, $\phi(\rr)\sim\exp(-\xi|\rr|^2/L_M^2)$ with $\xi\sim3$ which confines the wave function around the $AA$-stacked regions centred at $\rr=0$. This is in line with the heavy-fermion model for TBG.\cite{Song22}

%$|\psi(\rr)\rangle=\sum_\ell(f_{A,K,\ell}(\rr),f_{B,K,\ell}(\rr),f_{A,K',\ell}(\rr),f_{B,K',\ell}(\rr))^T$.
%Since the wave function is confined around the $AA$-stacked regions centred at $\rr=0$, the internal states $|\varphi\rangle$ are decoupled for adjacent moir\'e unit cells which is in line with the heavy-fermion model for TBG.\cite{Song22} We thus expect the angle $\varphi$ to be uniformly distributed over the sample due to slight disorder. Only for uniform strain, the phase should be coherent on a length scale beyond the moir\'e length.\cite{KwanX23} Our analysis should also approximately apply to $\nu=\pm2$ and due to real-space localization, valley-coherence should be limited to a moir\'e unit-cell if disorder dominates over, e.g., strain.  

{\it Nematicity.}
For $(\nu=-2$, $U=4$eV, $\epsilon=60,30,20)$ and $(\nu=2$, $U=0.5$eV, $\epsilon=60,30,20)$, the band structure lacks $C_3$ symmetry and only displays one mirror symmetry. Interestingly, the $SU(4)$-reduced HF density matrix does then not depend on $\epsilon$ and simply reads $\rho=\frac{1}{2}\mathcal{P}_K$ and $\rho=\frac{1}{2}({\bf 1}+\rho_{\rm{KIVC}})$, respectively, where $\mathcal{P}_K$ denotes the projection operator on valley $K$. This is in line with Refs. \onlinecite{Liu21,Jaoui22}. This universality of the {\it even} $SU(4)$-reduced HF density matrix based on Eq. (\ref{ReducedDM}) has to be contrasted to the {\it odd} $SU(4)$-reduced HF density matrix for which the contributions of the two layers are subtracted, see SM.\cite{SM}

{\it Superconductivity.} We will now discuss the pairing instability of the ground-state in the weak coupling regime $\epsilon\sim60$ and for $U=4$eV which should be a realistic value in twisted bilayer graphene.\cite{Sanchez24,Gonzalez21} This demands the analysis of the Cooper pair vertex $V$ for electrons with zero total momentum.\cite{Kohn65,Baranov92} The vertex can be parameterized in terms of the angles $\phi $ and $\phi'$ of the respective momenta of spin-up incoming and outgoing electrons at given energy $\varepsilon$. The instabilities of the vertex show up by solving the equation encoding the iteration of the scattering of Cooper pairs
\begin{align}
&V(\phi, \phi')= V_0 (\phi, \phi')-   \nonumber        \\ 
 & \frac{1}{(2\pi )^2} \int^{\Lambda_0} \frac{d \varepsilon }{\varepsilon } 
   \int_0^{2\pi } d \phi'' 
  \frac{\partial k_\perp }{\partial \varepsilon}  
      \frac{\partial k_\parallel }{\partial \phi''} 
  V_0 (\phi, \phi'')    
         V(\phi'', \phi'),
\label{pp}
\end{align}
where $k_\parallel ,k_\perp $ are the longitudinal and transverse components of the momentum for each energy contour line while $V_0 (\phi, \phi') $ is the bare vertex at a high-energy cutoff $\Lambda_0$.

Eq. (\ref{pp}) can be simplified by differentiating with respect to the cutoff which leads to
\begin{align}
\varepsilon \: \frac{\partial \widehat{V}(\phi, \phi' )}{\partial \varepsilon } 
 =   \frac{1}{2\pi }  \int_0^{2\pi } d \phi''  
 \widehat{V} (\phi, \phi'' )  \widehat{V}(\phi'', \phi' )\;,
\label{scaling}
\end{align}
with $\widehat{V} (\phi, \phi' ) = F(\phi ) F(\phi' ) V (\phi, \phi' )$ and $F(\phi ) = \sqrt{ (\partial k_\perp / \partial \varepsilon )
  (\partial k_\parallel / \partial \phi )/2\pi  }$. Eq. (\ref{scaling}) implies that the vertex is a function of the variable $\varepsilon/\Lambda_0$. If the initial condition $V_0 (\phi, \phi')$ has a negative eigenvalue for any of its harmonics projected onto the Fermi line, the solution of Eq. (\ref{scaling}) will display a divergence at a critical energy scale $\varepsilon_c$ as $\varepsilon \rightarrow 0$, i.e., the signature of the pairing instability given by 
\begin{align}
\label{EnergyScale}
\varepsilon_c = \Lambda_0 \: e^{-1/|\lambda |}\;,
\end{align}
where $\lambda$ denotes the negative eigenvalue and $\Lambda_0$ the effective band width. The initial vertex $V_0 (\phi, \phi')$ at the high-energy cutoff is dressed by the iteration of particle-hole scattering processes 
\begin{align}
V_0 (\phi, \phi') =  v_{\k-\k^\prime}
    + \frac{\overline{v}^2 \: \widetilde{\chi}_{\k+\k^\prime }}{1 - \overline{v} \: \widetilde{\chi}_{\k+\k^\prime }}   \;,
\label{init}
\end{align} 
where $\k, \k^\prime$ are the respective momenta for angles $\phi, \phi'$, $\overline{v}$ is the average potential in momentum space ($\approx 5$ meV $\times a^2$), and $\widetilde{\chi}_{\q}$ denotes the particle-hole susceptibility.\cite{Scalapino87,Gonzalez23} The final step is to project the vertex onto the harmonics $\cos (n\phi),\sin (n\phi)$ which build up the different contributions to $\widehat{V} (\phi, \phi' )$ at the high-energy cutoff.

We have carried out this operation along the Fermi lines of our model at filling fraction $\nu = -2.4$ shown in Fig. \ref{one} (A) and (A') and at filling fraction $\nu=2.3$ shown in Fig. \ref{one} (B). The eigenvalues for the different harmonics can be grouped according to the irreducible representations of the approximate symmetry groups $C_{h}$ and $C_{6v}$, respectively. 

\begin{table}[t!]
\begin{tabular}{c|c|c}
Eigenvalue $\lambda$  &   harmonics  &   Irr. Rep.\\
\hline \hline
2.12  &     $1$         &                  \\
\hline
0.51   & $\sin (\phi)$    & $A^{\prime\prime}$  \\
\hline
0.38  &     $\cos (\phi)$         &       $A^\prime$            \\
\hline
-0.19  &     $\sin (4\phi)$   &    $A^{\prime\prime}$ \\
\hline
0.18   &     $\sin (2\phi) $       & $A^{\prime\prime}$            \\                
\hline
-0.12   &    $\cos (6\phi) $    &  $A^\prime$
\end{tabular}
\caption{Eigenvalues of the Cooper-pair vertex with largest magnitude and dominant harmonics grouped according to the irreducible representations of the approximate $C_{h}$ symmetry, for the Fermi lines as shown in Fig. \ref{one} (A) and (A') .} 
  \label{table}
\end{table} 
\begin{table}[t!]
\begin{tabular}{c|c|c}
Eigenvalue $\lambda$  &   harmonics  &   Irr. Rep.\\
\hline \hline
3.47  &     $1$         &                  \\
\hline
0.89   &    \multirow{2}{*}{ $\{\cos (\phi),\sin (\phi)\}$ }    & \multirow{2}{*}{$E_2$}  \\
0.82   &                                  &                                      \\
\hline
$0.30$  &     \multirow{2}{*}{ $\{\cos (2\phi),\sin (2\phi)\}$ }    &  \multirow{2}{*}{$E_1$}  \\
$0.29$   &                                 &                                         \\
\hline
0.18   &     $\sin (3\phi)$       &     $B_2$               \\                
\hline
-0.17   &     $\cos (3\phi) $       &     $B_1$            
\end{tabular}
\caption{Eigenvalues of the Cooper-pair vertex with largest magnitude and dominant harmonics grouped according to the irreducible representations of the approximate $C_{6v}$ symmetry, for the Fermi line as shown in Fig. \ref{one} (B).} 
  \label{table2}
\end{table}   

The results are shown in Tables \ref{table} and \ref{table2}. In both cases, there is a negative coupling with relatively large magnitude $|\lambda| \approx0.19$ and $|\lambda| \approx0.17$, respectively, leading to a divergence in that channel at the energy scale of Eq. (\ref{EnergyScale}). In TBG, the magnitude of $\Lambda_0 $ is constrained by the reduced bandwidth of the second valence and conduction bands, respectively, as shown in Fig. \ref{one}. We can assign to $\Lambda_0 $ the value of half the bandwidth, so that $\Lambda_0 \sim 10$ meV. This leads to a critical temperature $T_c \sim 1$ K in both cases. However, the nature of the superconducting pairing is very different. Since the bands at $\nu=-2.4$ are nematic and spin-split, we predict Ising SC similarly to what happens in twisted trilayer graphene.\cite{Gonzalez23} For $\nu=2.3$, the bands are spin-degenerate and due to the odd pairing function with irrep $B_1$, the spinor-wave function needs to transform as a triplet state.

{\it Summary and Conclusions.}
We have discussed TBG structures under hydrostatic pressure that display flat bands and competition between different symmetry breaking patterns. Notably, for the normal state at 2-hole doping and $U=4$eV, we find a valley polarized state with nematicity for weak coupling as observed in Ref. \onlinecite{Cao21Nematic}. At 2-electron doping, we find a mixed intervalley coherent state with $T$-symmetric IVC components that give rise to a Kekul\'e charge-density wave order for strong coupling as observed in Refs. \onlinecite{Nuckolls23,Kwan23,Kim23}. At charge neutrality, we find a \cite{pure}{} K-IVC state that should be phase-disordered beyond the moir\'e scale.

For the superconducting phase, we expect SC only in the weak-coupling regime, as for the strong-coupling regime the Fermi line becomes more and more isotropic and thus no strong pairing-instability can develop. Moreover, we predict Ising SC for hole-doping and conventional triplet Cooper-pairing for electron doping in the weak coupling. This can be tested by induced spin-orbit coupling by proximity which would enhance the critical temperature only in the hole-doped regime as done in Bernal bilayer graphene.\cite{Zhang23,Holleis24}  

{\it Acknowledgement.}
The work was supported by grant PID2020-113164GBI00 funded by MCIN/AEI/10.13039/501100011033, PID2023-146461NB-I00 funded by Ministerio de Ciencia, Innovaci\'on y  Universidades as well as by the CSIC Research Platform on Quantum Technologies PTI-001. The access to computational resources of CESGA (Centro de Supercomputaci\'on de Galicia) is also gratefully acknowledged.

\newpage
\begin{widetext}
\begin{center}
{\Large\bf SUPPLEMENTAL MATERIAL}
\end{center}

\section{Microscopic real-space model}
\label{sec:model}
We consider commensurate twisted bilayer graphene (TBG) structures with a relative twist angle $\theta_i$, parametrized by an integer $i$ with the formula \cite{Lopes_dos_Santos_2007}: 
\begin{equation}
\cos(\theta_i)= \frac{3i^2 + 3i +\frac{1}{2}}{3i^2 + 3i +1}.
\end{equation}
The new basis vectors of the superlattice, in terms of the graphene basis $\a_1=a(1/2,\sqrt{3}/2)$, $\a_2=a(-1/2,\sqrt{3}/2)$ with $a\approx 0.246$nm being the graphene's lattice constant, are
\begin{align}
{\bf t}_1 = i \a_1 + (i +1) \a_2 \, \, \, \, \,    {\bf t}_2 = -(i+1) \a_1 + (2i +1) \a_2 , 
\end{align}
moreover, the number of sites inside a moiré unit cell is also given in terms of $i$
\begin{equation}
\label{ParticleNumber}
N= 4 \left( (i+1)^2 + i^2 + i(i+1) \right),
\end{equation}
and the new lattice constant can be calculated with
\begin{equation}
L_M = \sqrt{3i^2 +3i +1} a\;.
\end{equation}

To make contact to actual TBG experiments, the system  will be placed between two metallic gates, each at a distance $\xi/2$ to the bilayer. This additional consideration will be relevant when considering the effect of electron-electron interactions, since it will induce an external screening in the otherwise bare Coulomb potential. 

According to our previous remarks, the real first magic angle is then given by $i=31$ under ambient conditions. In our calculations we find that the bands become remarkably flat for $i=28$ ($\theta \sim 1.16$) within the non-interacting tight-binding model Hamiltonian $H_0$, defined in Section 2.1 . The reason for this discrepancy is the absence of in-plane relaxation in our model which notably alters the general shape of the bands at small twist angles \cite{Moon13}. For smaller twist angles, though, in-plane relaxation can be neglected and we will use $i=28$ as our benchmark for the first magic angle of equilibrium TBG. 

In this paper, we consider a TBG structure with $ i=9 $ ($ \theta_9 \sim 3.48^\circ $) with a modified reduced interlayer distance achieved by applying hydrostatic pressure of the GPa order (see \cite{jarillopressure} for numerical details on the required pressures) perpendicular to the plane of the crystal. In particular, we require a reduction of $d_{\perp,9}=\frac{1}{1.134} d_{\perp} ^0 $ with the equilibrium interlayer distance for of turbostratic graphene being  $d_\perp ^0=3.34$\r{A} to achieve flat bands near CNP similar to those of $i=28$.

For comparison, in Figure \ref{nointer} we show the band structures of $i=9 $ and $i=28$ calculated under a real space tight-binding scheme with no electron-electron interactions (see Eq. (6) in Sec. 2.1). It can be seen that in both cases, the bandwidth $ E_W $ in the $ \Gamma K $ high-symmetry line  is of the order of a few meV, and so we can expect electron correlations to play an important role due to the Coulomb potential being of the same order as the kinetic energy, i.e.  $ E_W \sim \frac{e^2}{4\pi\epsilon_0\epsilon L_M}$ with $\epsilon$ being the permittivity of the system and $ e $ the charge of the electron. Typical moire lattice constants for these structures are of the order of 13 nm. However, exerting pressure, the moire lattice scale may become considerably smaller of the order of 5 nm.

Let us finally note that we obtain almost identical bands and correlated insulator phases also for $i=12$ ($\theta_12 \sim 2.64^\circ $) with interlayer reduction $d_{\perp,12} =\frac{1}{1.095}d_\perp ^{0} $. Our results should might thus also closely resemble the phase diagram of magic angle bilayer graphene under ambient pressure. 

\begin{figure}[!h]
    \centering
    \includegraphics[scale=0.3]{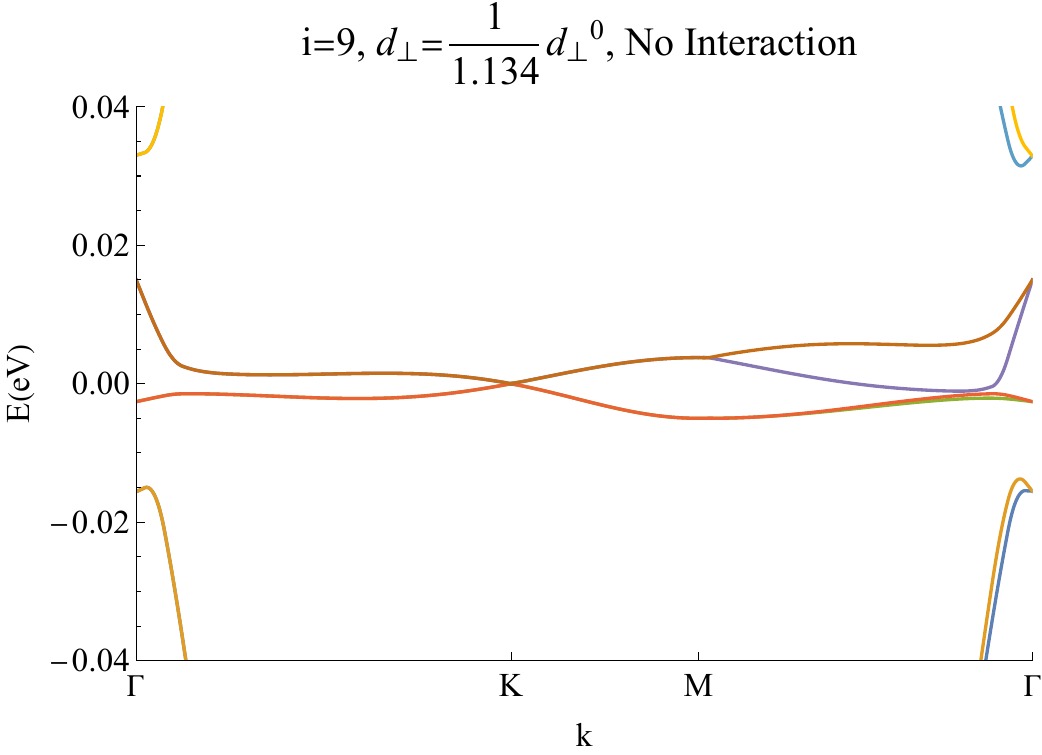}
    \includegraphics[scale=0.3]{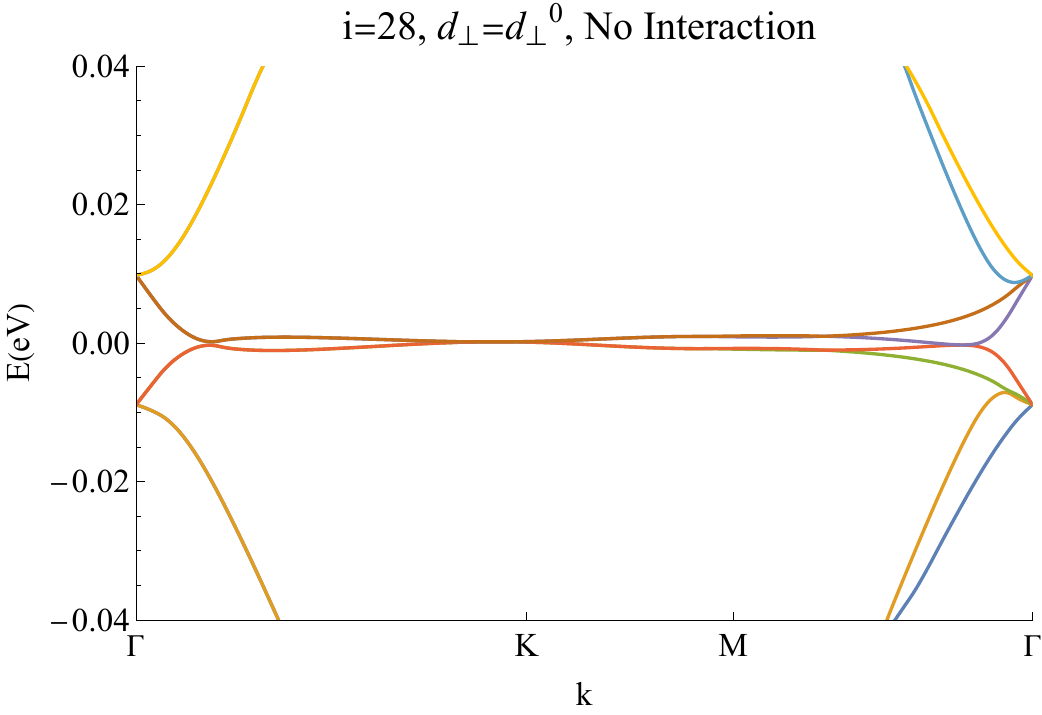}
    \caption{Low energy bands for $i=9$ and $i=28$ calculated using a free real space tight-binding model without $e^{-} $ - $e^{-} $ interaction. For $i=9$, a modified interlayer distance has been used. Both, the general shape and the flatness are remarkably similar in all cases. Here we take $d_{\perp}^0 = 3.34$ \AA, the equilibrium interlayer distance for AB stacking in TBG.  }
    \label{nointer}
\end{figure}

\subsection{Hartree-Fock theory}
\label{sec:hf}
In the main text, the general Hamiltonian $H$ was written as the sum of a non-interacting Hamiltonian $H_0$ and a term $H_{\rm int}$ containing the Hubbard and Coulomb interactions
\begin{align}
H = H_0 + H_{\rm int}\;.
\end{align}
The tight-binding Hamiltonian reads
\begin{align}
H_0=-\sum_{n,m}\sum_{i,j}\sum_{\sigma}t_{n,m}^{i,j}\psi^\dagger_{n,i,\sigma}\psi_{m,j,\sigma}
\end{align}
where the $n,m$ run over different unit cell, $i,j$ run over the different lattice sites within the unit cell and $\sigma$ denotes the spin projection. We assume that the hopping matrix element $t_{n,m}^{i,j}$ depends on the distance between lattice sites, i.e., $t_{n,m}^{i,j}=t(|\R_n-\R_m+\d_i-\d_j|)$ where $\R_n$ denotes the lattice vector of unit cell $n$ and $\d_i$ the position of lattice site $i$ with respect to the unit cell.  

The Coulomb interaction is given by 
\begin{align}  
H_{int} = \frac{1}{2} \sum_{n,m} \sum_{i,j}\sum_{\sigma,\sigma'} V_{n,m}^{i,j}\psi_{n,i,\sigma}^{\dagger}\psi_{m,j,\sigma'}^{\dagger}\psi_{m,j,\sigma'}\psi_{n,i,\sigma}\;.
\end{align}
Again, the interaction potential shall only depend on the distance between lattice sites, i.e., $V_{n,m}^{i,j}=V(|\R_n-\R_m+\d_i-\d_j|)$. We will further split the interaction term into a long-ranged Coulomb interaction and a short-ranged on-site Hubbard term to avoid singularities. We therefore have $H_{int} =H_V+H_U$ with
\begin{align}  
H_V&=\frac{1}{2} \sum_{n,m} \sum_{i,j}\sum_{\sigma,\sigma'} V_{n,m}^{i,j}\psi_{n,i,\sigma}^{\dagger}\psi_{m,j,\sigma'}^{\dagger}\psi_{m,j,\sigma'}\psi_{n,i,\sigma}\;,\\
H_U& =\frac{U}{2} \sum_{n,i}\sum_{\sigma}\psi_{n,i,\sigma}^{\dagger}\psi_{n,i,\bar\sigma}^{\dagger}\psi_{n,i,\bar\sigma}\psi_{n,i,\sigma} \;,
\end{align}
where $\bar\sigma$ denotes the opposite spin-projection to $\sigma$.

Again, the Coulomb potential shall only depend on the distance between lattice sites, $V_{n,m}^{i,j}= v(|\R_n - \R_m +\d_i-\d_j|) $, and is implemented by the double-gated potential
\begin{align}
v(|\rr|)=\frac{e^2}{4\pi \epsilon_0 \epsilon}\sum_{n} \frac{(-1)^n}{|\rr+ n\xi \boldsymbol{{z}}|} \xrightarrow[r \gg \xi]{} \frac{e^2}{4\pi\epsilon_0\epsilon} \frac{2\sqrt{2}e^{{-\pi r/\xi}}}{ \xi\sqrt{r/\xi}}\;,
    \label{potential}
 \end{align}
where $\epsilon$ stands for the intrinsic dielectric constant of the system.

The double-gated potential applies for the experimental setups where two metallic plates are placed at $z=\pm \xi/2$. For large distances, the interaction is thus effectively screened on a distance of $\xi/\pi$.\cite{Throckmorton12} We will choose $\xi=10$nm, a value consistent with several TBG experiments \cite{Saito_2020,Stepanov_2020} that needs to be compared to the moir\'e length $L_M\approx5$nm. For the exchange interaction (Fock), it then suffices to consider the central and the 19 surrounding moir\'e cells. For the direct interaction (Hartree), more than 100 surrounding moir\'e cells are included. Let us finally note that the final results do not significantly depend on $\xi$.

We can now perform a Fourier transform by
\begin{align}
\psi_{n,i,\sigma}=\frac{1}{\sqrt{N_c}}\sum_{\k\in1.BZ}\psi_{\k,i,\sigma}e^{i\k(\R_n+\zeta\d_i)}\;,
\end{align}
where $N_c$ is the number of unit cells. For sake of generality, we included an additional phase within the unit cell which is characterized by the position $\d_i$ of lattice site $i$ if we set $\zeta=1$. However, $\zeta$ can also be set to zero such there is the same phase for the whole unit cell.

We now define new variables by $\R_\ell=\R_n-\R_m$ and $\widetilde\R_p=\frac{1}{2}(\R_n+\R_m)$. The factor $\frac{1}{2}$ guarantees that the Jacobian of the mapping is norm-conserving and by choosing periodic boundary conditions, there are no finite size effects to take care of. We can thus replace $\sum_{n,m}\to\sum_{\ell,p}$ and since the hopping matrix element and the Coulomb potential only depend on $\R_\ell$, i.e., $t_{n,m}^{i,j}\to t_{\ell}^{i,j}$ and $V_{n,m}^{i,j}\to V_{\ell}^{i,j}=V(\R_\ell +\d_i-\d_j)$, we can use the following identity ($\widetilde\R_p=\frac{1}{2}\R_p$): 
\begin{align}
\frac{1}{N_c}\sum_pe^{-i\frac{\R_p}{2}(\k-\k')}=\delta_{\k,\k'+2\G}\;,
\end{align}
where $\G$ denotes a reciprocal lattice vector. Since the sum of the wave vectors is confined to the first Brillouin zone, there are no Umklapp processes to be taken care of. The interaction hamiltonian thus reads
\begin{align}
H_V&= \frac{1}{2}\sum_{\ell}\sum_{i,j}V_\ell^{i,j}\frac{1}{N_c}\sum_{\k,\k'}\sum_{\sigma,\sigma'}\psi^\dagger_{\k,i,\sigma}\psi_{\k',j,\sigma'}^\dagger\psi_{\k',j,\sigma'}\psi_{\k,i,\sigma}\;,\\
H_U&= \frac{U}{2}\sum_{\ell}\sum_{i} \frac{1}{N_c}\sum_{\k,\k'}\sum_{\sigma}\psi^\dagger_{\k,i,\sigma}\psi_{\k',i,\Bar{\sigma}}^\dagger\psi_{\k',i,\Bar{\sigma}}\psi_{\k,i,\sigma}\ .
\end{align}

The interacting system shall be treated within the restricted Hartree-Fock approach, i.e., we will only consider spin-symmetric solutions and the spin-quantum number $\sigma$ shall be suppressed from now on. The on-site interaction $H_U$ would then only lead to a constant energy shift if there was a homogeneous density distribution. However, already the non-interacting model shows strong localization around the AA-stacked regions and we also find sublattice polarization, each one being three-fold rotationally symmetric. In fact, the final results significantly depend on $U$ which can not easily be discussed within a continuum approach.

With the Hartree-Fock approximation, we thus have the following effective one-particle Hamilton operator $H^{HF}$$=H_0+H_V^{HF}+H_U^{HF}$ with:
\begin{align}
H_0&=-\sum_{\ell;i,j}t_\ell^{i,j}\sum_{\k,\sigma}\psi^\dagger_{\k,i,\sigma}\psi_{\k,j,\sigma}e^{-i\k(\R_\ell+\d_i-\d_j)}\\
H_V^{HF}&=\sum_{\ell}\sum_{i,j}V_\ell^{i,j}\left[\langle\mathcal{O}_H^j\rangle\mathcal{O}_H^i-\sum_{\sigma}\langle\mathcal{O}_F^{\ell,i,j,\sigma,\sigma}\rangle^*\mathcal{O}_F^{\ell,i,j,\sigma,\sigma}\right]\\
H_U^{HF}&=\frac{U}{N_c}\sum_{i}\sum_{\sigma}\psi_{\k,i,\sigma}^{\dagger}\psi_{\k,i,\sigma}\langle\psi_{\k,i,\bar\sigma}^{\dagger}\psi_{\k,i,\bar\sigma}\rangle.
\end{align}
In the Hartree-Fock decoupling we have assumed translation invariance, i.e, $\langle \psi^\dagger_{\k,i,\sigma}\psi_{\k',j,\sigma'} \rangle = 0$ if $\k \neq \k'$, and  constrained the spin structure to be diagonal, i.e., $\langle \psi^\dagger_{\k,i,\sigma}\psi_{\k,j,\Bar{\sigma}}\rangle = 0$. The operators $\mathcal{O}_H^i$ and $\mathcal{O}_F^{\ell,i,j,\sigma,\sigma'}$ are defined by
\begin{align}
\mathcal{O}_H^i&=\frac{1}{\sqrt{N_c}}\sum_{\k,\sigma}\psi^\dagger_{\k,i,\sigma}\psi_{\k,i,\sigma}\;,\\
\mathcal{O}_F^{\ell,i,j,\sigma,\sigma'}&=\frac{1}{\sqrt{N_c}}\sum_{\k}\psi^\dagger_{\k,i,\sigma}\psi_{\k,j,\sigma'}e^{-i\k(\R_\ell+\d_i-\d_j)}\;.
\end{align}

The Hartree-Fock equations are solved on the moir\'e Brillouin zone with a $12\times12$ grid. For each parameter set $(\nu,U,\epsilon)$, we usually perform 600 iterations in order to reach convergence; only close to the phase transition at $(\nu=0, U=0.5eV, \epsilon=15)$, 2000 iterations are necessary. Moreover, for each parameter set we perform the calculation for ten different initial conditions, see Sec. \ref{Roothaan}. This is important as the Hartree-Fock equations are non-linear and can lead to more than one stable solution. In fact, for ($\nu=-2$, $U=0.5$eV, $\epsilon=60$), ($\nu=-2$, $U=4$eV, $\epsilon=12,15,20,30,60$), ($\nu=2$, $U=0.5$eV, $\epsilon=15,20,30$), and ($\nu=2$, $U=4$eV, $\epsilon=15,20$), we find two different solutions/phases. Then, the solution with lowest ground-state energy $E_0$ is chosen which will give rise to a phase transition at ($\nu=-2$, $U=4$eV, $\epsilon\approx 20 $) and ($\nu=2$, $U=0.5$eV, $\epsilon\approx 20$).

The ground-state energy is given by 
\begin{align}
E_0&=E_{kin}+\frac{1}{2}E_{Hartree}+\frac{1}{2}E_{Fock}+\frac{1}{2}E_{U}\;.
\end{align}
where we defined the kinetic, Fock, and Hartree energy:
\begin{align}E_{kin}&=-\sum_{\ell;i,j}t_\ell^{i,j}\sum_{\k,\sigma}\langle\psi^\dagger_{\k,i,\sigma}\psi_{\k,j,\sigma}\rangle e^{-i\k(\R_\ell+\d_i-\d_j)}\\
E_{Fock}&=-\sum_{\ell;i,j} V^{i,j}_\ell \sum_{\sigma} |\langle\mathcal{O}_F^{\ell,i,j,\sigma,\sigma}\rangle|^2\\
E_{Hartree}&=\sum_{\ell; i,j} V_\ell^{i,j}\langle\mathcal{O}_H^i\rangle \langle\mathcal{O}_H^j\rangle \\
E_{U}&=\frac{U}{N_c}\sum_{i}\sum_{\k,\sigma}\langle\psi_{\k,i,\sigma}^{\dagger}\psi_{\k,i,\sigma}\rangle \langle\psi_{\k,i,\bar\sigma}^{\dagger}\psi_{\k,i,\bar\sigma}\rangle
\end{align}
In the following we will discuss the energy density per particle $E/A_M/N$, with $A_M=\frac{3}{2}a^2\sqrt{3i^2+3i+1}$ the area of the moir\'e lattice and $N$ the total number of electrons in the system, or simply the energy per particle $E/N$.

\subsection{Roothaan equations}
\label{Roothaan}
The self-consistent Hartree-Fock equations are also called Roothaan equations which can be written as a (generalized) eigenvalue problem
\begin{align}
F[\psi]\psi=\epsilon\psi
\end{align}
where $F$ denotes the so-called Fock matrix which depends on the wave functions $\psi$. $\psi$ shall denote the matrix of the eigenfunctions, and $\epsilon$ is the diagonal matrix of orbital energies. It is a set of nonlinear equation and the Fock matrix $F$ is actually an approximation of the true Hamiltonian operator of the quantum system, i.e., $F=H^{HF}\approx H$. It includes the effects of electron-electron repulsion only on an average level and because the Fock operator is a one-electron operator, it does not include the electron correlation energy.

This set of non-linear equations cannot be uniquely solved and might even possess several solutions. We will solve it iteratively by starting from a particular density distribution $n_i$ where the atomic positions $i=(\alpha,\ell)$ shall be parameterized by sublattice $\alpha=A,B$ and layer $\ell=1,2$. We thus have $n_i=n_0+\delta n_i$ where $n_0=1/2$ is the density of the neutral system and $\delta n_i=0.1\xi_i$. With $\xi_i=\mathcal{O}(1)$, we make sure that the system is pushed far from equilibrium. 

We consider four cases $\xi_{A,1}=\xi_{B,1}=\xi_{A,1}=\xi_{B,1}=1$, $\xi_{A,1}=-\xi_{B,1}=\xi_{A,1}=-\xi_{B,1}=1$, $\xi_{A,1}=-\xi_{B,1}=-\xi_{A,1}=\xi_{B,1}=1$, and $\xi_{A,1}=\xi_{B,1}=-\xi_{A,1}=-\xi_{B,1}=1$. We also consider the symmetric case with negative sign $n_i=n_0+0.2(0.5-\xi)$ and $\xi=1$. The initial condition can further be generalized by using arbitrary densities, but including the same sublattice and layer (im)balances, i.e., $\xi_i\in[0,1]$ is now a random number which gives another five initial conditions. 

After the first iteration, we adjust the chemical potential such that $\sum_i n_i=1/2+\nu/A_M$ with $A_M$ the area of the moir\'e unit cell. Depending on the filling factor and interaction strength, we may obtain more than one solution. We then choose the one with lower energy. This procedure also allows us to extract the energy difference between different phases.  

\section{Internal screening and effective dielectric constant}

One of the aims of our investigation is to provide a general analysis of the phases which may appear in twisted bilayer graphene, and which may apply to different screening conditions of the experimental samples. Yet, it is worthwhile to discuss the internal screening of the twisted bilayer, arising from particle-hole excitations of the electron system, and which is therefore independent of the screening by the external environment.

We can assess the magnitude of the internal screening by computing the contribution of the flat bands to the dielectric function $\epsilon ({\bf q}, \omega)$. In the 2D system, this is given in terms of the particle-hole susceptibility $\chi({\bf q}, \omega)$ by the expression 
\begin{align}
\epsilon ({\bf q}, \omega) = 1 + \frac{e^2}{2 \epsilon_0 |{\bf q}|} \chi({\bf q}, \omega)
\label{eps}
\end{align}
Provided this dielectric function has a smooth behavior, we can encode the internal screening into an effective dielectric constant $\epsilon $, obtained from the average of $\epsilon ({\bf q}, 0)$ over momenta ${\bf q}$ in the Brillouin zone of the twisted bilayer.

\iffalse
\begin{figure}[h]
\includegraphics[width=0.4\columnwidth]{bands004.pdf}

\caption{Low-energy bands of the interacting twisted bilayer with magic angle and interlayer distance as given in the main text, computed by means of a self-consistent Hartree-Fock approximation with on-site Hubbard repulsion $U = 4$ eV and $e^2/4\pi \epsilon = 0.18$ eV$\times a_0$ (where $a_0$ is the C-C distance). The dashed line stands for the Fermi level of the corresponding filling factor $\nu = -2$.}
\label{bands}
\end{figure}

In order to estimate the particle-hole susceptibility, it is convenient to consider a situation with a gap at the Fermi energy $\varepsilon_F$. We choose the case in which the phase of twisted bilayer graphene corresponds to the valley-polarized (nematic) phase described in the main text at $\nu = -2$ (see e.g. Fig. \ref{BandsU4.0}). Then a gap opens up between the three flat unoccupied bands ($i=1,2,3$) and the remaining flat occupied band ($j=4$). As already mention, we focus on internal screening from particle-hole excitations between these low-energy bands, which lead to partial contributions we can label respectively by $\chi_{14}, \chi_{24}, \chi_{34}$. The computation of these susceptibilities in the static limit gives the result shown in Fig. \ref{suscep}, along the $\Gamma M$ line of reflection symmetry of the twisted bilayer in the nematic phase.
\fi

In order to estimate the particle-hole susceptibility, it is convenient to consider a situation with a gap at the chemical potential. We further choose the parameters which correspond to the valley-polarized (nematic) phase described in the main text at $\nu = -2$ and with $U=4$eV and $\alpha=0.11$, a gap opens up between the three flat unoccupied bands ($i=1,2,3$) and the remaining flat occupied band ($j=4$). As already mention, we focus on internal screening from particle-hole excitations between these low-energy bands, which lead to partial contributions we can label respectively by $\chi_{14}, \chi_{24}, \chi_{34}$. The computation of these susceptibilities in the static limit gives the result shown on the left hand side of Fig. \ref{suscep}, along the $\Gamma M$ line of reflection symmetry of the twisted bilayer in the nematic phase.

\begin{figure}[h]
\includegraphics[width=0.3\columnwidth]{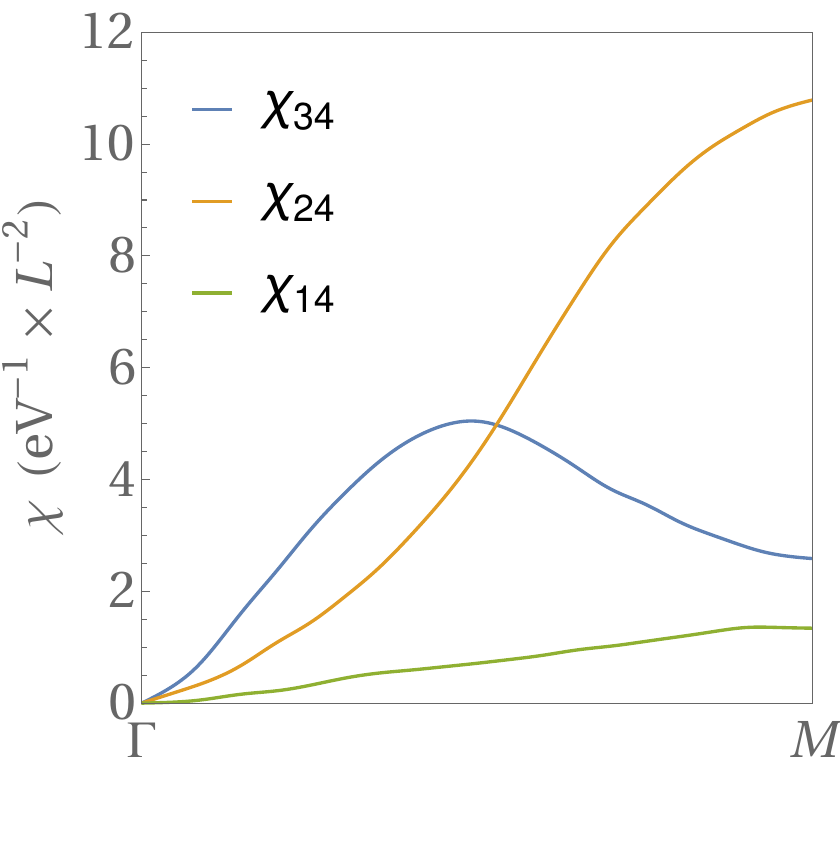}
\includegraphics[width=0.29\columnwidth]{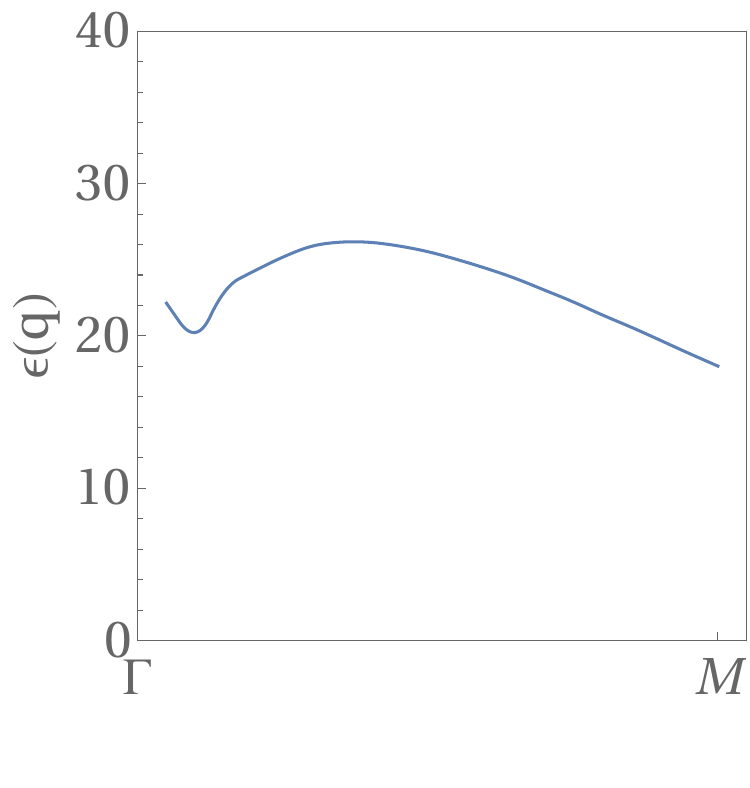}
\caption{Left: Particle-hole susceptibilities computed with the flat bands of the interacting theory represented in Fig. \ref{bands}. The plot shows the partial contributions $\chi_{ij}$ between the three flat unoccupied bands ($i=1,2,3$) and the flat occupied band ($j=4$). The susceptibilities are measured in units of eV$^{-1} \times L^{-2}$, where $L$ is the lattice constant of the moir\'e superlattice. Right: Dielectric function in the static limit computed from the flat bands of the interacting theory represented in Fig. \ref{bands}, along the $\Gamma M$ line of reflection symmetry of the twisted bilayer in the nematic phase.}
\label{suscep}
\end{figure}

From the particle-hole susceptibilities $\chi_{ij}$ and taking into account a factor 2 due to the spin degeneracy, we can compute the contribution of the flat bands to the dielectric function in Eq. (\ref{eps}) in the static limit. Such a function is represented on the right hand side of Fig. \ref{suscep}, where we observe that the behavior is smooth enough to extract a suitable lower bound for the effective dielectric constant. We can assign the estimate $\epsilon \gtrsim 20$, bearing in mind that this is just from the contribution of the flat bands.

%\begin{figure}[h]
%\includegraphics[width=0.3\columnwidth]{dielec004.pdf}
%
%\caption{Dielectric function in the static limit computed from the flat bands of the interacting theory represented in Fig. \ref{bands}, along the $\Gamma M$ line of reflection symmetry of the twisted bilayer in the nematic phase.}
%\label{dielec}
%\end{figure}

The remote bands are very large in number and should, therefore, sensibly increase the value of $\epsilon $. We can make a very rough estimate of their contribution up to not very large energies (of order of $\sim 1$ eV) by considering the particle-hole susceptibility of a graphene compound with a multiplicity $N=8$ accounting for spin, valley, and layer degrees of freedom. Taking $e^2/\epsilon_0 \approx 17.7$ eV nm and the Fermi velocity of graphene $\hbar v_F \approx 0.6$ eV nm, we get the order of magnitude of the contribution in Eq. (\ref{eps}) from remote bands
\begin{align}
\Delta \epsilon \sim \frac{e^2}{2 \epsilon_0 |{\bf q}|} \frac{N |{\bf q}|}{16 \hbar v_F} \sim  7
\end{align}  
which should be added to the larger and more precise estimate from the flat bands.

\section{$SU(4)$-reduced single-particle density matrix}
From the ansatz for the single-particle wave function $|\psi\rangle=\sum_i\psi_i|\rr_i\rangle$,
\begin{align}
|\psi\rangle&=\sum_{\ell,\gamma}\sum_{\rr_i\in \gamma, \ell}\!\!\!\left[e^{i\K_\ell\cdot\rr_i}f_{\gamma,K,\ell}(\rr_i)+e^{i\K_\ell'\cdot\rr_i}f_{\gamma,K',\ell}(\rr_i)\right]|\rr_i\rangle\;,
\end{align}
where $\gamma=A,B$ stands for the sublattice and $\ell=1,2$ the two layers, we can define the {\it $SU(4)$-reduced} single-particle density matrix
\begin{align}
\label{ReducedDM}
\rho_{\gamma,\delta;\gamma',\delta'}=\int_{A_M}d\rr \sum_\ell f_{\gamma,\delta,\ell}^*(\rr)f_{\gamma',\delta',\ell}(\rr)\;,
\end{align}
where $\delta=K,K'$ denotes the valley degree of freedom. Notice that we perform the integral over the "coarse-grained" position vectors $\rr$ which are inside the moir\'e unit cell $A_M$ and sum over  both layers. The underlying model is, of course, still a discrete lattice model, and the precise definition of the integral/sum can be found below. Also notice that starting from the usual definition $\rho=|f\rangle\langle f|$ with $|f\rangle$ denoting a four-component spinor, we will actually discuss the Hermitian conjugate of the $SU(4)$-reduced single-particle density matrix which does, though, not change any of our conclusions.

We can decompose $\rho$ into Hermitian components, $\rho_{\gamma,\delta;\gamma',\delta'} = \sum_{\alpha \beta} \rho_{\alpha,\beta} [\sigma_\alpha \tau_\beta]_{\gamma, \delta; \gamma' \delta'} = \sum_{\alpha \beta} \langle\sigma_\alpha\tau_\beta \rangle [\sigma_\alpha \tau_\beta]_{\gamma, \delta; \gamma' \delta'} $ where $\sigma_\alpha, \tau_\beta$ denote the Pauli matrices for sublattice and valley degree of freedom including the unity matrix with $\alpha,\beta=0,x,y,z$. The expectation values of the sixteen generators that define the $SU(4)$ symmetry group can be grouped into four categories. 

The first one describes intra-sublattice and intra-valley terms. They read
\begin{align}
\langle1\rangle&=\rho_{AK,AK}+\rho_{BK,BK}+\rho_{AK',AK'}+\rho_{BK',BK'}\;,\\
\langle\sigma_z\rangle&=\rho_{AK,AK}-\rho_{BK,BK}+\rho_{AK',AK'}-\rho_{BK',BK'}\;,\\
\langle\tau_z\rangle&=\rho_{AK,AK}+\rho_{BK,BK}-\rho_{AK',AK'}-\rho_{BK',BK'}\;,\\
\langle\sigma_z\tau_z\rangle&=\rho_{AK,AK}-\rho_{BK,BK}-\rho_{AK',AK'}+\rho_{BK',BK'}\;.
\end{align}
The second group describes intra-sublattice and inter-valley terms. They read
\begin{align}
\langle\sigma_x\rangle&=2\Re(\rho_{AK,BK}+\rho_{AK',BK'})\;,\\
\langle\sigma_y\rangle&=2\Im(\rho_{AK,BK}+\rho_{AK',BK'})\;,\\
\langle\sigma_x\tau_z\rangle&=2\Re(\rho_{AK,BK}-\rho_{AK',BK'})\;,\\
\langle\sigma_y\tau_z\rangle&=2\Im(\rho_{AK,BK}-\rho_{AK',BK'})\;.
\end{align}
The third group describes inter-sublattice and intra-valley terms. They read
\begin{align}
\langle\tau_x\rangle&=2\Re(\rho_{AK,AK'}+\rho_{BK,BK'})\;,\\
\langle\tau_y\rangle&=2\Im(\rho_{AK,AK'}+\rho_{BK,BK'})\;,\\
\langle\sigma_z\tau_x\rangle&=2\Re(\rho_{AK,AK'}-\rho_{BK,BK'})\;,\\
\langle\sigma_z\tau_y\rangle&=2\Im(\rho_{AK,AK'}-\rho_{BK,BK'})\;.
\end{align}
The last group describes inter-sublattice and inter-valley terms. They read
\begin{align}
\langle\sigma_x\tau_x\rangle&=2\Re(\rho_{AK,BK'}+\rho_{BK,AK'})\;,\\
\langle\sigma_x\tau_y\rangle&=2\Im(\rho_{AK,BK'}+\rho_{BK,AK'})\;,\\
\langle\sigma_y\tau_x\rangle&=2\Im(\rho_{AK,BK'}-\rho_{BK,AK'})\;,\\
\langle\sigma_y\tau_y\rangle&=-2\Re(\rho_{AK,BK'}-\rho_{BK,AK'})\;.
\end{align}

In the main text, we discuss the order parameters related to valley polarization $\langle\tau_z\rangle$, valley Hall effect $\langle\sigma_z\rangle$, and quantum Hall effect $\langle\sigma_z\tau_z\rangle$. Moreover, we discuss the order parameter for Kramers intervalley coherence $|\langle\sigma_x\tau_{x/y}\rangle|\equiv\sqrt{\langle\sigma_x\tau_x\rangle^2+\langle\sigma_x\tau_y\rangle^2}$ and time-reversal invariant intervalley coherence $|\langle\sigma_y\tau_{x/y}\rangle|\equiv\sqrt{\langle\sigma_y\tau_x\rangle^2+\langle\sigma_y\tau_y\rangle^2}$.
\section{Order parameters for a non-interacting lattice model}
\begin{figure}[!h]
    \centering
    \includegraphics[scale=0.35]{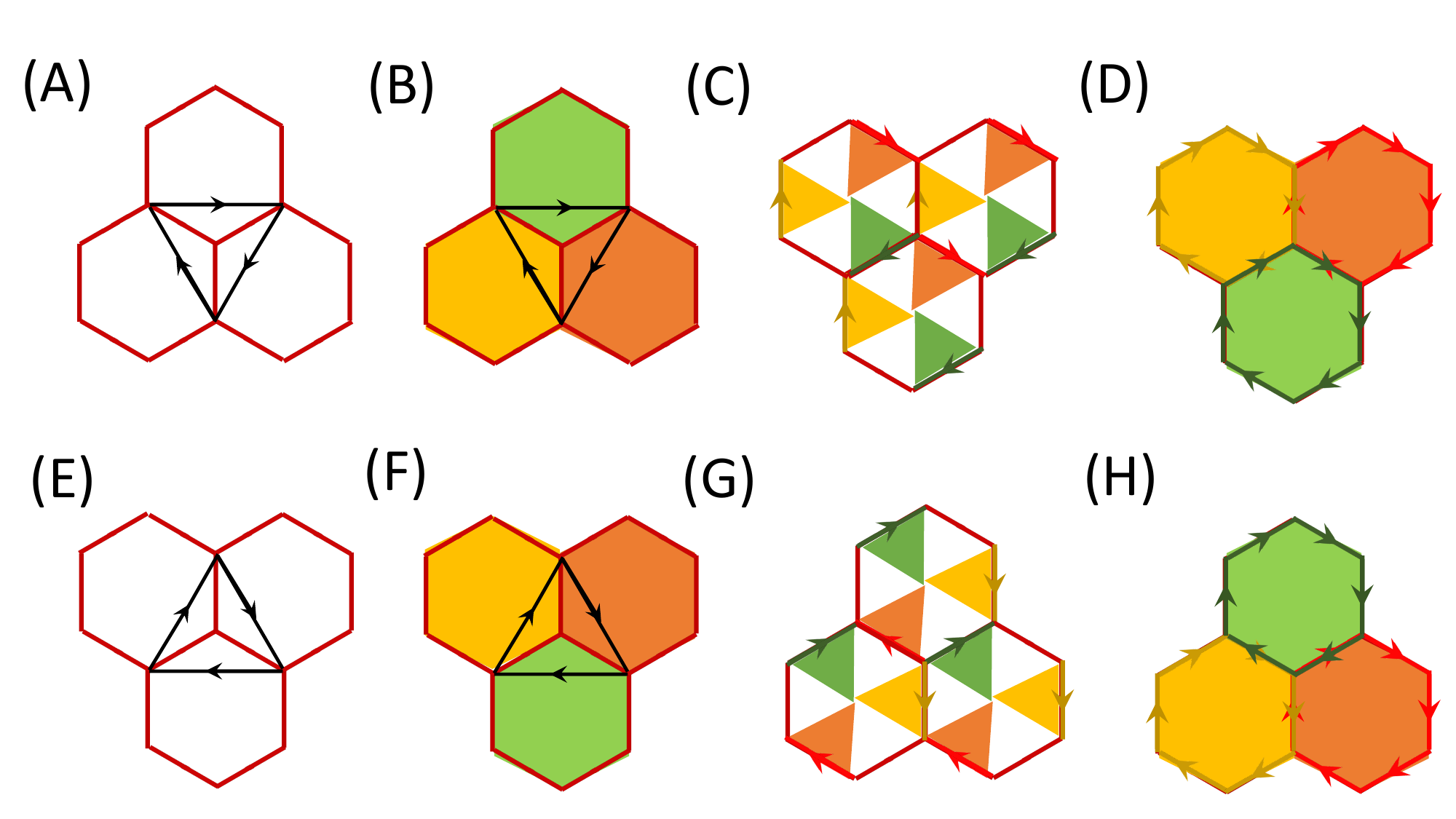}
    \caption{Triangular and hexagonal loops on the lattice in order to determine the valley order parameters. (A), (B) show the triangular loop with the central atom belonging to  the A-sublattice, yielding intra-sublattice contributions; (C) and (D) show the three hexagonal loops with the central atom belonging to the A-sublattice, yielding inter-sublattice contributions. The different colors stand for additional phases, e.g., $e^{i2(n-1)\pi/3}$ with $n=0$ (green), $n=1$ (red), and $n=2$ (yellow). The corresponding loops of (E), (F), (G), and (H) are related to the B-sublattice.}
    \label{flux}
\end{figure}

We will now define the 16 order parameters related to the $SU(4)$-symmetry group of the sublattice and valley degrees of freedom that define the $SU(4)$-reduced single-particle density matrix. For this, we have to fix the geometry of single layer graphene which we repeat here to ease readability. To simplify the discussion, we will further suppress the layer index $\ell$.

The hexagonal lattice shall be described by two basis vectors $\a_1=a_0(\sqrt{3}/2,3/2)$, $\a_2=a_0(-\sqrt{3}/2,3/2)$ with $a_0=1.42$\r{A}. The nearest neighbour sites are defined by ${\bm \delta}_1=a_0(\sqrt{3}/2,-1/2)$, ${\bm \delta}_2=a_0(0,1)$, ${\bm \delta}_0={\bm \delta}_3=a_0(-\sqrt{3}/2,-1/2)$. Also note that $\a_1=\d_2-\d_3$ and $\a_2=\d_2-\d_1$. The Brillouin zone is spanned by $\b_1=\frac{2\pi}{3a_0}(\sqrt{3},1)$ and $\b_2=\frac{2\pi}{3a_0}(-\sqrt{3},1)$ that defines the two $K$-points $\K=(\frac{4\pi}{3\sqrt{3}a_0},0)$ and $\K'=-\K$. 

In order to derive the long-wavelength coefficients, we will frequently use $e^{\pm i\K\cdot\a}=e^{\pm i\frac{2\pi}{3}}$ with $\a=\a_1,\a_2-\a_1,-\a_2$ and $e^{\pm i\K\cdot\d_1}+e^{\pm i\K\cdot\d_2}+e^{\pm i\K\cdot\d_3}=0$ which can also be written as $\sum_{n=0,1,2}e^{in\frac{2\pi}{3}}=0$. Note that this is precisely the condition for the appearance of the Dirac cone in single-layer graphene. 

In the long-wavelength limit, the wave function can be separated into a rapidly oscillating part and a slowly oscillating envelope function. A general wave function $|\psi\rangle=\sum_{\rr_i}\psi(\rr_i)|\rr_i\rangle$ contains contributions from both valleys and our tight-binding model differentiates between A- and B-sublattice. We thus make an ansatz which discriminates sublattice as well as valley degree of freedom,
\begin{align}
|\psi\rangle&=\sum_{\rr_i\in A}\left[e^{i\K\cdot\rr_i}f_{AK}^{}(\rr_i)+e^{-i\K\cdot\rr_i}f_{AK'}^{}(\rr_i)\right]|\rr_i\rangle
+\sum_{\rr_i\in B}\left[e^{i\K\cdot\rr_i}f_{BK}^{}(\rr_i)+e^{-i\K\cdot\rr_i}f_{BK'}^{}(\rr_i)\right]|\rr_i\rangle\;.
\label{AnsatzWFAB}
\end{align}
The envelope function $f_{\gamma,\delta}$ shall be smooth on the moir\'e scale $L_M$, and
we will aproximate
\begin{align}
f_{\gamma,\delta}(\rr+\a)=f_{\gamma,\delta}(\rr)+\mathcal{O}(a/L_M)\;.
\end{align}
We can now perform suitable loops on the lattice that will yield the components of the single-particle density matrix due to constructive and destructive interference, see Fig. \ref{flux}. These components can be grouped into four categories and involve overlap functions of the form $\psi^*(\rr)\psi(\rr+\a)$. They are a property of the initial wave function $|\psi\rangle$ and the expression is thus gauge-invariant since the global phase of $|\psi\rangle$ drops out. 

In the following, we will divide the honeycomb lattice into three adjacent hexagons with central site $\rr_i^\Delta$ belonging to the $A$- and $B$-lattices, respectively. In fact, the loops generally depend on the central site  due to the spacial dependence of the envelope function. However, some of the expressions involving inter-valley scattering processes will also depend on the space-dependent phase factors $e^{\pm i2\K\cdot\rr_i}$. In order to also uniformly discuss these terms, we will arrange the moire lattice such that on both layers there is an $A$-atom at $\rr_i=0$. With $\rr_{i(n_1^A,n_2^A)}^\Delta=n_1^A\a_1+n_2^A\a_2$, the central site on the $A$-lattice is given by $n_1^A=n_2^A=m$ and $n_1^A=-n_2^A=3m$ with $m\in\mathbb{Z}$. With $\rr_{i(n_1^B,n_2^B)}^\Delta=n_1^B\a_1+n_2^B\a_2+\d_1$, the central site on the $B$-lattice is given by $n_1^B=n_1^A+1$ and $n_2^B=n_2^A-1$. The space-dependent phase factors will thus be one for all cases, $e^{\pm i2\K\cdot\rr_i}=1$.

Let us finally comment on the ambiguity of choosing the point of origin. For this, consider that $\rr_i^\Delta=n_1\a_1+n_2\a_2+\d$ which gives rise to an additional phase $e^{\pm i2\K\cdot\d}$. The order parameters that depend on the space-dependent phase factors $e^{\pm i2\K\cdot\rr_i}$ are thus only defined up to a global phase that is precisely the $U(1)$ valley phase of TBG. Also, note that $\d=\pm\a_1$ amounts to a cyclic permutation of the sublattice labels.

\subsection{Intra-sublattice, intra-valley channel}
To define the order parameters of the intra-sublattice, intra-valley channel within the tight-binding model, we define $\Delta_{++}$ that is characterized by the wave function overlap of adjacent atoms of the same sublattice that form a triangle,
\begin{align}
\Delta_{++}^{}(\rr_i)=\sum_{n=0,1,2}\psi^*(\rr_i\pm\d_{n})\psi(\rr_i\pm\d_{n+1})\;,
\end{align}
where the upper sign refers to $\rr_i\in A$, the lower sign to $\rr_i\in B$. Inserting the ansatz of Eq. (\ref{AnsatzWFAB}) in the above equation, we obtain for $\rr_i\in A$:
\begin{align}
\Delta_{++}^{}(\rr_i)&=\left[e^{-i\K\cdot(\rr_i+\d_0)}f_{BK}^*(\rr_i)+e^{i\K\cdot(\rr_i+\d_0)}f_{BK'}^*(\rr_i)\right]\left[e^{i\K\cdot(\rr_i+\d_1)}f_{BK}^{}(\rr_i)+e^{-i\K\cdot(\rr_i+\d_1)}f_{BK'}^{}(\rr_i)\right]\notag\\
&+\left[e^{-i\K\cdot(\rr_i+\d_1)}f_{BK}^*(\rr_i)+e^{i\K\cdot(\rr_i+\d_1)}f_{BK'}^*(\rr_i)\right]\left[e^{i\K\cdot(\rr_i+\d_2)}f_{BK}^{}(\rr_i)+e^{-i\K\cdot(\rr_i+\d_2)}f_{BK'}^{}(\rr_i)\right]\notag\\
&+\left[e^{-i\K\cdot(\rr_i+\d_2)}f_{BK}^*(\rr_i)+e^{i\K\cdot(\rr_i+\d_2)}f_{BK'}^*(\rr_i)\right]\left[e^{i\K\cdot(\rr_i+\d_3)}f_{BK}^{}(\rr_i)+e^{-i\K\cdot(\rr_i+\d_3)}f_{BK'}^{}(\rr_i)\right]\;.
\end{align}
This can be grouped together to read
\begin{align}
\Delta_{++}^{}(\rr_i)&=
\left[e^{i\K\cdot(\d_1-\d_0)}+e^{i\K\cdot(\d_2-\d_1)} +e^{i\K\cdot(\d_3-\d_2)}\right]f_{BK}^*(\rr_i)f_{BK}^{}(\rr_i)\notag\\
&+\left[e^{-i\K\cdot(\d_1-\d_0)}+e^{-i\K\cdot(\d_2-\d_1)} +e^{-i\K\cdot(\d_3-\d_2)}\right]f_{BK'}^*(\rr_i)f_{BK'}^{}(\rr_i)\notag\\
&+e^{-i2\K\cdot\rr_i}\left[e^{-i\K\cdot(\d_1+\d_0)}+e^{-i\K\cdot(\d_2+\d_1)} +e^{-i\K\cdot(\d_3+\d_2)}\right]f_{BK}^*(\rr_i)f_{BK'}^{}(\rr_i)\notag\\
&+e^{i2\K\cdot\rr_i}\left[e^{i\K\cdot(\d_1+\d_0)}+e^{i\K\cdot(\d_2+\d_1)} +e^{i\K\cdot(\d_3+\d_2)}\right]f_{BK'}^*(\rr_i)f_{BK}^{}(\rr_i)\;.
\end{align}
We now use the fact that differences between nearest-neighbour vectors are related to lattice vectors $\a_i$ and that sums of nearest-neighbour vectors are related to negative nearest-neighbour vectors $\d_i$. The last two lines of the above equation is thus zero due to the "Dirac-cone condition" $e^{\pm i\K\cdot\d_1}+e^{\pm i\K\cdot\d_2}+e^{\pm i\K\cdot\d_3}=0$ and we get with $\alpha=e^{i\frac{2\pi}{3}}$
\begin{align}
\Delta_{++}^{}(\rr_i)=3\alpha^*f_{BK}^*(\rr_i)f_{BK}^{}(\rr_i)+3\alpha f_{BK'}^*(\rr_i)f_{BK'}^{}(\rr_i)\;.
\end{align}
This yields the expression for the components of the single-particle density matrix:
\begin{align}
f_{BK}^*(\rr_i)f_{BK}^{}(\rr_i)&=\frac{1}{3}\left(-\Re\Delta_{++}^{}(\rr_i)-\frac{1}{\sqrt{3}}\Im\Delta_{++}^{}(\rr_i)\right)\\
f_{BK'}^*(\rr_i)f_{BK'}^{}(\rr_i)&=\frac{1}{3}\left(-\Re\Delta_{++}^{}(\rr_i)+\frac{1}{\sqrt{3}}\Im\Delta_{++}^{}(\rr_i)\right)\;.
\end{align}
Equivalent formulas are obtained for  $\rr_i\in B$: 
\begin{align}
\Delta_{++}^{}(\rr_i)=3\alpha f_{AK}^*(\rr_i)f_{AK}^{}(\rr_i)+3\alpha^* f_{AK'}^*(\rr_i)f_{AK'}^{}(\rr_i)\;.
\end{align}
This yields the expression for the components of the single-particle density matrix:
\begin{align}
f_{AK}^*(\rr_i)f_{AK}^{}(\rr_i)&=\frac{1}{3}\left(-\Re\Delta_{++}^{}(\rr_i)+\frac{1}{\sqrt{3}}\Im\Delta_{++}^{}(\rr_i)\right)\\
f_{AK'}^*(\rr_i)f_{AK'}^{}(\rr_i)&=\frac{1}{3}\left(-\Re\Delta_{++}^{}(\rr_i)-\frac{1}{\sqrt{3}}\Im\Delta_{++}^{}(\rr_i)\right)\;.
\end{align}

As indicated in Fig. \ref{flux}, we will only sum over the center of the Kekul\'e unit cell indicated by $\rr_i^\Delta$ which takes care of the factor $1/3$. Summing over all sites of the moir\'e unit cell, we then have
\begin{align}
\rho_{AK,AK}&=\sum_{\rr_i^\Delta\in B}\left(-\Re\Delta_{++}^{}(\rr_i)+\frac{1}{\sqrt{3}}\Im\Delta_{++}^{}(\rr_i)\right)\;,\\
\rho_{AK',AK'}&=\sum_{\rr_i^\Delta\in B}\left(-\Re\Delta_{++}^{}(\rr_i)-\frac{1}{\sqrt{3}}\Im\Delta_{++}^{}(\rr_i)\right)\;,\\
\rho_{BK,BK}&=\sum_{\rr_i^\Delta\in A}\left(-\Re\Delta_{++}^{}(\rr_i)-\frac{1}{\sqrt{3}}\Im\Delta_{++}^{}(\rr_i)\right)\;,\\
\rho_{BK',BK'}&=\sum_{\rr_i^\Delta\in A}\left(-\Re\Delta_{++}^{}(\rr_i)+\frac{1}{\sqrt{3}}\Im\Delta_{++}^{}(\rr_i)\right)\;.
\end{align}
\subsection{Intra-sublattice, inter-valley channel}
In order to discuss the intra-sublattice, inter-valley channels within the tight-binding model, we need to consider the triangular loop dressed by three phases $\alpha^{n-1}$ with $\alpha=e^{i2\pi/3}$ and $n=0,1,2$. We thus define
\begin{align}
\Delta_{+-}^{}(\rr_i)=\sum_{n=0,1,2}\alpha^{n-1}\psi^*(\rr_i\pm\d_{n})\psi(\rr_i\pm\d_{n+1})\;,
\end{align}
where again the upper sign refers to $\rr_i\in A$, the lower sign to $\rr_i\in B$. 

Inserting the ansatz of Eq. (\ref{AnsatzWFAB}) in the above equation, we obtain for $\rr_i\in A$
\begin{align}
\Delta_{+-}^{}(\rr_i)&=\alpha^*\left[e^{-i\K\cdot(\rr_i+\d_0)}f_{BK}^*(\rr_i)+e^{i\K\cdot(\rr_i+\d_0)}f_{BK'}^*(\rr_i)\right]\left[e^{i\K\cdot(\rr_i+\d_1)}f_{BK}^{}(\rr_i)+e^{-i\K\cdot(\rr_i+\d_1)}f_{BK'}^{}(\rr_i)\right]\notag\\
&+\left[e^{-i\K\cdot(\rr_i+\d_1)}f_{BK}^*(\rr_i)+e^{i\K\cdot(\rr_i+\d_1)}f_{BK'}^*(\rr_i)\right]\left[e^{i\K\cdot(\rr_i+\d_2)}f_{BK}^{}(\rr_i)+e^{-i\K\cdot(\rr_i+\d_2)}f_{BK'}^{}(\rr_i)\right]\notag\\
&+\alpha\left[e^{-i\K\cdot(\rr_i+\d_2)}f_{BK}^*(\rr_i)+e^{i\K\cdot(\rr_i+\d_2)}f_{BK'}^*(\rr_i)\right]\left[e^{i\K\cdot(\rr_i+\d_3)}f_{BK}^{}(\rr_i)+e^{-i\K\cdot(\rr_i+\d_3)}f_{BK'}^{}(\rr_i)\right]\;.
\end{align}
This can be grouped together to read
\begin{align}
\Delta_{+-}^{}(\rr_i)&=
\left[\alpha^*e^{i\K\cdot(\d_1-\d_0)}+e^{i\K\cdot(\d_2-\d_1)} +\alpha e^{i\K\cdot(\d_3-\d_2)}\right]f_{BK}^*(\rr_i)f_{BK}^{}(\rr_i)\notag\\
&+\left[\alpha^*e^{-i\K\cdot(\d_1-\d_0)}+e^{-i\K\cdot(\d_2-\d_1)} +\alpha e^{-i\K\cdot(\d_3-\d_2)}\right]f_{BK'}^*(\rr_i)f_{BK'}^{}(\rr_i)\notag\\
&+e^{-i2\K\cdot\rr_i}\left[\alpha^*e^{-i\K\cdot(\d_1+\d_0)}+e^{-i\K\cdot(\d_2+\d_1)} +\alpha e^{-i\K\cdot(\d_3+\d_2)}\right]f_{BK}^*(\rr_i)f_{BK'}^{}(\rr_i)\notag\\
&+e^{i2\K\cdot\rr_i}\left[\alpha^*e^{i\K\cdot(\d_1+\d_0)}+e^{i\K\cdot(\d_2+\d_1)} +\alpha e^{i\K\cdot(\d_3+\d_2)}\right]f_{BK}^*(\rr_i)f_{BK'}^{}(\rr_i)\;.
\end{align}
Now, the first two lines of the above equation become zero due to the "Dirac-cone condition" and also the forth line for the same reason. We thus simply have
\begin{align}
\Delta_{+-}^{}(\rr_i)&=3\alpha^*e^{-i2\K\cdot\rr_i}f_{BK}^*(\rr_i)f_{BK'}^{}(\rr_i).
\end{align}
For $\rr_i\in B$, we get analogously
\begin{align}
\Delta_{+-}^{}(\rr_i)&=3\alpha^*e^{i2\K\cdot\rr_i}f_{AK'}^*(\rr_i)f_{AK}^{}(\rr_i)\;.
\end{align}
As shown above, the space-dependent phase factor is always one, i.e., $e^{\pm i2\K\cdot\rr_i}=1$. Summing only over the central sites then yields
\begin{align}
\rho_{AK',AK}&=\rho_{AK,AK'}^*=\alpha\sum_{\rr_i^\Delta\in B}\Delta_{+-}^{}(\rr_i)\;,\notag\\
\rho_{BK,BK'}&=\rho_{BK',BK}^*=\alpha\sum_{\rr_i^\Delta\in A}\Delta_{+-}^{}(\rr_i)\;.
\end{align}
\subsection{Inter-sublattice, intra-valley channel}
To define the order parameters for the inter-sublattice, intra-valley channels based on a tight-binding model, we now define the following quantities related to the three overlap functions around a hexagon:
\begin{align}
\Delta_{-+}^{1,n}(\rr_i)&=\psi^*(\rr_i\pm\d_1)\psi(\rr_i)+\alpha^n\psi^*(\rr_i\mp\a_2\pm\d_3)\psi(\rr_i\mp\a_2)+\alpha^{-n}\psi^*(\rr_i\mp\a_1\pm\d_2)\psi(\rr_i\mp\a_1)\;,\\
\Delta_{-+}^{2,n}(\rr_i)&=\alpha^{-n}\psi^*(\rr_i\pm\d_2)\psi(\rr_i)+\psi^*(\rr_i\pm\a_1\pm\d_1)\psi(\rr_i\pm\a_1)+\alpha^{n}\psi^*(\rr_i\pm\a_1\mp\a_2\pm\d_3)\psi(\rr_i\pm\a_1\mp\a_2)\;,\\
\Delta_{-+}^{3,n}(\rr_i)&=\alpha^{n}\psi^*(\rr_i\pm\d_3)\psi(\rr_i)+\alpha^{-n}\psi^*(\rr_i\mp\a_1\pm\a_2\pm\d_2)\psi(\rr_i\mp\a_1\pm\a_2)+\psi^*(\rr_i\pm\a_2\pm\d_1)\psi(\rr_i\pm\a_2)\;,
\end{align}
where again the upper sign refers to $\rr_i\in A$, the lower sign to $\rr_i\in B$ and $\alpha=\exp(i\frac{2\pi}{3})$ with $n=0,1,2$. 

Inserting the ansatz of Eq. (\ref{AnsatzWFAB}) in the above equation for $\Delta_{+-}^{1,n}$, we obtain for $\rr_i\in A$
\begin{align}
&\Delta_{-+}^{1,n}(\rr_i)=\left[e^{-i\K\cdot(\rr_i+\d_1)}f_{BK}^*(\rr_i)+e^{i\K\cdot(\rr_i+\d_1)}f_{BK'}^*(\rr_i)\right]\left[e^{i\K\cdot\rr_i}f_{AK}^{}(\rr_i)+e^{-i\K\cdot\rr_i}f_{AK'}^{}(\rr_i)\right]\notag\\
&+\alpha^{n}\left[e^{-i\K\cdot(\rr_i-\a_2+\d_3)}f_{BK}^*(\rr_i)+e^{i\K\cdot(\rr_i-\a_2+\d_3)}f_{BK'}^*(\rr_i)\right]\left[e^{i\K\cdot(\rr_i-\a_2)}f_{AK}^{}(\rr_i)+e^{-i\K\cdot(\rr_i-\a_2)}f_{AK'}^{}(\rr_i)\right]\notag\\
&+\alpha^{-n}\left[e^{-i\K\cdot(\rr_i-\a_1+\d_2)}f_{BK}^*(\rr_i)+e^{i\K\cdot(\rr_i-\a_1+\d_2)}f_{BK'}^*(\rr_i)\right]\left[e^{i\K\cdot(\rr_i-\a_1)}f_{AK}^{}(\rr_i)+e^{-i\K\cdot(\rr_i-\a_1)}f_{AK'}^{}(\rr_i)\right]\;.
\end{align}
This can be grouped together to read
\begin{align}
\Delta_{-+}^{1,n}(\rr_i)&=
\left[e^{-i\K\cdot\d_1}+\alpha^{n}e^{-i\K\cdot\d_3} +\alpha^{-n}e^{-i\K\cdot\d_2}\right]f_{BK}^*(\rr_i)f_{AK}^{}(\rr_i)\notag\\
&+\left[e^{i\K\cdot\d_1}+\alpha^{n}e^{i\K\cdot\d_3} +\alpha^{-n}e^{i\K\cdot\d_2}\right]f_{BK'}^*(\rr_i)f_{AK'}^{}(\rr_i)\notag\\
&+e^{-i2\K\cdot\rr_i}\left[e^{-i\K\cdot\d_1}+\alpha^{n}e^{i\K\cdot(2\a_2-\d_3)}+\alpha^{-n}e^{i\K\cdot(2\a_1-\d_2)}\right]f_{BK}^*(\rr_i)f_{AK'}^{}(\rr_i)\notag\\
&+e^{i2\K\cdot\rr_i}\left[e^{i\K\cdot\d_1}+\alpha^{n}e^{-i\K\cdot(2\a_2-\d_3)}+\alpha^{-n}e^{-i\K\cdot(2\a_1-\d_2)}\right]f_{BK'}^*(\rr_i)f_{AK}^{}(\rr_i)\;.
\end{align}
For $n=0$, the first and second line vanish and for $n=1$ and $n=2$, only the first and second line contribute, respectively. This yields
\begin{align}
\Delta_{-+}^{1,n=0}(\rr_i)&=3\alpha^*e^{-i2\K\cdot\rr_i}f_{BK}^*(\rr_i)f_{AK'}^{}(\rr_i)+3\alpha e^{i2\K\cdot\rr_i}f_{BK'}^*(\rr_i)f_{AK}^{}(\rr_i)\;,\\
\Delta_{-+}^{1,n=1}(\rr_i)&=3\alpha^*f_{BK}^*(\rr_i)f_{AK}^{}(\rr_i)\;,\\
\Delta_{-+}^{1,n=2}(\rr_i)&=3\alpha f_{BK'}^*(\rr_i)f_{AK'}^{}(\rr_i)\;.
\end{align}
For $\Delta_{-+}^{2,n}$, we get with $\Delta_{-+}^{2,n}(\rr_i)=\Delta_{-+}^{1,n}(\rr_i-\a_1)$
\begin{align}
\Delta_{-+}^{2,n=0}(\rr_i)&=3e^{-i2\K\cdot\rr_i}f_{BK}^*(\rr_i)f_{AK'}^{}(\rr_i)+3e^{i2\K\cdot\rr_i}f_{BK'}^*(\rr_i)f_{AK}^{}(\rr_i)\;,\\
\Delta_{-+}^{2,n=1}(\rr_i)&=3\alpha^*f_{BK}^*(\rr_i)f_{AK}^{}(\rr_i)\;,\;\\
\Delta_{-+}^{2,n=2}(\rr_i)&=3\alpha f_{BK'}^*(\rr_i)f_{AK'}^{}(\rr_i)\;.
\end{align}
For $\Delta_{-+}^{3,n}$, we get with $\Delta_{-+}^{3,n}(\rr_i)=\Delta_{-+}^{1,n}(\rr_i-\a_2)$
\begin{align}
\Delta_{-+}^{3,n=0}(\rr_i)&=3\alpha e^{-i2\K\cdot\rr_i}f_{BK}^*(\rr_i)f_{AK'}^{}(\rr_i)+3\alpha^*e^{i2\K\cdot\rr_i}f_{BK'}^*(\rr_i)f_{AK}^{}(\rr_i)\;,\\
\Delta_{-+}^{3,n=1}(\rr_i)&=3\alpha^*f_{BK}^*(\rr_i)f_{AK}^{}(\rr_i)\;,\\
\Delta_{-+}^{3,n=2}(\rr_i)&=3\alpha f_{BK'}^*(\rr_i)f_{AK'}^{}(\rr_i)\;.
\end{align}

Inserting the ansatz of Eq. (\ref{AnsatzWFAB}) in the above equation, we obtain for $\rr_i\in B$:
\begin{align}
&\Delta_{-+}^{1,n}(\rr_i)=\alpha^{-n}\left[e^{-i\K\cdot(\rr_i-\d_2)}f_{AK}^*(\rr_i)+e^{i\K\cdot(\rr_i-\d_2)}f_{AK'}^*(\rr_i)\right]\left[e^{i\K\cdot\rr_i}f_{BK}^{}(\rr_i)+e^{-i\K\cdot\rr_i}f_{BK'}^{}(\rr_i)\right]\notag\\
&+\left[e^{-i\K\cdot(\rr_i+\a_1-\d_1)}f_{AK}^*(\rr_i)+e^{i\K\cdot(\rr_i+\a_1-\d_1)}f_{AK'}^*(\rr_i)\right]\left[e^{i\K\cdot(\rr_i+\a_1)}f_{BK}^{}(\rr_i)+e^{-i\K\cdot(\rr_i+\a_1)}f_{BK'}^{}(\rr_i)\right]\notag\\
&+\alpha^{n}\left[e^{-i\K\cdot(\rr_i+\a_1-\a_2-\d_3)}f_{AK}^*(\rr_i)+e^{i\K\cdot(\rr_i+\a_1-\a_2-\d_3)}f_{AK'}^*(\rr_i)\right]\left[e^{i\K\cdot(\rr_i+\a_1-\a_2)}f_{BK}^{}(\rr_i)+e^{-i\K\cdot(\rr_i+\a_1-\a_2)}f_{BK'}^{}(\rr_i)\right]\;.
\end{align}
This can be grouped together to read
\begin{align}
\Delta_{-+}^{1,n}(\rr_i)&=
\left[\alpha^{-n}e^{i\K\cdot\d_2}+e^{i\K\cdot\d_1} +\alpha^ne^{i\K\cdot\d_3}\right]f_{AK}^*(\rr_i)f_{BK}^{}(\rr_i)\notag\\
&+\left[\alpha^{-n}e^{-i\K\cdot\d_2}+e^{-i\K\cdot\d_1} +\alpha^ne^{-i\K\cdot\d_3}\right]f_{AK'}^*(\rr_i)f_{BK'}^{}(\rr_i)\notag\\
&+e^{-i2\K\cdot\rr_i}\left[\alpha^{-n}e^{-i\K\cdot\d_2}+e^{i\K\cdot(2\a_1-\d_1)}+\alpha^ne^{i\K\cdot(2\a_1-2\a_2-\d_3)}\right]f_{AK}^*(\rr_i)f_{BK'}^{}(\rr_i)\notag\\
&+e^{i2\K\cdot\rr_i}\left[\alpha^{-n}e^{i\K\cdot\d_2}+e^{-i\K\cdot(2\a_1-\d_1)}+\alpha^ne^{-i\K\cdot(2\a_1-2\a_2-\d_3)}\right]f_{AK'}^*(\rr_i)f_{BK}^{}(\rr_i)\;.
\end{align}
This gives
\begin{align}
\Delta_{-+}^{1,n=0}(\rr_i)&=3\alpha e^{-i2\K\cdot\rr_i}f_{AK}^*(\rr_i)f_{BK'}^{}(\rr_i)+3\alpha^*e^{i2\K\cdot\rr_i}f_{AK'}^*(\rr_i)f_{BK}^{}(\rr_i)\;,\\
\Delta_{-+}^{1,n=1}(\rr_i)&=3\alpha^*f_{AK'}^*(\rr_i)f_{BK'}^{}(\rr_i)\;,\\
\Delta_{-+}^{1,n=2}(\rr_i)&=3\alpha f_{AK}^*(\rr_i)f_{BK}^{}(\rr_i)\;.
\end{align}
For $\Delta_{-+}^{2,n}$, we get with $\Delta_{-+}^{2,n}(\rr_i)=\Delta_{-+}^{1,n}(\rr_i+\a_1)$
\begin{align}
\Delta_{-+}^{2,n=0}(\rr_i)&=3e^{-i2\K\cdot\rr_i}f_{AK}^*(\rr_i)f_{BK'}^{}(\rr_i)+3e^{i2\K\cdot\rr_i}f_{AK'}^*(\rr_i)f_{BK}^{}(\rr_i)\;,\\
\Delta_{-+}^{2,n=1}(\rr_i)&=3\alpha^*f_{AK'}^*(\rr_i)f_{BK'}^{}(\rr_i)\;,\\
\Delta_{-+}^{2,n=2}(\rr_i)&=3\alpha f_{AK}^*(\rr_i)f_{BK}^{}(\rr_i)\;.
\end{align}
For $\Delta_{-+}^{3,n}$, we get with $\Delta_{-+}^{3,n}(\rr_i)=\Delta_{-+}^{1,n}(\rr_i+\a_2)$
\begin{align}
\Delta_{-+}^{3,n=0}(\rr_i)&=3\alpha^*e^{-i2\K\cdot\rr_i}f_{AK}^*(\rr_i)f_{BK'}^{}(\rr_i)+3\alpha e^{i2\K\cdot\rr_i}f_{AK'}^*(\rr_i)f_{BK}^{}(\rr_i)\;,\\
\Delta_{-+}^{3,n=1}(\rr_i)&=3\alpha^*f_{AK'}^*(\rr_i)f_{BK'}^{}(\rr_i)\;,\\
\Delta_{-+}^{3,n=2}(\rr_i)&=3\alpha f_{AK}^*(\rr_i)f_{BK}^{}(\rr_i)\;.
\end{align}
Adding the three channels for separate $n$, we get for $\rr_i\in A$:
\begin{align}
\sum_{m=1,2,3}\Delta_{-+}^{m,0} (\rr_i)&=0\\
\sum_{m=1,2,3}\Delta_{-+}^{m,1} (\rr_i)&=9\alpha^*f_{BK}^*(\rr_i)f_{AK}^{}(\rr_i)\\
\sum_{m=1,2,3}\Delta_{-+}^{m,2} (\rr_i)&=9\alpha f_{BK'}^*(\rr_i)f_{AK'}^{}(\rr_i)
\end{align}
Adding the three channels for separate $n$, we get for $\rr_i\in B$:
\begin{align}
\sum_{m=1,2,3}\Delta_{-+}^{m,0}(\rr_i)&=0\\
\sum_{m=1,2,3}\Delta_{-+}^{m,1}(\rr_i)&=9\alpha^*f_{AK'}^*(\rr_i)f_{BK'}^{}(\rr_i)\\
\sum_{m=1,2,3}\Delta_{-+}^{m,2}(\rr_i)&=9\alpha f_{AK}^*(\rr_i)f_{BK}^{}(\rr_i)
\end{align}
Now summing over every third lattice site, we obtain
\begin{align}
\rho_{AK,BK}=\frac{\alpha^*}{3}&\sum_{m=1,2,3}\left(\sum_{\rr_i\in A^\Delta}\left[\Delta_{-+}^{m,1}(\rr_i)\right]^*+\sum_{\rr_i\in B^\Delta}\Delta_{-+}^{m,2}(\rr_i)\right)\;,\\
\rho_{AK',BK'}=\frac{\alpha^{}}{3}&\sum_{m=1,2,3}\left(\sum_{\rr_i\in A^\Delta}\left[\Delta_{-+}^{m,2}(\rr_i)\right]^*+\sum_{\rr_i\in B^\Delta}\Delta_{-+}^{m,1}(\rr_i)\right)\;.
\end{align}

\subsection{Inter-band, inter-valley channel}
To define the order parameters for the inter-sublattice, inter-valley channels based on a tight-binding model, we now define the following quantities related to the six overlap functions around a hexagon:
\begin{align}
\Delta_{--}^{1}(\rr_i)&=\psi^*(\rr_i\pm\d_1)\psi(\rr_i)+\psi^*(\rr_i\mp\a_2)\psi(\rr_i\pm\d_1)+\psi^*(\rr_i\mp\a_2\pm\d_3)\psi(\rr_i\mp\a_2)\notag\\&+\psi^*(\rr_i\mp\a_1)\psi(\rr_i\mp\a_2\pm\d_3)+\psi^*(\rr_i\mp\a_1\pm\d_2)\psi(\rr_i\mp\a_1)+\psi^*(\rr_i)\psi(\rr_i\mp\a_1\pm\d_2)\\
\Delta_{--}^2(\rr_i)&=\psi^*(\rr_i\pm\d_2)\psi(\rr_i)+\psi^*(\rr_i\pm\a_1)\psi(\rr_i\pm\d_2)
+\psi^*(\rr_i\pm\a_1\pm\d_1)\psi(\rr_i\pm\a_1)\notag\\
&+\psi^*(\rr_i\pm\a_1\mp\a_2)\psi(\rr_i\pm\a_1\pm\d_1)
+\psi^*(\rr_i\pm\d_1)\psi(\rr_i\pm\a_1\mp\a_2)+\psi^*(\rr_i)\psi(\rr_i\pm\d_1)\\
\Delta_{--}^{3}(\rr_i)&=\psi^*(\rr_i\pm\d_3)\psi(\rr_i)+\psi^*(\rr_i\mp\a_1\pm\a_2)\psi(\rr_i\pm\d_3)+\psi^*(\rr_i\mp\a_1\pm\a_2\pm\d_2)\psi(\rr_i\mp\a_1\pm\a_2)\notag\\&+\psi^*(\rr_i\pm\a_2)\psi(\rr_i\mp\a_1\pm\a_2\pm\d_2)+\psi^*(\rr_i\pm\a_2\pm\d_1)\psi(\rr_i\pm\a_2)+\psi^*(\rr_i)\psi(\rr_i\pm\a_2\pm\d_1)
\end{align}
where again the upper sign refers to $\rr_i\in A$, the lower sign to $\rr_i\in B$. 

Inserting the ansatz of Eq. (\ref{AnsatzWFAB}) in the above equation, we obtain for $\rr_i\in A$:
\begin{align}
&\Delta_{--}^1(\rr_i)=\left[e^{-i\K\cdot(\rr_i+\d_1)}f_{BK}^*(\rr_i)+e^{i\K\cdot(\rr_i+\d_1)}f_{BK'}^*(\rr_i)\right]\left[e^{i\K\cdot\rr_i}f_{AK}^{}(\rr_i)+e^{-i\K\cdot\rr_i}f_{AK'}^{}(\rr_i)\right]\notag\\
&+\left[e^{-i\K\cdot(\rr_i-\a_2)}f_{AK}^*(\rr_i)+e^{i\K\cdot(\rr_i-\a_2)}f_{AK'}^*(\rr_i)\right]\left[e^{i\K\cdot(\rr_i+\d_1)}f_{BK}^{}(\rr_i)+e^{-i\K\cdot(\rr_i+\d_1)}f_{BK'}^{}(\rr_i)\right]\notag\\
&+\left[e^{-i\K\cdot(\rr_i-\a_2+\d_3)}f_{BK}^*(\rr_i)+e^{i\K\cdot(\rr_i-\a_2+\d_3)}f_{BK'}^*(\rr_i)\right]\left[e^{i\K\cdot(\rr_i-\a_2)}f_{AK}^{}(\rr_i)+e^{-i\K\cdot(\rr_i-\a_2)}f_{AK'}^{}(\rr_i)\right]\notag\\
&+\left[e^{-i\K\cdot(\rr_i-\a_1)}f_{AK}^*(\rr_i)+e^{i\K\cdot(\rr_i-\a_1)}f_{AK'}^*(\rr_i)\right]\left[e^{i\K\cdot(\rr_i-\a_2+\d_3)}f_{BK}^{}(\rr_i)+e^{-i\K\cdot(\rr_i-\a_2+\d_3)}f_{BK'}^{}(\rr_i)\right]\notag\\
&+\left[e^{-i\K\cdot(\rr_i-\a_1+\d_2)}f_{BK}^*(\rr_i)+e^{i\K\cdot(\rr_i-\a_1+\d_2)}f_{BK'}^*(\rr_i)\right]\left[e^{i\K\cdot(\rr_i-\a_1)}f_{AK}^{}(\rr_i)+e^{-i\K\cdot(\rr_i-\a_1)}f_{AK'}^{}(\rr_i)\right]\notag\\
&+\left[e^{-i\K\cdot\rr_i}f_{AK}^*(\rr_i)+e^{i\K\cdot\rr_i}f_{AK'}^*(\rr_i)\right]\left[e^{i\K\cdot(\rr_i-\a_1+\d_2)}f_{BK}^{}(\rr_i)+e^{-i\K\cdot(\rr_i-\a_1+\d_2)}f_{BK'}^{}(\rr_i)\right]
\;.
\end{align}
This can be grouped together to read
\begin{align}
\Delta_{--}^{1}(\rr_i)&=
\left[e^{-i\K\cdot\d_1}+e^{-i\K\cdot\d_3} +e^{-i\K\cdot\d_2}\right]f_{BK}^*(\rr_i)f_{AK}^{}(\rr_i)\notag\\
&+\left[e^{i\K\cdot\d_1}+e^{i\K\cdot\d_3} +e^{i\K\cdot\d_2}\right]f_{BK'}^*(\rr_i)f_{AK'}^{}(\rr_i)\notag\\
&+e^{-i2\K\cdot\rr_i}\left[e^{-i\K\cdot\d_1}+e^{i\K\cdot(2\a_2-\d_3)}+e^{i\K\cdot(2\a_1-\d_2)}\right]f_{BK}^*(\rr_i)f_{AK'}^{}(\rr_i)\notag\\
&+e^{i2\K\cdot\rr_i}\left[e^{i\K\cdot\d_1}+e^{-i\K\cdot(2\a_2-\d_3)}+e^{-i\K\cdot(2\a_1-\d_2)}\right]f_{BK'}^*(\rr_i)f_{AK}^{}(\rr_i)\notag\\
&+\left[e^{i\K\cdot(\a_2+\d_1)}+e^{i\K\cdot(\a_1-\a_2+\d_3)} +e^{i\K\cdot(-\a_1+\d_2)}\right]f_{AK}^*(\rr_i)f_{BK}^{}(\rr_i)\notag\\
&+\left[e^{-i\K\cdot(\a_2+\d_1)}+e^{-i\K\cdot(\a_1-\a_2+\d_3)} +e^{-i\K\cdot(-\a_1+\d_2)}\right]f_{AK'}^*(\rr_i)f_{BK'}^{}(\rr_i)\notag\\
&+e^{-i2\K\cdot\rr_i}\left[e^{i\K\cdot(\a_2-\d_1)}+e^{i\K\cdot(\a_1+\a_2-\d_3)}+e^{i\K\cdot(\a_1-\d_2)}\right]f_{AK}^*(\rr_i)f_{BK'}^{}(\rr_i)\notag\\
&+e^{i2\K\cdot\rr_i}\left[e^{-i\K\cdot(\a_2-\d_1)}+e^{-i\K\cdot(\a_1+\a_2-\d_3)}+e^{-i\K\cdot(\a_1-\d_2)}\right]f_{AK'}^*(\rr_i)f_{BK}^{}(\rr_i)\;.
\end{align}

Now, the intra-valley contribution vanish due to the node of the Dirac dispersion at the $K$-point and only the inter-valley contributions survive. With $\alpha=e^{i2\pi/3}$ and $\Delta_{--}^2(\rr_i)=\Delta_{--}^1(\rr_i-\a_1)$ and $\Delta_{--}^3(\rr_i)=\Delta_{--}^1(\rr_i-\a_2)$, we then obtain (omitting the space-dependence of the envelope functions)
\begin{align}
\Delta_{--}^1(\rr_i)&=3\left(e^{-i2\K\cdot\rr_i}\alpha^* f_{BK}^* f_{AK'}+e^{i2\K\cdot\rr_i}\alpha f_{BK'}^*f_{AK}+ e^{-i2\K\cdot\rr_i}\alpha f_{AK}^*f_{BK'}+e^{i2\K\cdot\rr_i}\alpha^* f_{AK'}^*f_{BK}\right)\;,\\
\Delta_{--}^2(\rr_i)&=3\left(e^{-i2\K\cdot\rr_i}f_{BK}^*f_{AK'}+e^{i2\K\cdot\rr_i}f_{BK'}^*f_{AK}+ e^{-i2\K\cdot\rr_i}\alpha^*f_{AK}^*f_{BK'}+ e^{i2\K\cdot\rr_i}\alpha f_{AK'}^*f_{BK}\right)\;,\\
\Delta_{--}^3(\rr_i)&=3\left(e^{-i2\K\cdot\rr_i} \alpha f_{BK}^*f_{AK'}+e^{i2\K\cdot\rr_i} \alpha^*f_{BK'}^*f_{AK}+ e^{-i2\K\cdot\rr_i}f_{AK}^*f_{BK'}+ e^{i2\K\cdot\rr_i}f_{AK'}^*f_{BK}\right)\;.
\end{align}

For $\rr_i\in B$, we get the equivalent expressions by interchanging $A$ and $B$ and $\alpha$ and $\alpha^*$. This yields (again omitting the space-dependence of the envelope functions)
\begin{align}
\Delta_{--}^1(\rr_i)&=3\left(e^{-i2\K\cdot\rr_i}\alpha f_{AK}^* f_{BK'}+e^{i2\K\cdot\rr_i}\alpha^* f_{AK'}^*f_{BK}+ e^{-i2\K\cdot\rr_i}\alpha^* f_{BK}^*f_{AK'}+e^{i2\K\cdot\rr_i}\alpha f_{BK'}^*f_{AK}\right)\;,\\
\Delta_{--}^2(\rr_i)&=3\left(e^{-i2\K\cdot\rr_i}f_{AK}^*f_{BK'}+e^{i2\K\cdot\rr_i}f_{AK'}^*f_{BK}+ e^{-i2\K\cdot\rr_i}\alpha f_{BK}^*f_{AK'}+ e^{i2\K\cdot\rr_i}\alpha^* f_{BK'}^*f_{AK}\right)\;,\\
\Delta_{--}^3(\rr_i)&=3\left(e^{-i2\K\cdot\rr_i} \alpha^* f_{AK}^*f_{BK'}+e^{i2\K\cdot\rr_i} \alpha f_{AK'}^*f_{BK}+ e^{-i2\K\cdot\rr_i}f_{BK}^*f_{AK'}+ e^{i2\K\cdot\rr_i}f_{BK'}^*f_{AK}\right)\;.
\end{align}
We thus find the same expressions for the two different sublattices. We also see that $\Delta_{--}^1(\rr_i)+\Delta_{--}^2(\rr_i)+\Delta_{--}^3(\rr_i)=0$ which is related to current conservation.

As discussed above, the space-dependent phase factor can be chosen to be one for all cases, $e^{\pm i2\K\cdot\rr_i}=1$. The above expressions can thus be combined in the following way:
\begin{align}
\Delta_a(\rr_i)\equiv\alpha\Delta_{--}^1(\rr_i)+\Delta_{--}^2(\rr_i)+\alpha^*\Delta_{--}^3(\rr_i)&=9(f_{BK}^*(\rr_i) f_{AK'}(\rr_i)+\alpha^*f_{AK}^*(\rr_i)f_{BK'}(\rr_i))\;,\\
\Delta_b(\rr_i)\equiv\alpha^*\Delta_{--}^1(\rr_i)+\Delta_{--}^2(\rr_i)+\alpha\Delta_{--}^3(\rr_i)&=9(f_{BK'}^* (\rr_i)f_{AK}(\rr_i)+\alpha f_{AK'}^*(\rr_i)f_{BK}(\rr_i))\;.
\end{align}
Therefore, we get
\begin{align}
3f_{AK}^*(\rr_i)f_{BK'}(\rr_i)&=\frac{i}{3\sqrt{3}}\left[\Delta_a(\rr_i)-\alpha^*\Delta_b^*(\rr_i)\right]\;,\\
3f_{BK}^*(\rr_i)f_{AK'}(\rr_i)&=\frac{-i}{3\sqrt{3}}\left[\alpha\Delta_a(\rr_i)-\Delta_b^*(\rr_i)\right]\;.
\end{align}
Summing over the centers of the Kekul\'e structure and including a factor 3, we then finally get
\begin{align}
\rho_{AK,BK'}&=\frac{i}{3\sqrt{3}}\left(\sum_{\rr_i\in A^\Delta}+\sum_{\rr_i\in B^\Delta}\right)\left[\Delta_a(\rr_i)-\alpha^*\Delta_b^*(\rr_i)\right]\;,\label{AB}\\
\rho_{BK,AK'}&=\frac{-i}{3\sqrt{3}}\left(\sum_{\rr_i\in A^\Delta}+\sum_{\rr_i\in B^\Delta}\right)\left[\alpha\Delta_a(\rr_i)-\Delta_b^*(\rr_i)\right]\;.\label{BA}
\end{align}

%\subsection{Order parameter for the Hartree-Fock theory}}
\section{Order parameters for an interacting lattice model treated within the Hartree-Fock theory}
We are now in the position to define the order parameter for the tight-binding model based on the Hartree-Fock (HF) density matrix 
\begin{align}
\rho_{ij}^{HF}=\frac{1}{N_c}\sum_\sigma\sum_{\varepsilon_\k \le \varepsilon_{F,\sigma}}\langle \psi_{\k,i,\sigma}^\dagger\psi_{\k,j,\sigma}\rangle\;, 
%e^{-i \k \cdot (\boldsymbol{R}_i + \d_i - \boldsymbol{R}_j - \d_j)} ;,
\label{Fockappendix}
\end{align}
where the subscripts $i, j$ refers to the atom at position $\d_i, \d_j$ in the unit cell, with $N_c$ the number of moir\'e unit cells and $\varepsilon_{F,\sigma}$ the Fermi energy for spin-channel $\sigma$. Since we only consider spin-degenerate solutions, we will drop the spin-index $\sigma$ in the following.

Our procedure thus starts with a "many-body" density matrix defined by the many-body ground-state which in a Hartree-Fock scheme can be reduced to a "one-body" density matrix as correlations involving two or more particles can always be expressed by one-body correlation functions by virtue of Wick's theorem. The HF (one-body) density matrix of Eq. (\ref{Fockappendix}) thus represents a pure state as it is obtained from the many-body ground state which is approximated by a Slater-determinant within the Hartree-Fock scheme.

In the following, we will map the HF (one-body) density matrix to a SU(4)-reduced HF density matrix by "tracing out" the atomic degrees of freedom. The SU(4)-reduced HF density matrix thus only contains the long-wavelength degrees of freedom of the continuum model consisting of sublattice and valley degrees of freedom. The mapping procedure is guided by constructing the single-particle density matrix from the single-particle wave function of Eq. (\ref{AnsatzWFAB}). Our procedure thus involves the HF (one-body) density matrix, the reduced  HF (one-body) density matrix and the single-particle density matrix.

The mapping from a many-body to a single-particle system can be done because the Hartree-Fock theory describes {\it non-interacting} electrons. We thus partially trace out the HF density matrix $\rho_{ij}^{HF}$ to obtain a SU(4)-reduced HF density matrix by relating it to the single-particle density matrix via $\rho_{ij}=\psi_i^*\psi_j$ with the single-particle wave-function $|\psi\rangle=\sum_i\psi_i|\rr_i\rangle$ where $|\rr_i\rangle$ stands for the localized Wannier-orbital at lattice site $\rr_i$. Hence, we interpret the HF density matrix $\rho_{ij}^{HF}$ as the hopping amplitude from site $j$ to site $i$ of the interacting system.

To summarize, the SU(4)-reduced HF density matrix is obtained by first replacing the wave function overlap in the expressions of the various "$\Delta$"-functions defined in Sec. IV by the corresponding matrix elements of $\rho_{ij}^{HF}$ and then using the relations of the "$\Delta$"-functions to the reduced single-particle density matrix. The procedure is also schematically shown in Fig. \ref{FlowChart}.

\begin{figure}[!h]
    \centering
    \includegraphics[scale=0.35]{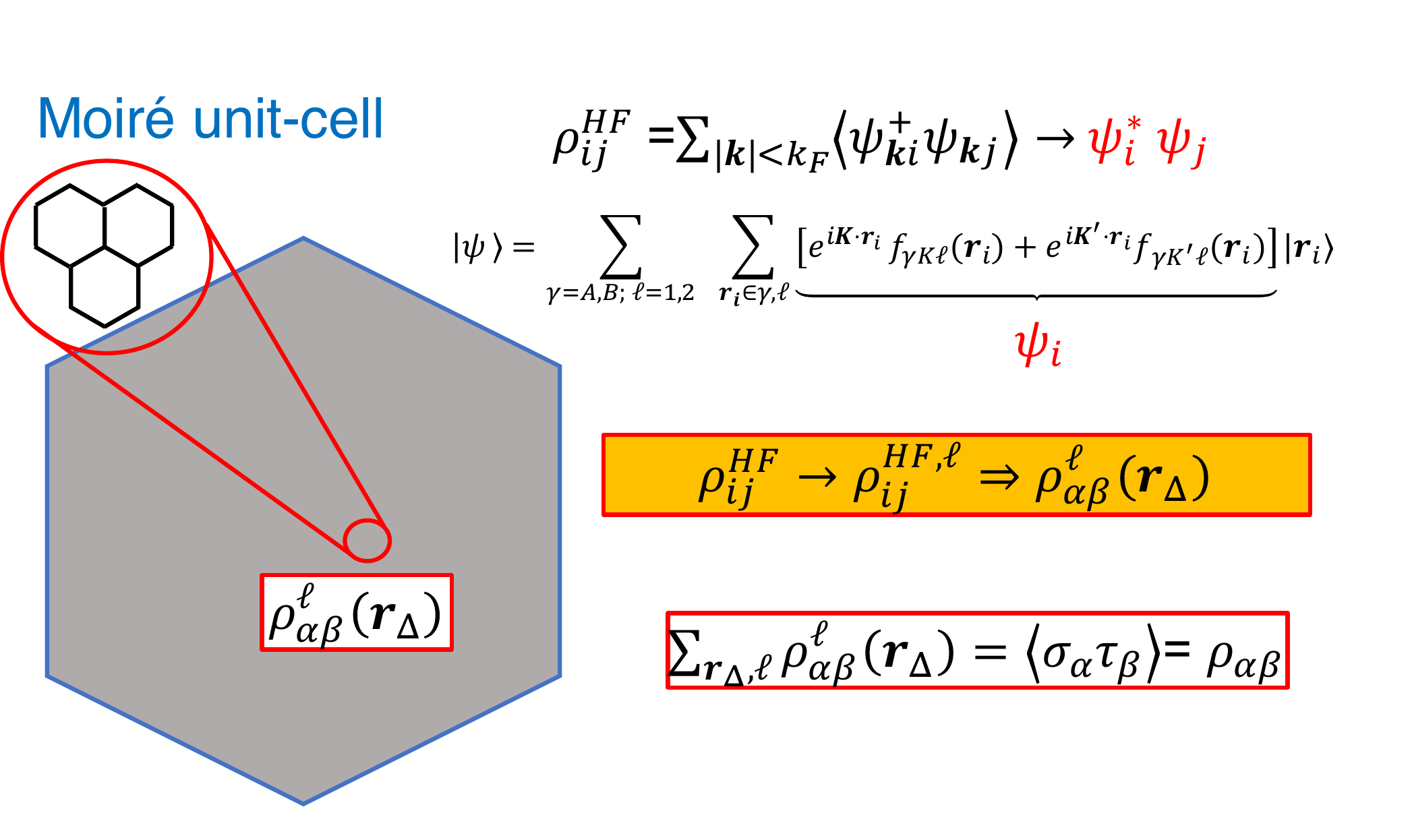}
    \caption{Schematic procedure how to obtain the SU(4)-reduced density matrix. Starting from the HF density matrix $\rho_{ij}^{HF}$, one focuses on the central atom $\rr_\Delta$ of three adjacent graphene hexagons on layer $\ell$. One then obtains $\rho_{\alpha\beta}^\ell(\rr_\Delta)$ following the procedure outlined in Fig. \ref{flux}, guided by the mapping of $\rho_{ij}^{HF}$ to the single-particle wave-function $|\psi\rangle=\sum_i\psi_i|\rr_i\rangle$. Summing over the two layers, one then obtains the SU(4)-reduced density matrix $\rho_{\alpha\beta}$.}
    \label{FlowChart}
\end{figure}

As is clear from the definition in Eq. (\ref{Fockappendix}), the HF density matrix $\rho_{ij}^{HF}$ includes the information of all occupied states, see Eq. (\ref{Fockappendix}). However, the procedure described above relies on the decomposition of the wave functions into the two valleys, Eq. (\ref{AnsatzWFAB}). This decomposition is only valid for states near charge neutrality and our approach cannot account for states at the bottom of the spectrum. Another important condition for our mapping to work is that the Brillouin zone of the moir\'e system is small compared to the Brillouin zone of the graphene lattice. Then, there is hardly any $\k$-dependence of the wave function (other than the valley dependence) and the ansatz of Eq. (\ref{AnsatzWFAB}) shall represent a good approximation for states close to the Fermi energy. 

As is evident from Fig. \ref{TraceRho}, the trace of $\rho$ displays particle-hole symmetry as it is centred around zero at charge neutrality and not - as might be expected - equal to the number of occupied states. The trace can, therefore, also be negative since only non-diagonal matrix elements of $\rho_{ij}^{HF}$ with $i\neq j$ are used in the reduction process. Nevertheless, the states that are occupied at the bottom of the Fermi sea should only contribute via a constant background that can be appropriately normalized as shown in the following section. 

\section{Normalization of the $SU(4)$-reduced HF density matrix}
\label{sec:sym}
\begin{figure}[!h]
    \centering
    \includegraphics[scale=0.35]{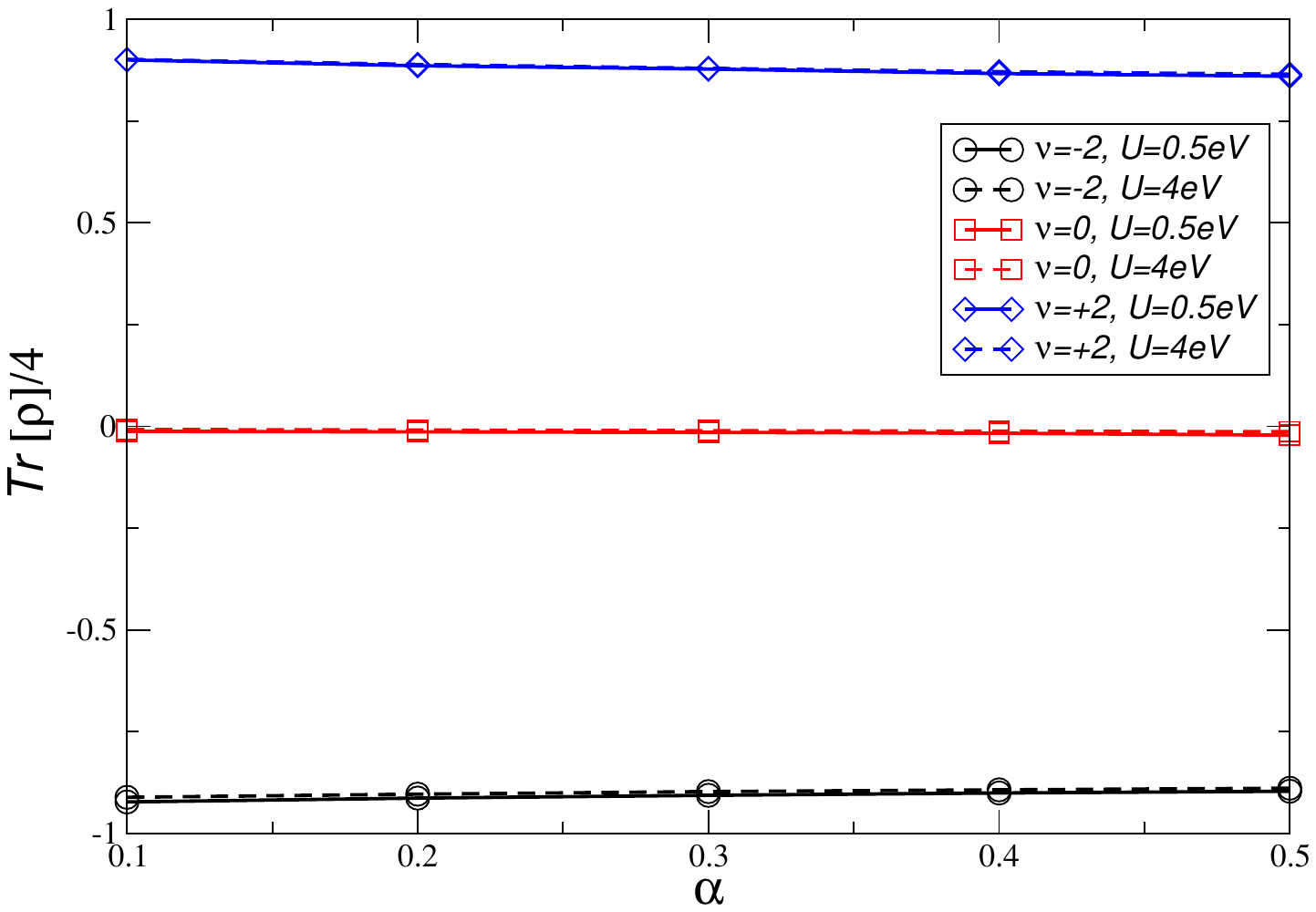}
    \caption{The trace of the (bare) reduced HF density matrix $\rho$ as obtained from our algorithm for various integer filling factor, Hubbard on-line interaction, and long-ranged coupling parameter $\alpha=\frac{e^2}{4\pi\epsilon_0\epsilon}$ in units of $eV\times a$.}
    \label{TraceRho}
\end{figure}

Let us now analyze the $SU(4)$-reduced HF density matrix for the different parameters and in Fig. \ref{TraceRho}, we show its trace. Independent of the interaction parameters, the trace only depends on the filling factor $\nu$, up to small deviations, and is centred around zero at charge neutrality. We thus have Tr$\rho\approx2\nu$.

This is to be compared to the trace of the (original) Hartree-Fock density matrix, Tr$\rho_{ij}^{HF}=N_e$, where $N_e$ is the number of electrons of the tight-binding model. With the number of lattice sites of a moir\'e cell, $N$, see Eq. (\ref{ParticleNumber}), we thus have Tr$\rho_{ij}^{HF}=N+\nu$ . To this moment, we do not have an explanation why the reduction process leads to a shift of the trace and why it displays particle-hole symmetry, i.e., why it becomes zero at charge neutrality.

In order to discuss the other 15 components of $\rho$, we introduce 
\begin{align}
\rho'=\sum_{(\alpha,\beta)\neq(0,0)}\rho_{\alpha,\beta}\sigma_\alpha\tau_\beta\;,
\end{align}
where $\alpha,\beta=0,x,y,z$ denote the different Pauli-matrices for sublattice and valley degree of freedom, including the unity matrix. In Fig. \ref{Weight}, we show ${\rho '}^2={\rho '}_{+,+}^2+\rho_{+,-}^2+\rho_{-,+}^2+\rho_{-,-}^2$ where the four components ${\rho}_{\nu,\mu}^2$ refer to the sum of intra-sublattice intra-valley (without the diagonal part indicated by the prime) ($\nu=+,\mu=+$), intra-sublattice inter-valley ($\nu=+,\mu=-$), inter-sublattice intra-valley ($\nu=-,\mu=+$), and inter-sublattice inter-valley ($\nu=-,\mu=-$) contributions. Again, the absolute value is almost independent of the parameters even thought the relative weight changes. Moreover, the weight of $\rho_{-,+}^2$ and especially of $\rho_{+,-}^2$ is negligible. They shall be discussed in more detail in Sec. VI.

\begin{figure}[!h]
    \centering
    \includegraphics[scale=0.35]{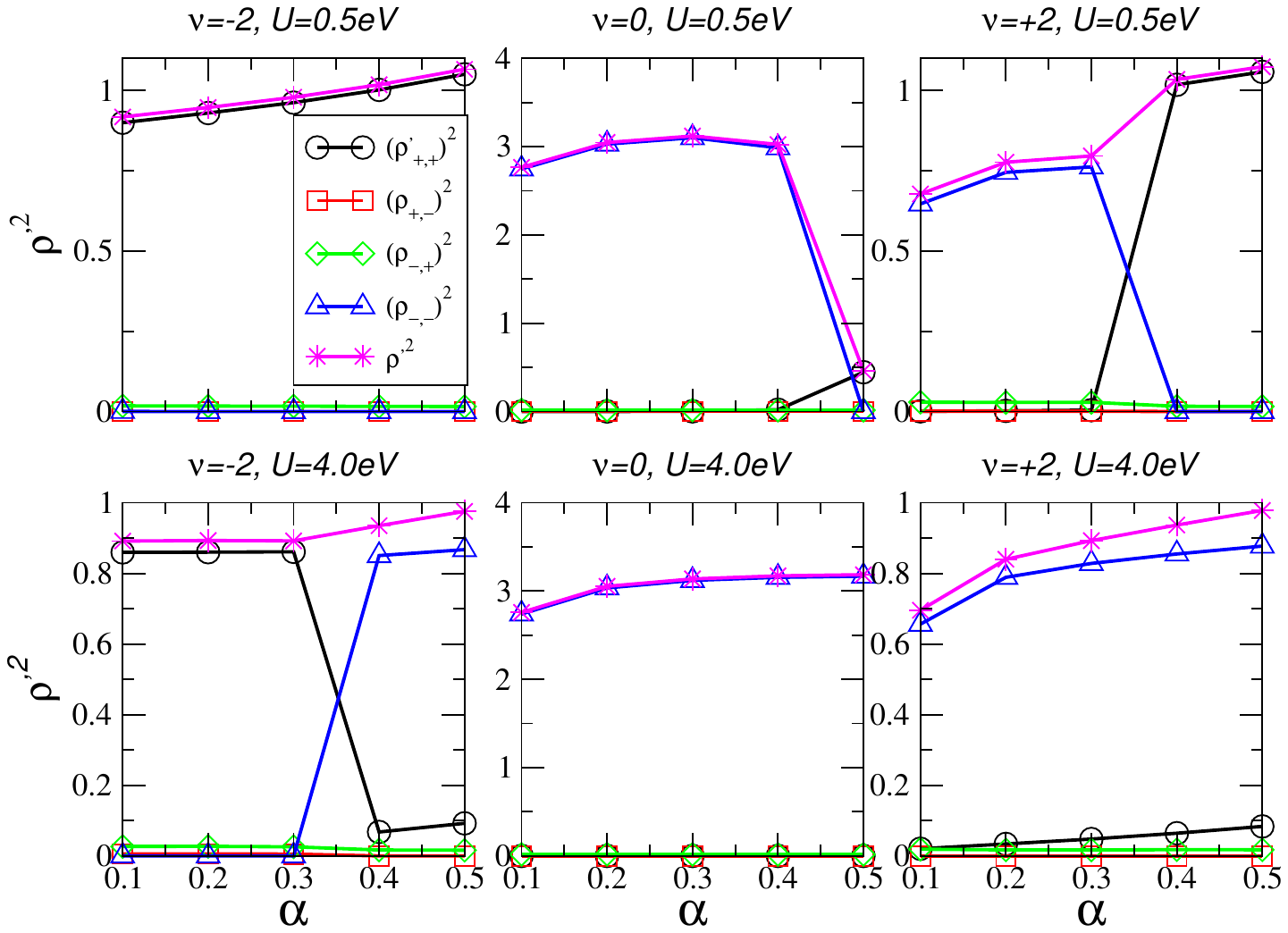}
    \caption{The weights of the (bare) reduced HF density matrix $\rho'$ as obtained from our algorithm for various integer filling factors, Hubbard on-line interactions, and long-ranged coupling parameters $\alpha=\frac{e^2}{4\pi\epsilon_0\epsilon}$ in units of $eV\times a$.}
    \label{Weight}
\end{figure}

Our algorithm should accurately account for the relative weight between the matrix elements. However, since we relate the (many-body) HF density matrix to a single-particle wave function, we may not expect to get the normalization right. We now define $Q=\lambda\rho$ and add a term proportional to the unity matrix. The normalized reduced HF density matrix shall thus be given by (we will keep the notation)
\begin{align}
\label{ScaledRho}
\rho \to \frac{1}{2}+Q\;,
\end{align}
For $\nu=\pm2$, the parameter $\lambda$ is now chosen such that $\Tr\rho=\frac{4+\nu}{2}$. This sets the normalization of $\rho$ to that of a reduced HF density matrix for $\frac{4+\nu}{2}$ electrons, i.e. the electrons in the flat bands. 

For $\nu=0$, we already have $\Tr\rho\approx2$, see Fig. \ref{TraceRho} and we choose $\lambda$ such that $Q^2=1/4$. As $\rho_{+,-}$ and $\rho_{-,+}$ are almost zero, we will only discuss the diagonal and off-diagonal matrix elements to simplify the discussion. The results are shown in Figs. \ref{Diagonal}, \ref{OffDiagonal}, and \ref{PhaseOff}.

\begin{figure}[!h]
    \centering
    \includegraphics[scale=0.35]{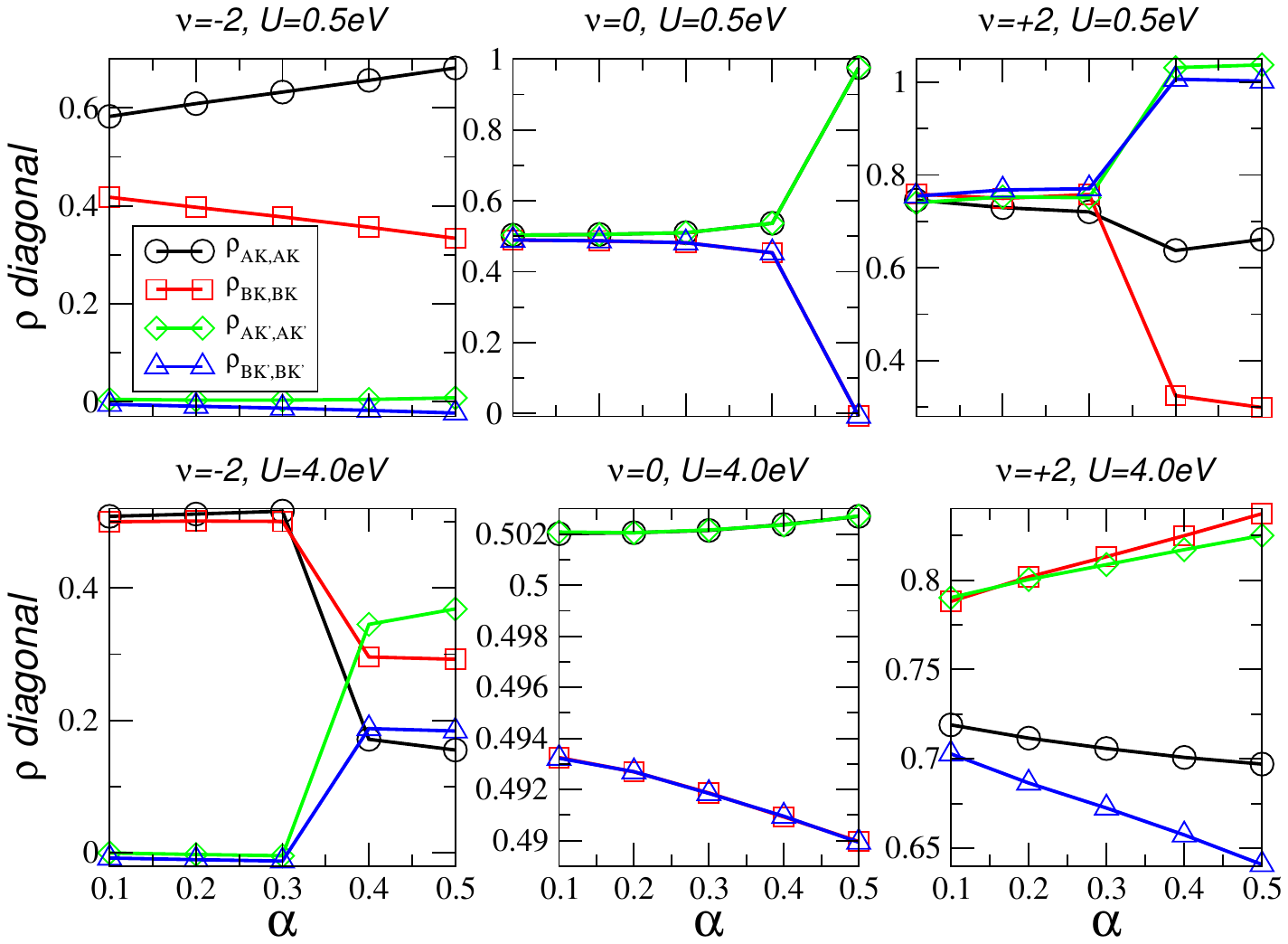}
    \caption{The diagonal matrix elements of the reduced HF density matrix as obtained from Eq. (\ref{ScaledRho}) for various integer filling factor, Hubbard on-line interaction, and long-ranged coupling parameter $\alpha=\frac{e^2}{4\pi\epsilon_0\epsilon}$ in units of $eV\times a$.}
    \label{Diagonal}
\end{figure}

\begin{figure}[!h]
    \centering
    \includegraphics[scale=0.35]{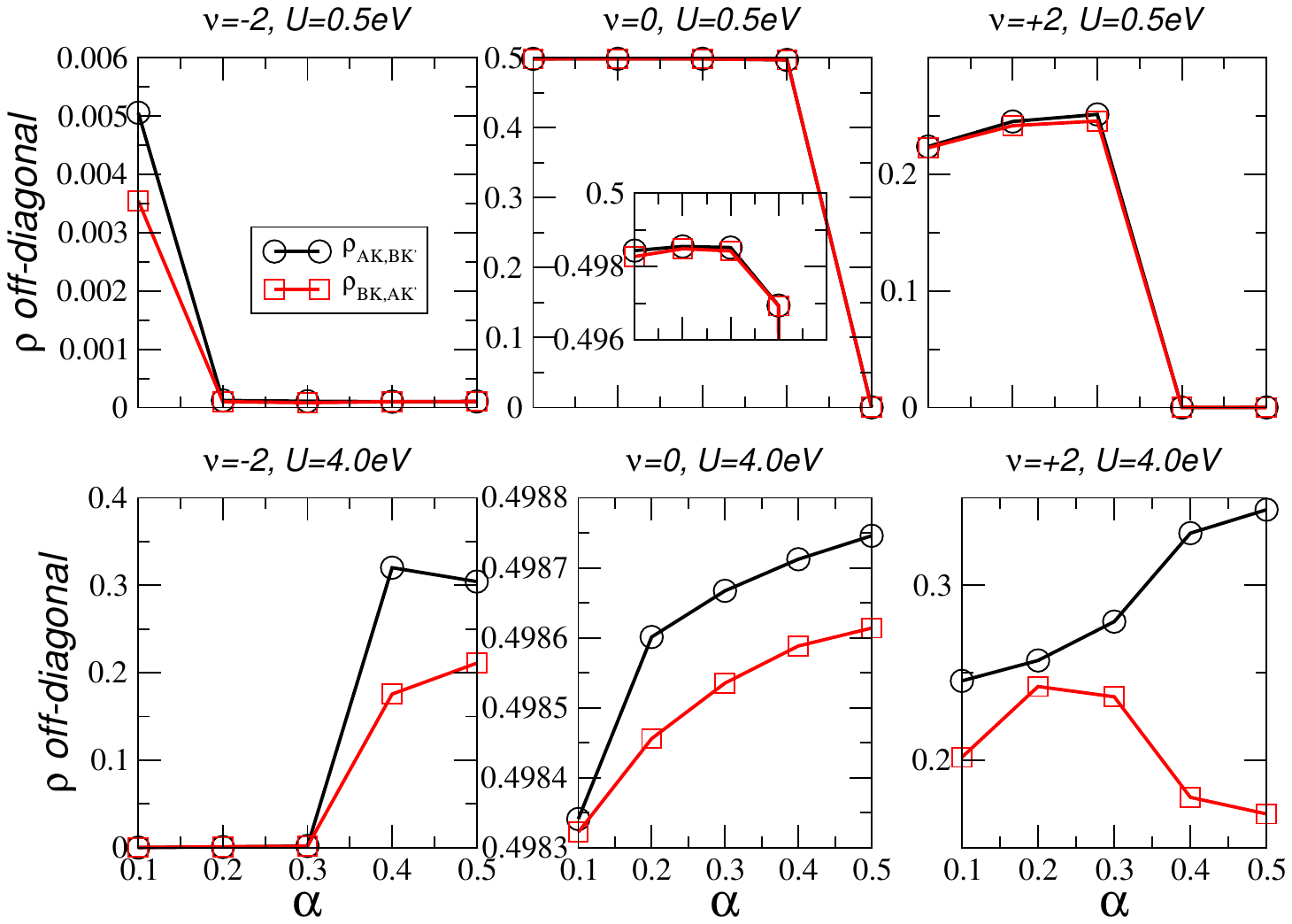}
    \caption{The absolute values of the off-diagonal matrix elements of the reduced HF density matrix as obtained from Eq. (\ref{ScaledRho}) for various integer filling factor, Hubbard on-line interaction, and long-ranged coupling parameter $\alpha=\frac{e^2}{4\pi\epsilon_0\epsilon}$ in units of $eV\times a$.}
    \label{OffDiagonal}
\end{figure}

\begin{figure}[!h]
    \centering
    \includegraphics[scale=0.35]{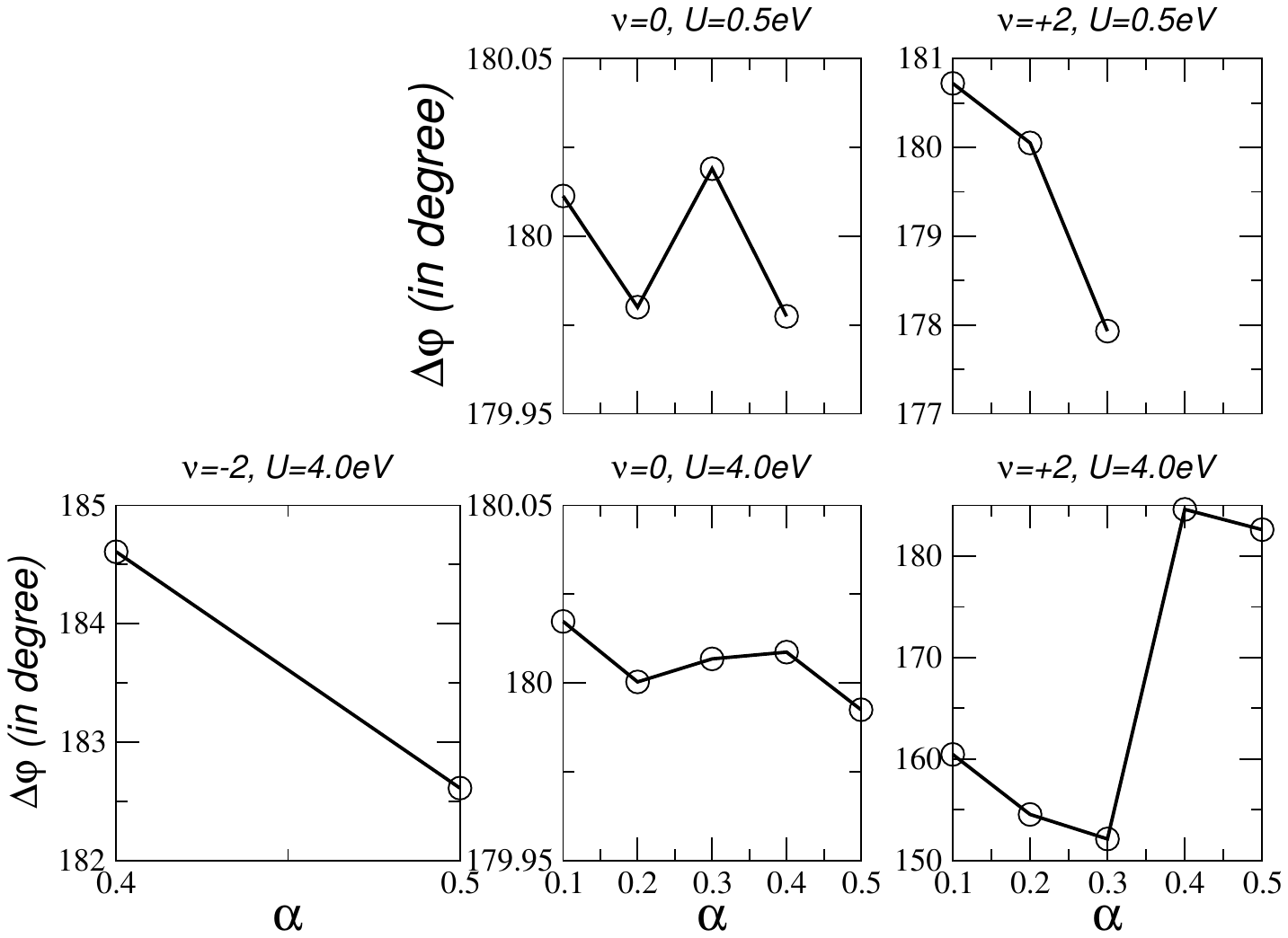}
    \caption{The phase-difference of the off-diagonal matrix elements of the reduced HF density matrix as obtained from Eq. (\ref{ScaledRho}) for various integer filling factor, Hubbard on-line interaction, and long-ranged coupling parameter $\alpha=\frac{e^2}{4\pi\epsilon_0\epsilon}$ in units of $eV\times a$.}
    \label{PhaseOff}
\end{figure}

\section{Discussion}
We are now in the position to discuss the phase diagram via the reduced density matrix. 
\subsection{Charge-neutrality point}
Let us first discuss the filling factor $\nu=0$. First, we note that for all parameters we find a pure state with $\rho^2=\rho$. Even though our normalization of $Q$ and the shift of Eq. (\ref{ScaledRho}) allows for a pure state, it is still remarkable as it shows up. For example, if we choose the same normalization $\lambda{\rho'}^2=1/4$ for $\nu=\pm2$, we do not obtain $\rho^2=\rho$. 

As can be seen from Figs. \ref{Diagonal}, \ref{OffDiagonal}, and \ref{PhaseOff}, the ground-state can be well approximated by a Kramer-intervalley-coherent (K-IVC) state
\begin{align}
\label{ValleyCoherent}
|\Delta\varphi\rangle_{\rm{KIVC}}=\frac{1}{\sqrt{2}}(c_1|AK\rangle+e^{i\varphi_1}c_2|BK'\rangle)\otimes\frac{1}{\sqrt{2}}(c_2|BK\rangle+e^{i\varphi_2}c_1|AK'\rangle)\;,
\end{align}
with $\rho\approx\rho_{\rm{KIVC}}=|\Delta\varphi\rangle\langle\Delta\varphi|$ with $\Delta\varphi=\varphi_1-\varphi_2\approx|\pi|$. However, for $\rho^2=\rho$, all matrix elements must contribute.

For $U=4$eV, $c_1=c_2=1$ and we have a "pure" valley-coherent state. For $U=0.5$eV, we only have $c_1c_2=1$ and for $\alpha\approx0.4$eV\r{A}, we predict a phase transition to a chiral insulator. 
\subsection{Hole doping at half-filling}
At $\nu=-2$, we find predominately a valley polarized state, i.e., there is no off-diagonal matrix elements. However, it is not a pure state, also by construction. 

For $U=0.5$eV, the diagonal matrix elements for one valley, say $K'$, are given by $\rho_{XK',XK'}\approx0$ (the small negative value is within our numerical accuracy). For the diagonal matrix elements of the other valley, we find $\rho_{XK,XK}=\rho^0_\pm \pm c\tilde\alpha$ with $\alpha=\tilde\alpha$eV$\times a$ and $c\approx0.2$ and $\rho^0_\pm\approx0.5\pm0.05$. The upper/lower sign applies for sublattice $X=A/B$.

For $U=4$eV, we find that the valley polarized state is only the ground-state for $\tilde\alpha\leq0.3$. In this case, we find again for one valley, say $K$, $\rho_{XK,XK}\approx0.5$ and for the other valley $\rho_{XK',XK'}\approx0$.

For $U=4$eV and $\tilde\alpha\geq0.4$, we find a valley coherent state with non-zero off-diagonal elements. However, this state cannot be simply written as a pure state of Eq. (\ref{ValleyCoherent}).

\subsection{Electron doping at half-filling}
At $\nu=2$, we find predominately a valley coherent state, i.e., there are off-diagonal matrix elements. However, it is not a pure state, again also by construction. 

For $U=0.5$eV, we find that the valley coherent state is only the ground-state for $\tilde\alpha\leq0.3$. Even though, it is not the K-IVC pure state of Eq. (\ref{ValleyCoherent}), we can still approximate $\rho=\frac{1}{2}(1+\rho_{\rm{KIVC}})$. 

For $U=0.5$eV and $\tilde\alpha\geq0.4$, we find a valley polarized state with zero off-diagonal elements. We can approximate this state by filling all electrons of one valley, say $K'$, with $\rho_{XK',XK'}=1$ and for the diagonal matrix elements of the other valley, we again find $\rho_{XK,XK}=\rho^0_\pm \pm c\tilde\alpha$ with $\alpha=\tilde\alpha$eV$\times a$ and $c\approx0.2$ and $\rho^0_\pm\approx0.5\pm0.05$. The upper/lower sign applies for sublattice $X=A/B$.

For $U=4$eV, we always find a valley coherent state with non-zero off-diagonal elements. The diagonal matrix elements show a linear behavior in $\tilde\alpha$, however, it cannot be associated to the previously identified K-IVC-state. In fact, we expect a phase-transition at for $\alpha\approx0.3$eV$\times a$ from a valley coherent state with $\Delta\varphi\neq\pi$ to a valley coherent state with $\Delta\varphi\approx\pi$. 

\subsection{Summary}
We can now summarize the density matrix for the different parameters.
\begin{itemize}
\item $\nu=-2$, $U=0.5$eV:
\begin{align}
\rho=\left(
\begin{array}{cccc}
x(\alpha)&0&0&0\\
0&y(\alpha)&0&0\\
0&0&0&0\\
0&0&0&0
\end{array}\right)\;,
\end{align}
where $x(\alpha)+y(\alpha)=1$ and we can approximate $x(\alpha)=0.56+0.21\tilde\alpha$ and $y(\alpha)=0.44-0.21\tilde\alpha$.
\item $\nu=0$, $U=0.5$eV, $\tilde\alpha\leq0.4$:
\begin{align}
\rho=\frac{1}{2}\left(
\begin{array}{cccc}
c_1^2&0&0&c_1c_2e^{i\varphi}\\
0&c_2^2&c_1c_2e^{-i\varphi}&0\\
0&c_1c_2e^{i\varphi}&c_1^2&0\\
c_1c_2e^{-i\varphi}&0&0&c_2^2
\end{array}\right)\;,
\end{align}
where $c_1\approx c_2\approx1$.
\item $\nu=0$, $U=0.5$eV, $\tilde\alpha\geq0.4$:
\begin{align}
\rho=\left(
\begin{array}{cccc}
1&0&0&0\\
0&0&0&0\\
0&0&1&0\\
0&0&0&0
\end{array}\right)\;.
\end{align} 
\item $\nu=2$, $U=0.5$eV, $\tilde\alpha\leq0.3$:
\begin{align}
\rho=\left(
\begin{array}{cccc}
x(\alpha)&0&0&0\\
0&y(\alpha)&0&0\\
0&0&1&0\\
0&0&0&1
\end{array}\right)\;,
\end{align}
where $x(\alpha)+y(\alpha)=1$ and we can approximate $x(\alpha)=0.56+0.21\tilde\alpha$ and $y(\alpha)=0.44-0.21\tilde\alpha$.
\item $\nu=2$, $U=0.5$eV, $\tilde\alpha\geq0.3$:
\begin{align}
\rho=\frac{1}{2}{\bf 1}+\frac{1}{4}\left(
\begin{array}{cccc}
1&0&0&e^{i\varphi}\\
0&1&e^{-i\varphi}&0\\
0&e^{i\varphi}&1&0\\
e^{-i\varphi}&0&0&1
\end{array}\right)\;.
\end{align} 
\item $\nu=-2$, $U=4$eV, $\tilde\alpha\leq0.3$:
\begin{align}
\rho=\frac{1}{2}\left(
\begin{array}{cccc}
1&0&0&0\\
0&1&0&0\\
0&0&0&0\\
0&0&0&0
\end{array}\right)\;,
\end{align}
\item $\nu=-2$, $U=4$eV, $\tilde\alpha\geq0.3$:
\begin{align}
\rho=\left(
\begin{array}{cccc}
x(\alpha)&0&0&a(\alpha)e^{i\varphi}\\
0&0.3&b(\alpha)e^{-i\varphi}&0\\
0&b(\alpha)e^{i\varphi}&y(\alpha)&0\\
a(\alpha)e^{-i\varphi}&0&0&0.2
\end{array}\right)\;,
\end{align} 
where $x(\alpha)+y(\alpha)=1/2$ and we can approximate $x(\alpha)=0.25-0.21\tilde\alpha$ and $y(\alpha)=0.25+0.21\tilde\alpha$. We further approximate $a(\alpha)=0.4-0.2\tilde\alpha$ and $b(\alpha)=0.1+0.2\tilde\alpha$.
\item $\nu=0$, $U=4$eV:
\begin{align}
\rho=\frac{1}{2}\left(
\begin{array}{cccc}
1&0&0&e^{i\varphi}\\
0&1&e^{-i\varphi}&0\\
0&e^{i\varphi}&1&0\\
e^{-i\varphi}&0&0&1
\end{array}\right)\;.
\end{align}
\item $\nu=2$, $U=4$eV:
\begin{align}
\rho=\left(
\begin{array}{cccc}
x(\alpha)&0&0&a(\alpha)e^{i\varphi_1}\\
0&y(\alpha)&b(\alpha)e^{i\varphi_2}&0\\
0&b(\alpha)e^{-i\varphi_2}&y'(\alpha)&0\\
a(\alpha)e^{-i\varphi_1}&0&0&x'(\alpha)
\end{array}\right)\;,
\end{align} 
where $x(\alpha)\approx x'(\alpha)\approx0.72-0.2\tilde\alpha$ and $y(\alpha)\approx y'(\alpha)\approx0.88+0.2\tilde\alpha$. For $\tilde\alpha\leq0.3$, $a(\alpha)\approx b(\alpha)\approx 0.25$ and $\Delta\varphi(\alpha)\to5\pi/6$ for $\tilde\alpha\to0.3$. For $\tilde\alpha\geq0.3$, $a(\alpha)\approx0.25-0.2\tilde\alpha$ and  $b(\alpha)\approx 0.25+0.2\tilde\alpha$ and $\Delta\varphi\approx\pi$.
\end{itemize}

\section{Subdominant and "odd" order parameters}
In this section, we will briefly discuss the subdominant order parameters which are related to intra-sublattice inter-valley and inter-sublattice intravalley scattering. We will also discuss the "odd" order parameters of these channels as they acquire large values in the nematic phases.
\subsection{Subdominant order parameters}
As commented above, the phases can be either characterized by a valley polarized (VP) or intervalley coherent (IVC) order parameter. However, also subdominant order parameter occur in the intra-sublattice inter-valley and inter-sublattice intra-valley channel.

In Fig. (\ref{SDOrderParameterPlus}), we show the absolute values of the order parameters related to intra-sublattice inter-valley and inter-sublattice intra-valley scattering. The largest value is given by $\langle\sigma_x\rangle$ and $\langle\sigma_y\rangle$, respectively; still, it is approximately one order of magnitude lower than the dominant order parameters related to either VP or IVC. Let us finally note that it is adiabatically related to the inter-Chern order parameter recently discussed in Ref. \onlinecite{Nuckolls23}.

\begin{figure}[!h]
    \centering
    \includegraphics[scale=0.35]{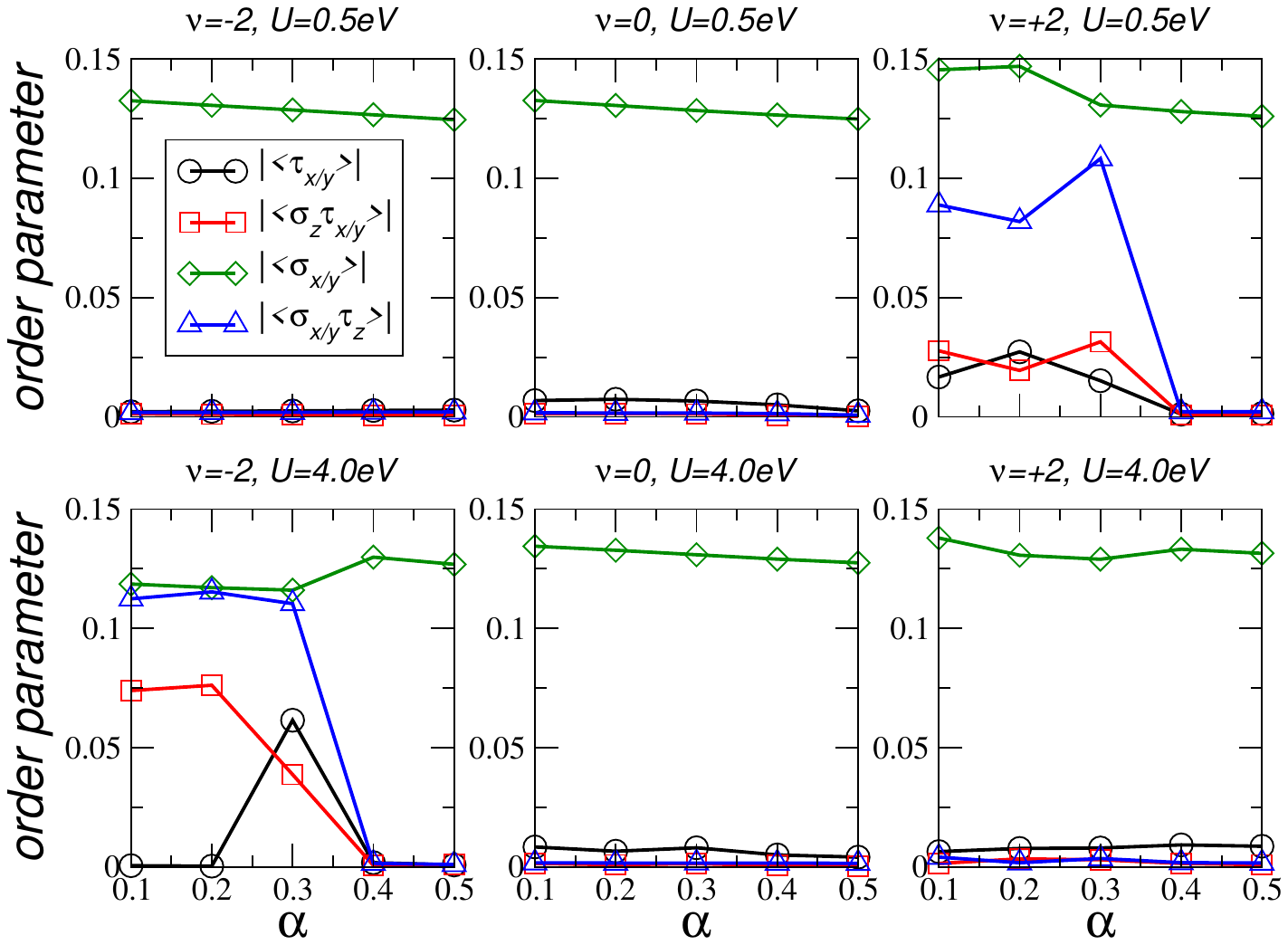}
    \caption{Absolute values of order parameters related to intra-sublattice inter-valley and inter-sublattice intra-valley scattering for various integer filling factor, Hubbard on-line interaction, and long-ranged coupling parameter $\alpha=\frac{e^2}{4\pi\epsilon_0\epsilon}$ in units of $eV\times a$. In the caption, we abbreviated $|\langle\tau_{x/y}\rangle|=\sqrt{\langle\tau_{x}\rangle^2+\langle\tau_{y}\rangle^2}$, $|\langle\sigma_z\tau_{x/y}\rangle|=\sqrt{\langle\sigma_z\tau_{x}\rangle^2+\langle\sigma_z\tau_{y}\rangle^2}$, $|\langle\sigma_{x/y}\rangle|=\sqrt{\langle\sigma_{x}\rangle^2+\langle\sigma_{y}\rangle^2}$, and $|\langle\sigma_{x/y}\tau_z\rangle|=\sqrt{\langle\sigma_{x}\tau_z\rangle^2+\langle\sigma_{y}\tau_z\rangle^2}$.}
    \label{SDOrderParameterPlus}
\end{figure}

\subsection{Dominant "odd" order parameters}
Here, we will briefly discuss extensions of the reduced density matrix as discussed in Eq. (\ref{ReducedDM}) by discussing the "odd" superpositions of the two layers. We thus define
\begin{align}
\label{ReducedDM}
\rho_{\alpha,\beta;\alpha',\beta'}^-=\int_{A_M}d\rr\sum_\ell (-1)^{\ell-1}f_{\alpha,\beta,\ell}^*(\rr)f_{\alpha',\beta',\ell}(\rr)\;.
\end{align}
This density matrix shall be related to the usual representation $\rho_{\alpha,\beta}^-=\langle\sigma_\alpha\tau_\beta\eta_z\rangle$ where $\eta_z$ may be interpreted as the $z$-component of the Pauli-matrix with respect to the two layers.

In Fig. (\ref{SDOrderParameterMinus}), we show the absolute values of the order parameters related to intra-sublattice inter-valley and inter-sublattice intra-valley scattering of this odd channel. The largest value is given by $\langle\sigma_x\eta_z\rangle$ and $\langle\sigma_y\eta_z\rangle$, respectively, in the phase where there is nematic order. Let us finally note that the corresponding order parameters related to VP and IVC are negligible. 

\begin{figure}[!h]
    \centering
    \includegraphics[scale=0.35]{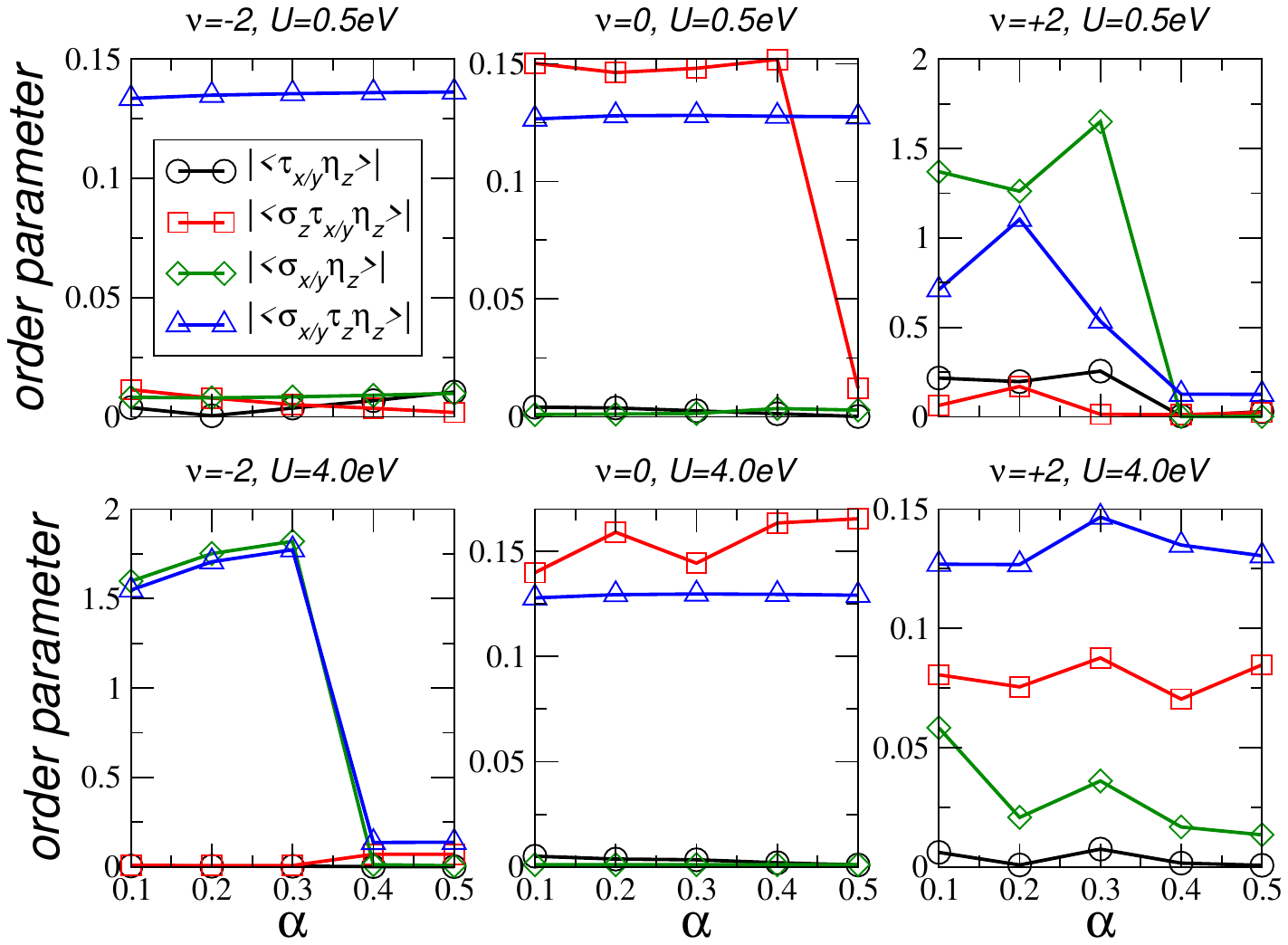}
    \caption{Absolute values of order parameters related to intra-sublattice inter-valley and inter-sublattice intra-valley scattering and odd layer-superposition for various integer filling factor, Hubbard on-line interaction, and long-ranged coupling parameter $\alpha=\frac{e^2}{4\pi\epsilon_0\epsilon}$ in units of $eV\times a$. In the caption, we abbreviated $|\langle\tau_{x/y}\eta_z\rangle|=\sqrt{\langle\tau_{x}\eta_z\rangle^2+\langle\tau_{y}\eta_z\rangle^2}$, $|\langle\sigma_z\tau_{x/y}\eta_z\rangle|=\sqrt{\langle\sigma_z\tau_{x}\eta_z\rangle^2+\langle\sigma_z\tau_{y}\eta_z\rangle^2}$, $|\langle\sigma_{x/y}\eta_z\rangle|=\sqrt{\langle\sigma_{x}\eta_z\rangle^2+\langle\sigma_{y}\eta_z\rangle^2}$, and $|\langle\sigma_{x/y}\tau_z\eta_z\rangle|=\sqrt{\langle\sigma_{x}\tau_z\eta_z\rangle^2+\langle\sigma_{y}\tau_z\eta_z\rangle^2}$.}
    \label{SDOrderParameterMinus}
\end{figure}

\section{Ground-state energies and phase transitions}
We observe several phase transitions, however, the only "real" phase transition occurs  for $\nu=0$, $U=0.5$eV and $\alpha=0.4$eV$\times a$. The other phase transitions result due to a competition between the valley polarized (VP) and inter-valley coherent (IVC) phase, both being a stable solution . 

Both phases emerge for different initial conditions and we choose the one with lower total energy. In the tables below, we list the total energy difference $\Delta E=E_{VP}-E_{IVC}$ for the parameters where we find the two phases. We also list the energy difference for the kinetic energy $E_{kin}$, the Fock and Hartree energy $E_{Fock}$ and $E_{Hartree}$ in units of $\tilde\alpha$, and finally the Hubbard energy $E_U$ in units of $\tilde U$, where we introduced the dimensional quantities $\tilde\alpha$ and $\tilde U$ by $\alpha=\tilde\alpha$eV$\times a$ and $U=\tilde U$eV. For $\nu=-2$ and $U=4.0$eV as well as for $\nu=2$ and $U=0.5$eV, there is a sign change for $\Delta E$ and thus a phase transition from valley polarized to valley coherent and valley coherent to valley polarized, respectively.
\begin{center}
\begin{tabular}{ |p{1.8cm}|p{1.3cm}|p{1.3cm}|p{1.6cm}|p{2.0cm}|p{1.5cm}|}
 \hline
 \multicolumn{6}{|c|}{\Large{\Large $\Delta E=E_{VP}-E_{IVC}$ in meV for $\nu = -2$, $U=0.5$eV}}\\
 \hline
 $\alpha$ in eV$\times a$ & $\Delta E$ & $\Delta E_{kin}$ & $\Delta E_{Fock}/\tilde\alpha$ & $\Delta E_{Hartree}/\tilde \alpha$ &  $\Delta E_{U}/\tilde U$    \\
 \hline
0.1 &$-0.001$& $-0.001$ & $0.025$ &  $0.061$ & $0.006$ \\
 \hline
\end{tabular}

\begin{tabular}{ |p{1.8cm}|p{1.3cm}|p{1.3cm}|p{1.6cm}|p{2.0cm}|p{1.5cm}|}
 \hline
 \multicolumn{6}{|c|}{\Large{$\Delta E=E_{VP}-E_{IVC}$ in meV for $\nu = -2$, $U=4.0$eV}} \\
 \hline
 $\alpha$ in eV$\times a$ & $\Delta E$ & $\Delta E_{kin}$ & $\Delta E_{Fock}/\tilde\alpha$ & $\Delta E_{Hartree}/\tilde \alpha\;$ &  $\Delta E_{U}/\tilde U$     \\
 \hline
0.1 &$-0.147$& $+0.0285$ & $+2.302$ &  $+1.933$ & $-0.266$ \\
0.2 &$-0.099$& $+0.124$ & $-0.230$ & $+1.279$ & $-0.214$ \\
0.3 &$-0.027$ & $+0.018$ & $+0.693$ & $+0.881$ & $-0.201$ \\
0.4 &$+0.131$ & $-0.146$ & $+1.783$ & $+0.459$ & $-0.191$ \\
0.5 &$+0.365$ & $-0.308$ & $+2.614$ & $-0.619$& $-0.086$\\
 \hline
\end{tabular}

\begin{tabular}{ |p{1.8cm}|p{1.3cm}|p{1.3cm}|p{1.6cm}|p{2.0cm}|p{1.5cm}|}
 \hline
 \multicolumn{6}{|c|}{\Large{\Large $\Delta E=E_{VP}-E_{IVC}$ in meV for $\nu = 2$, $U=0.5$eV}}\\
 \hline
 $\alpha$ in eV$\times a$ & $\Delta E$ & $\Delta E_{kin}$ & $\Delta E_{Fock}/\tilde\alpha$ & $\Delta E_{Hartree}/\tilde \alpha\;$ &  $\Delta E_{U}/\tilde U$    \\
 \hline
0.2 &$+0.240$& $+0.176$ & $+1.776$ &  $-5.826$ & $+0.460$ \\
0.3 &$+0.030$& $+0.356$ & $+1.200$ &  $-8.336$ & $+0.834$ \\
0.4 &$-0.438$& $+0.424$ & $-0.518$ &  $-5.656$ & $+0.212$ \\
 \hline
\end{tabular}

\begin{tabular}{ |p{1.8cm}|p{1.3cm}|p{1.3cm}|p{1.6cm}|p{2.0cm}|p{1.5cm}|}
 \hline
 \multicolumn{6}{|c|}{\Large{\Large $\Delta E=E_{VP}-E_{IVC}$ in meV for $\nu = 2$, $U=4.0$eV}} \\
 \hline
 $\alpha$ in eV$\times a$ & $\Delta E$ & $\Delta E_{kin}$ & $\Delta E_{Fock}/\tilde\alpha$ & $\Delta E_{Hartree}/\tilde \alpha\;$ &  $\Delta E_{U}/\tilde U$    \\
 \hline
0.2 &$+0.050$& $+0.193$ & $-0.009$ &  $+1.091$ & $-0.222$ \\
0.3 &$+0.149$& $+0.059$ & $+1.481$ &  $+0.087$ & $-0.213$ \\
 \hline
\end{tabular}
\end{center}

\section{Renormalized band structures}
Here, we present some typical Hartree-Fock bands along the high symmetry lines. In Fig. \ref{BandsU0.5} and \ref{BandsU4.0}, the bands are shown for filling factor $\nu=-2$. In all cases, there is one split-band indicated by the blue curve. However, also other remote valence bands may contribute for filling factors below $\nu=-2$.
\begin{figure}[h]
    \includegraphics[scale=0.2]{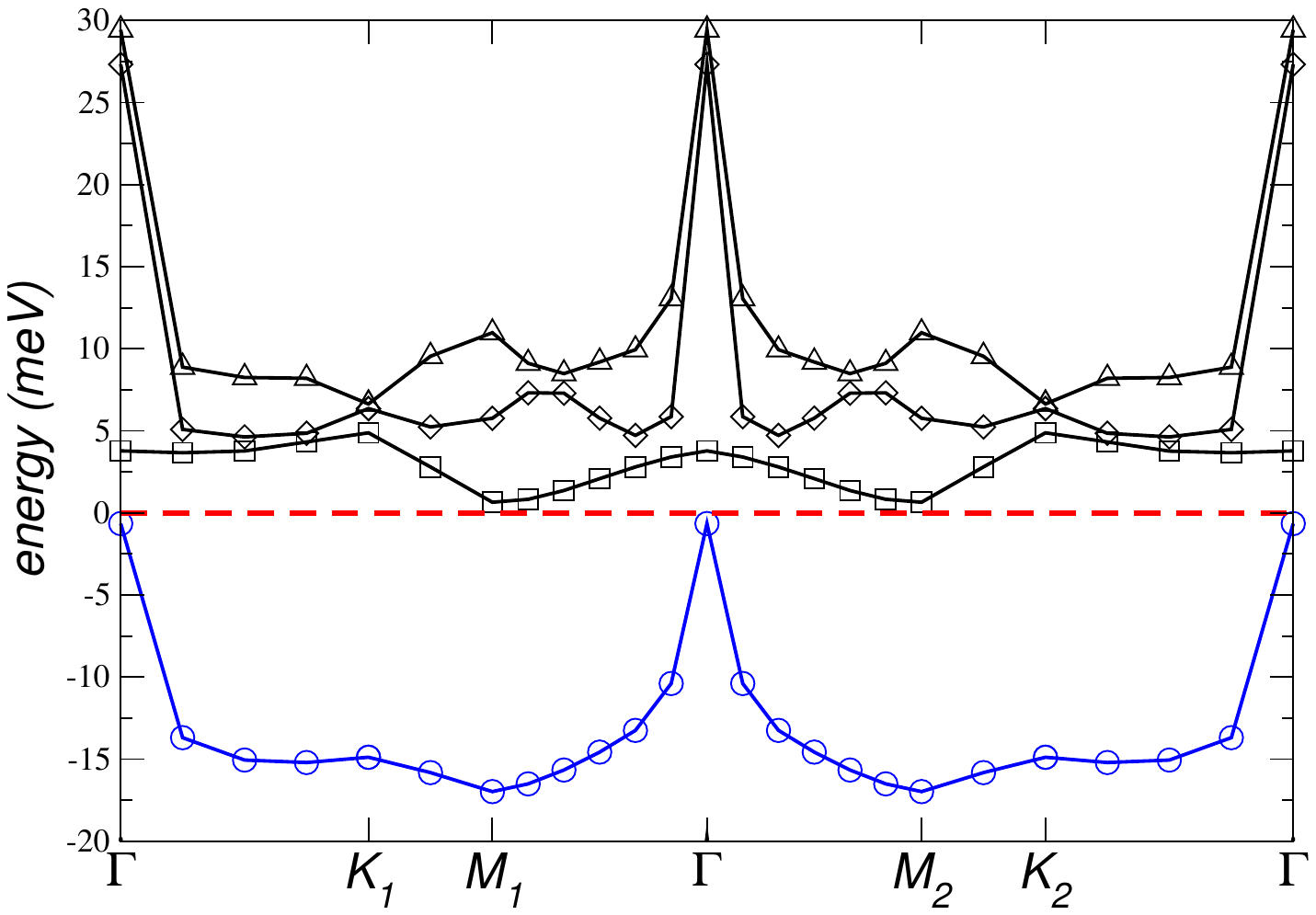}
    \includegraphics[scale=0.2]{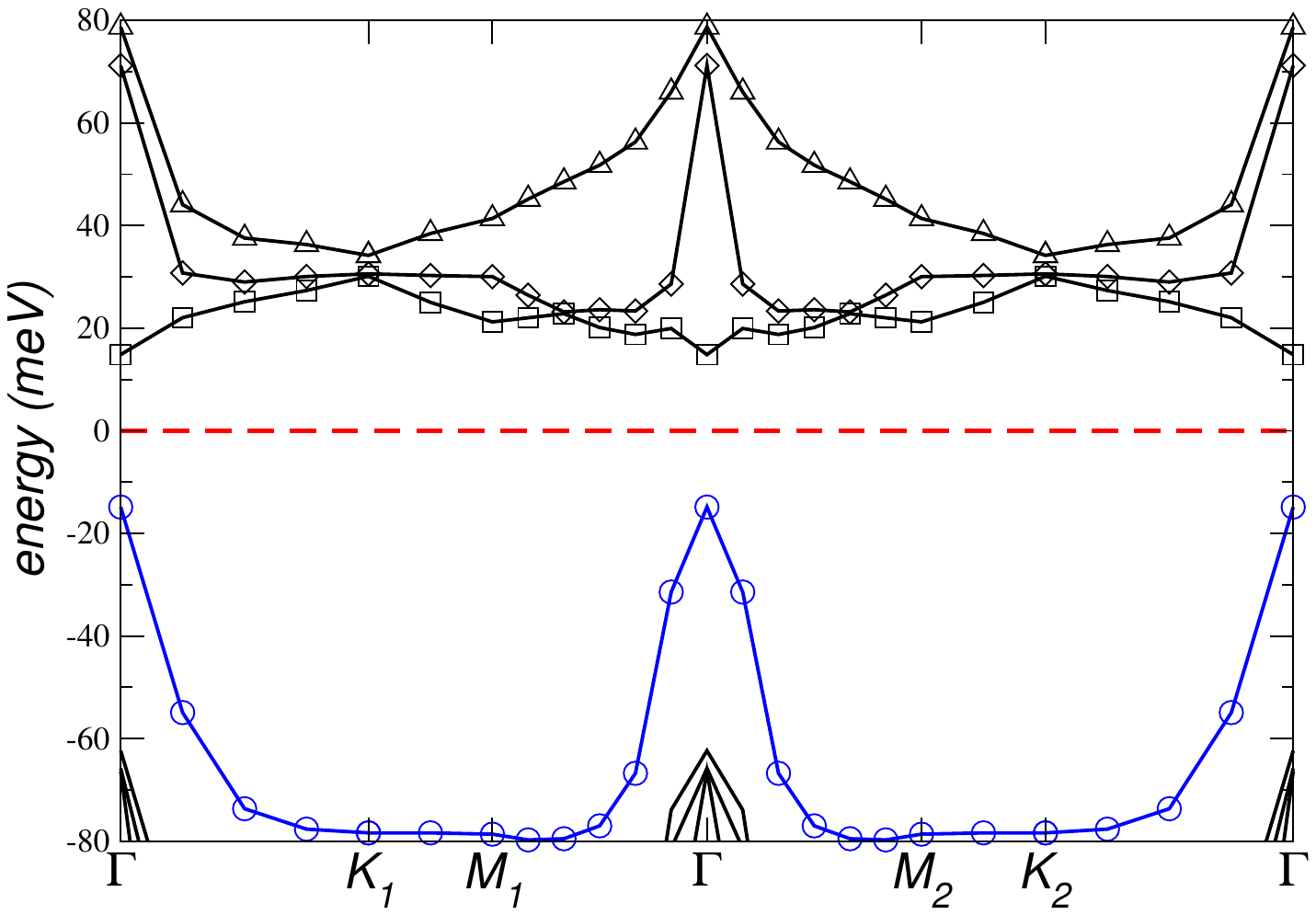}
    \caption{Hartree Fock bands for $\nu=-2$ and $U=0.5$eV at $\epsilon=60$ (left) and $\epsilon=12$ (right). The chemical potential is set to zero and indicated by the red dashed line.}
    \label{BandsU0.5}
\end{figure}
\begin{figure}[h]
    \includegraphics[scale=0.2]{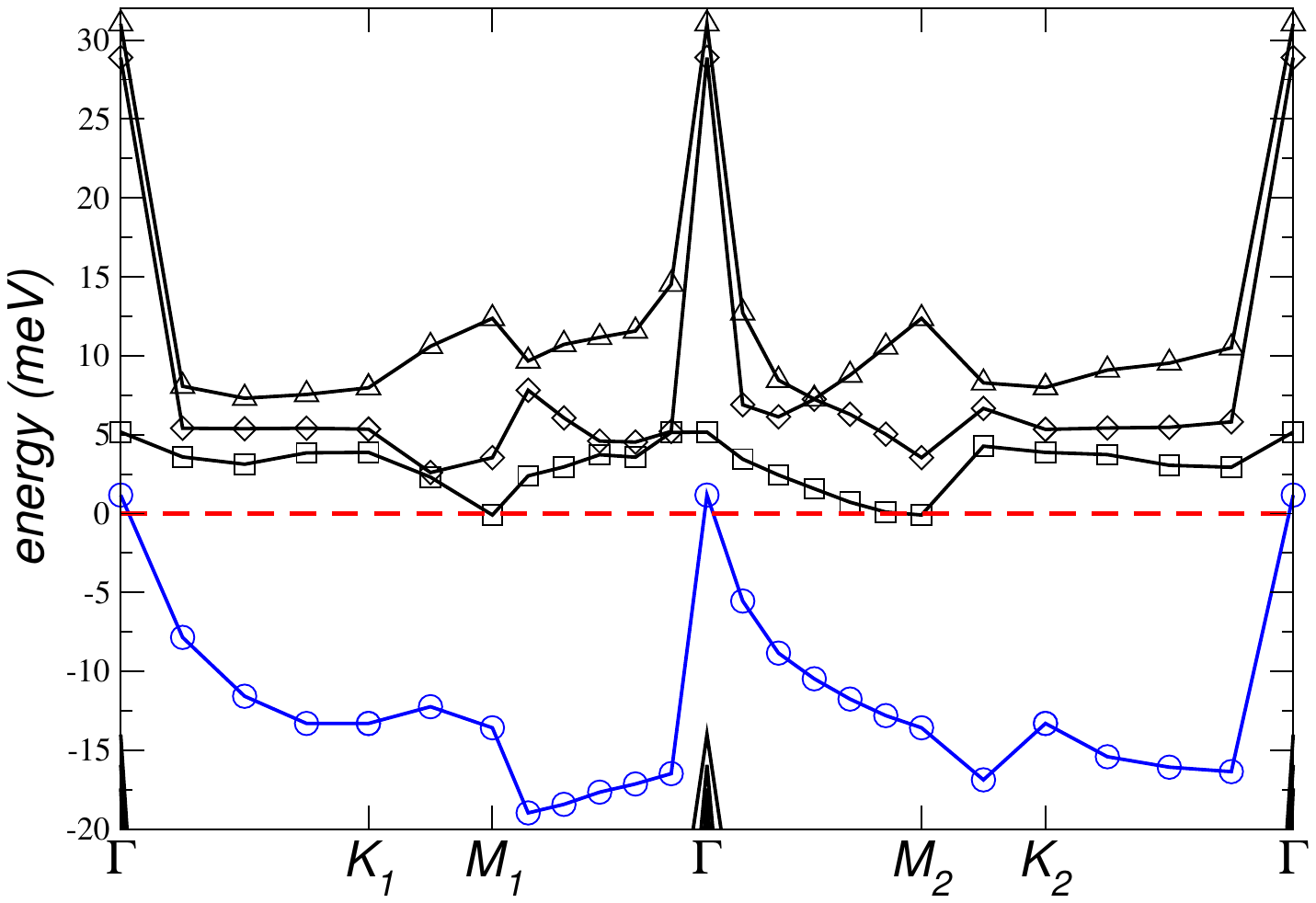}
    \includegraphics[scale=0.2]{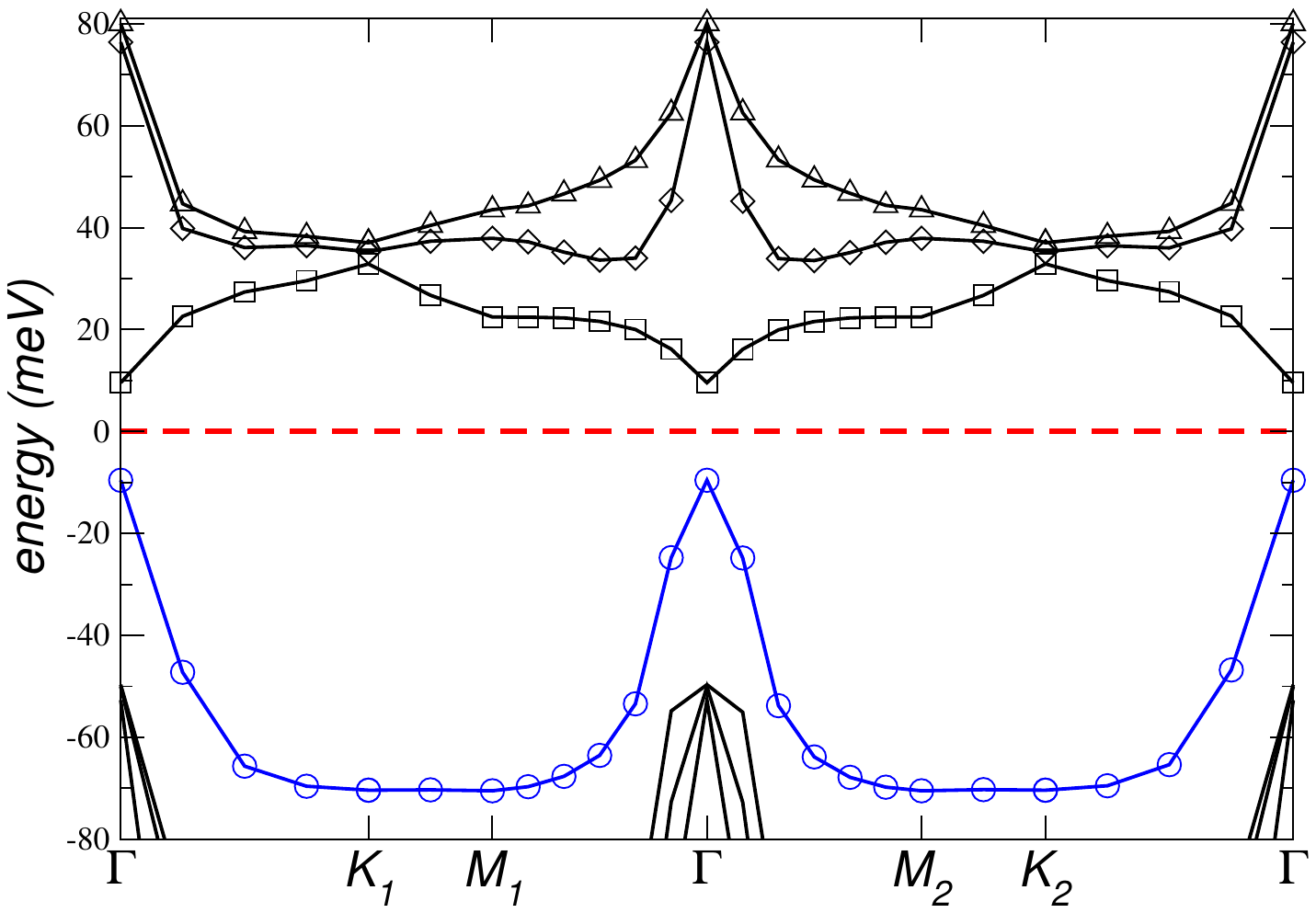}
    \caption{Hartree Fock bands for $\nu=-2$ and $U=4.0$eV at $\epsilon=60$ (left) and $\epsilon=12$ (right). The chemical potential is set to zero and indicated by the red dashed line.}
    \label{BandsU4.0}
\end{figure}

In Fig. \ref{BandsU0.5n2} and \ref{BandsU4.0n2}, the bands are shown for filling factor $\nu=2$. Again, in all cases there is one split-band indicated by the blue curve. In contrary to hole-doping, other remote conduction bands do not contribute for filling factors above $\nu=2$. 
\begin{figure}[h]
    \includegraphics[scale=0.2]{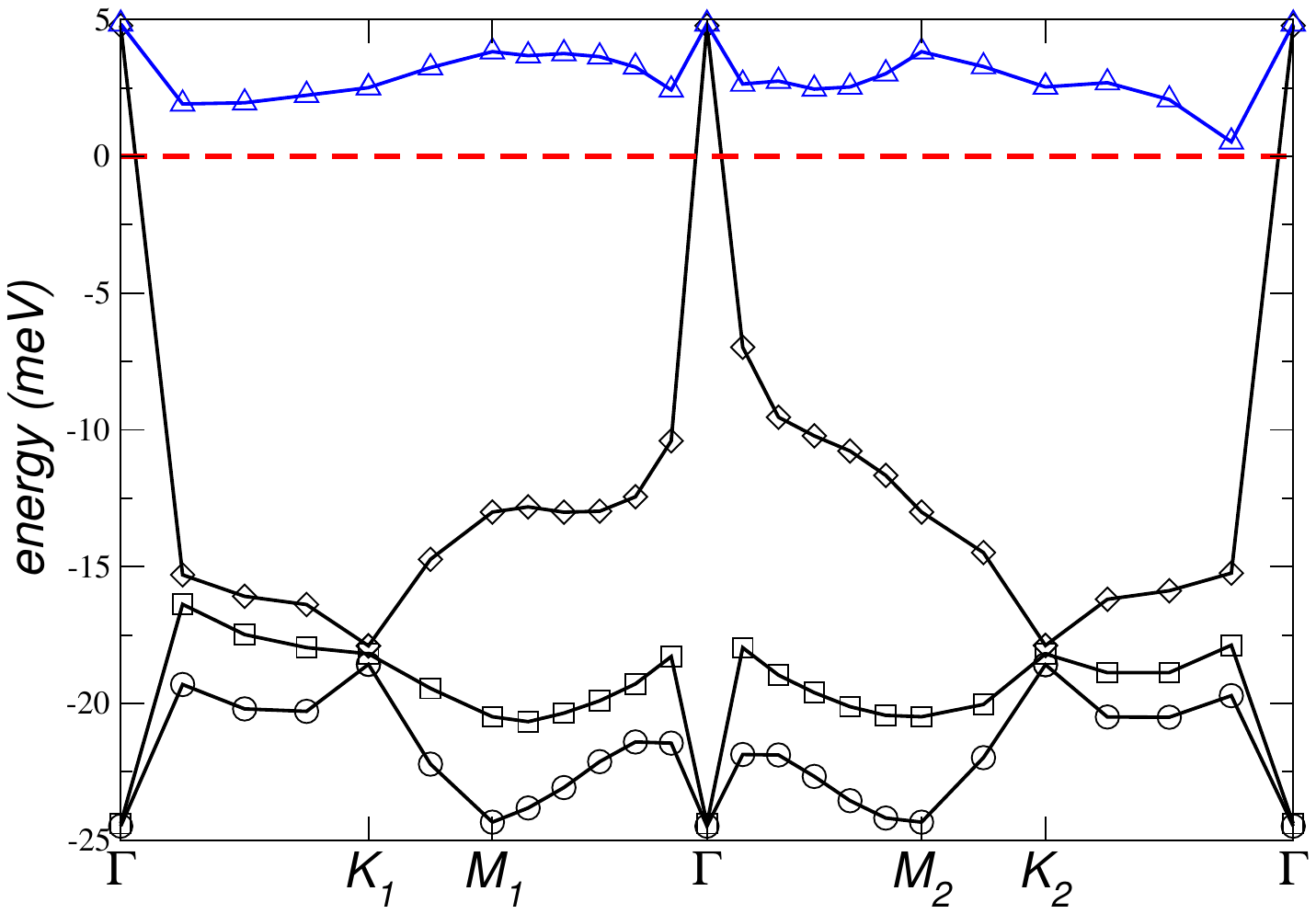}
    \includegraphics[scale=0.2]{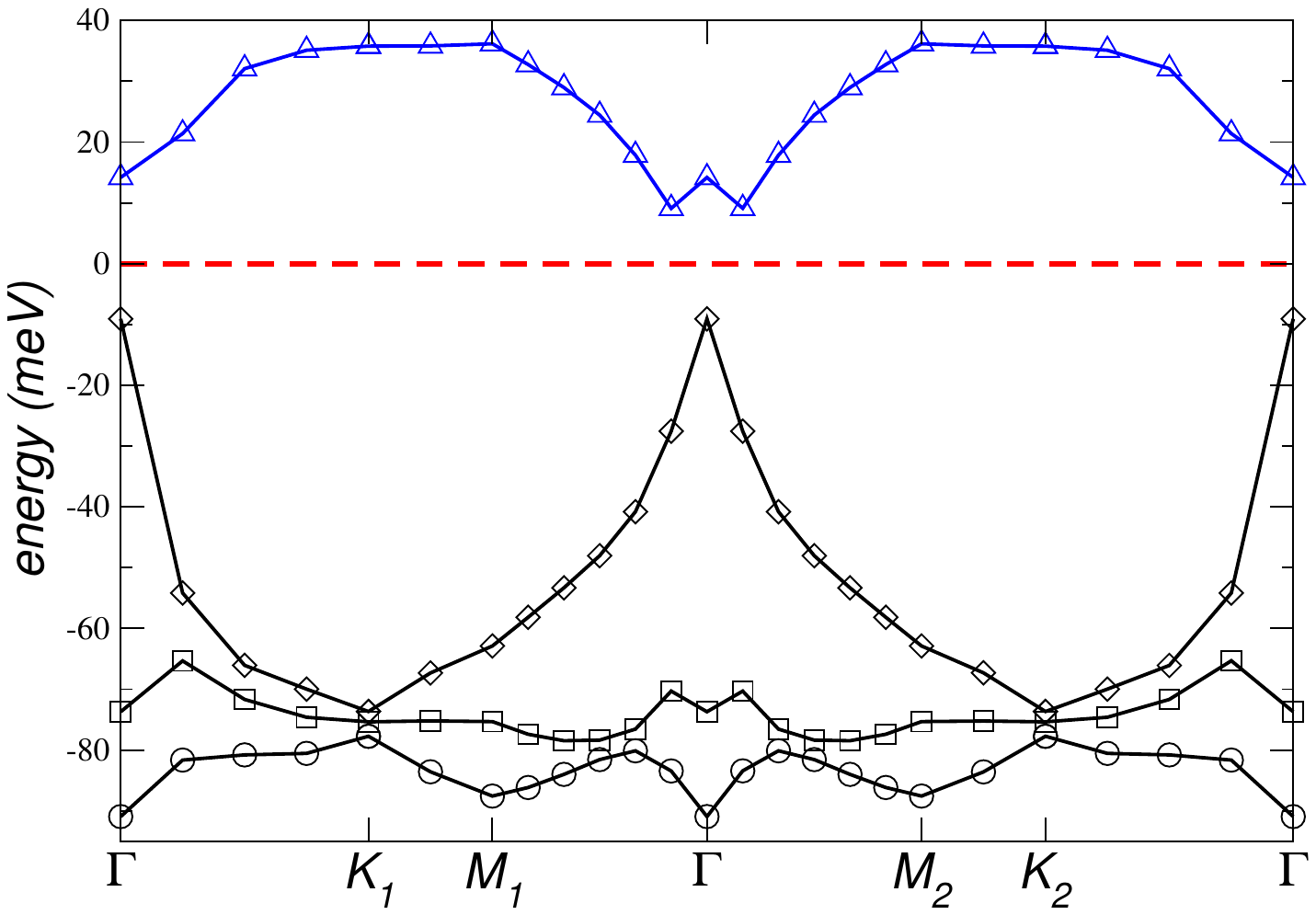}
    \caption{Hartree Fock bands for $\nu=2$ and $U=0.5$eV at $\epsilon=60$ (left) and $\epsilon=12$ (right). The chemical potential is set to zero and indicated by the red dashed line.}
    \label{BandsU0.5n2}
\end{figure}
\begin{figure}[h]
    \includegraphics[scale=0.2]{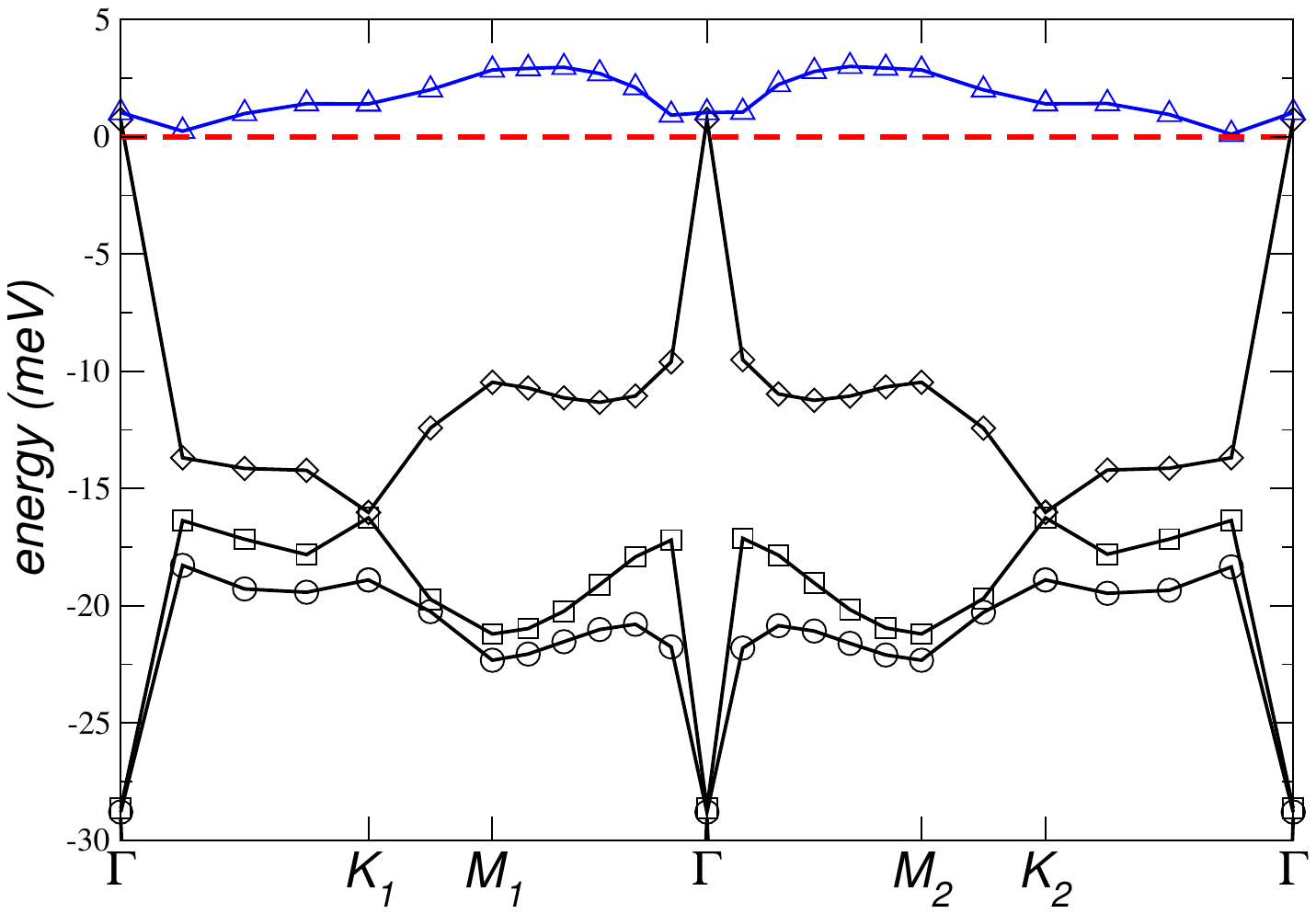}
    \includegraphics[scale=0.2]{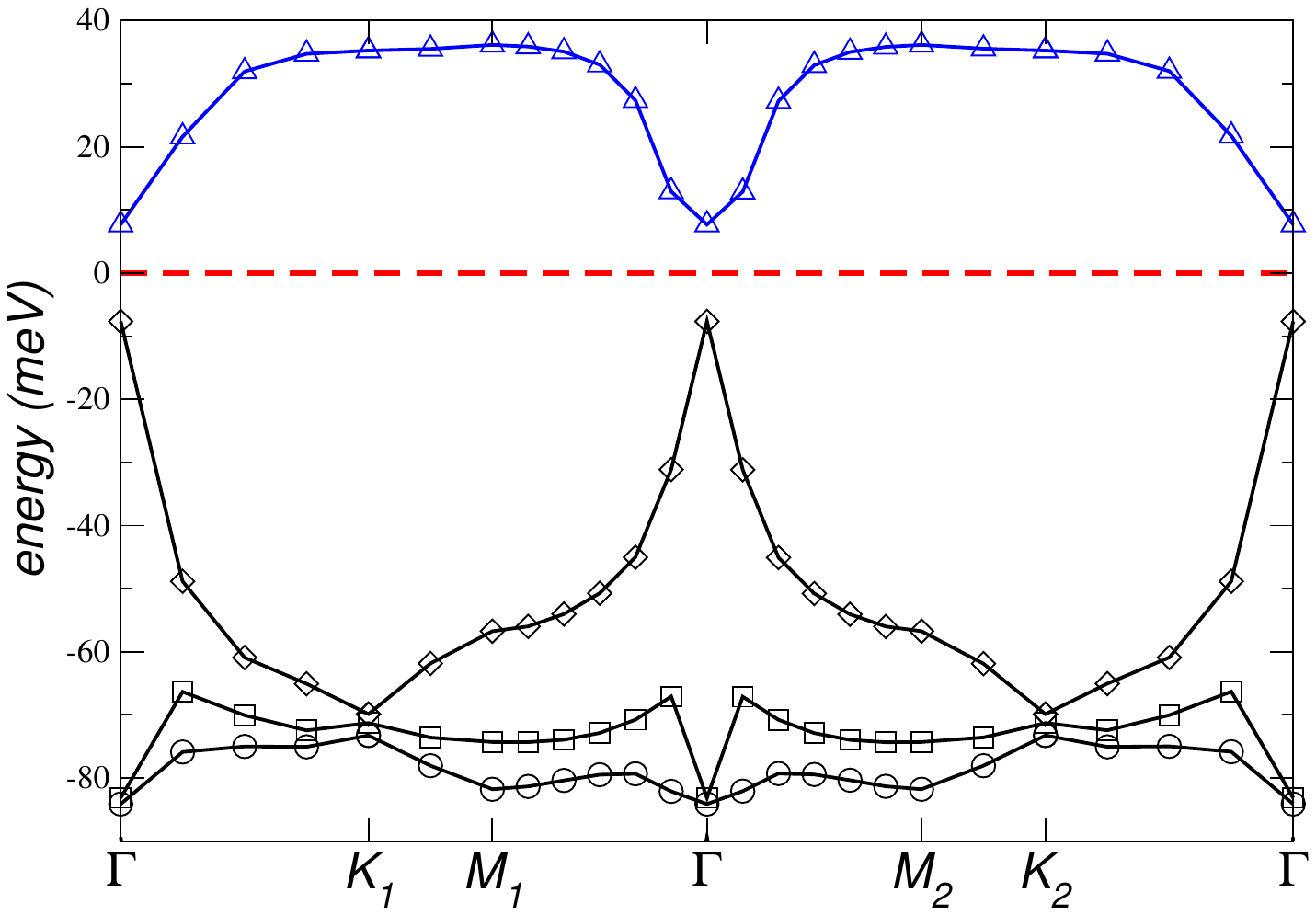}
    \caption{Hartree Fock bands for $\nu=2$ and $U=4.0$eV at $\epsilon=60$ (left) and $\epsilon=12$ (right). The chemical potential is set to zero and indicated by the red dashed line.}
    \label{BandsU4.0n2}
\end{figure}

The bands for $\nu=-2$ and $\nu=2$ do not display an obvious many-body particle-hole symmetry. Nevertheless, the phase diagram recovers the particle-hole symmetry in the strong-coupling limit. This is due to the fact that the interaction term which is particle-hole symmetric, becomes more dominant relative to the kinetic term which breaks the particle hole symmetry. 

Generally, the gap at $\nu=\pm2$ is topological with Chern number $|C|=1$. This changes at charge neutrality and where the gap becomes trivial. The band structure is shown in Figs. \ref{BandsU0.5n0} and \ref{BandsU4.0n0}.
\begin{figure}[h]
    \includegraphics[scale=0.2]{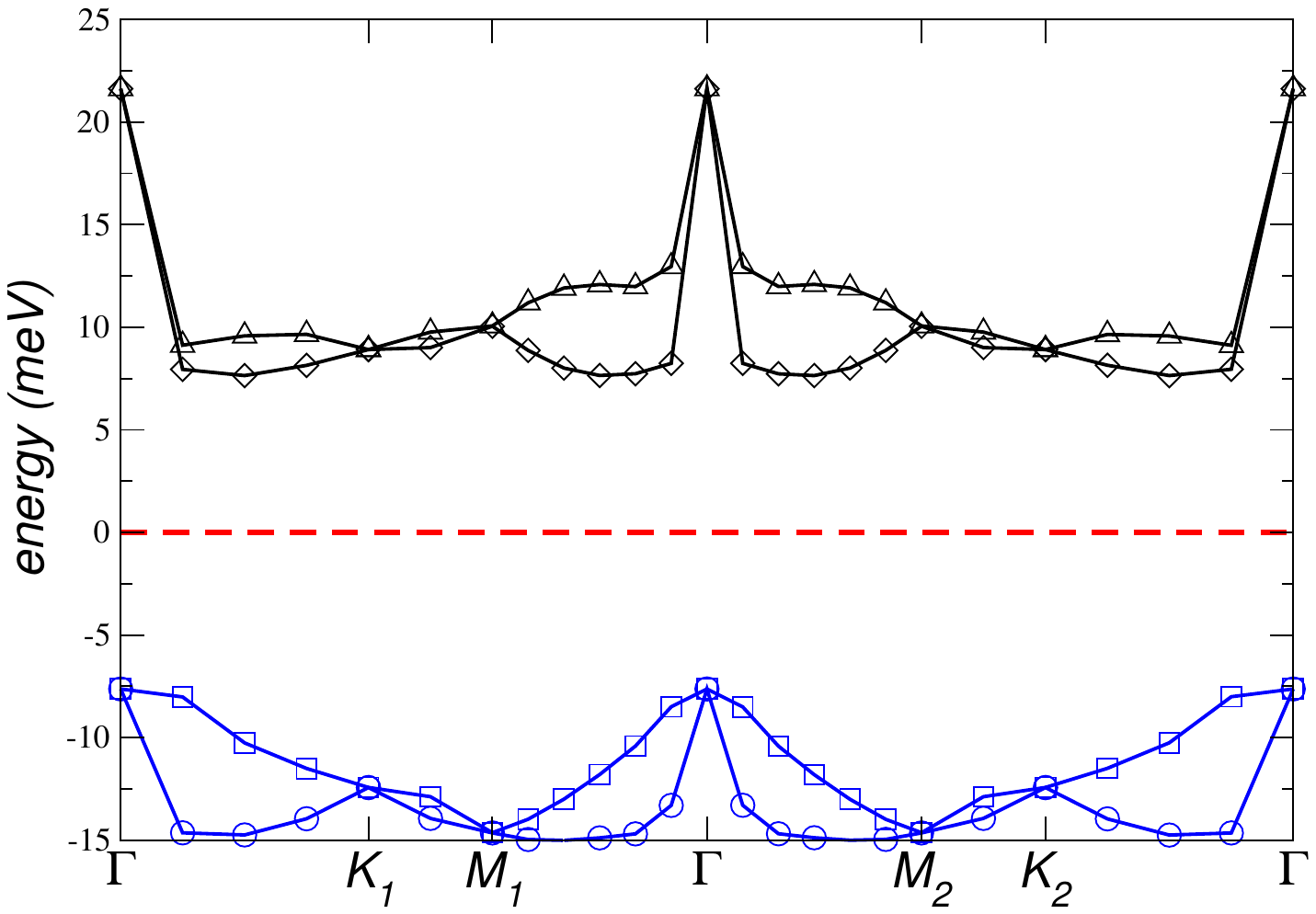}
    \includegraphics[scale=0.2]{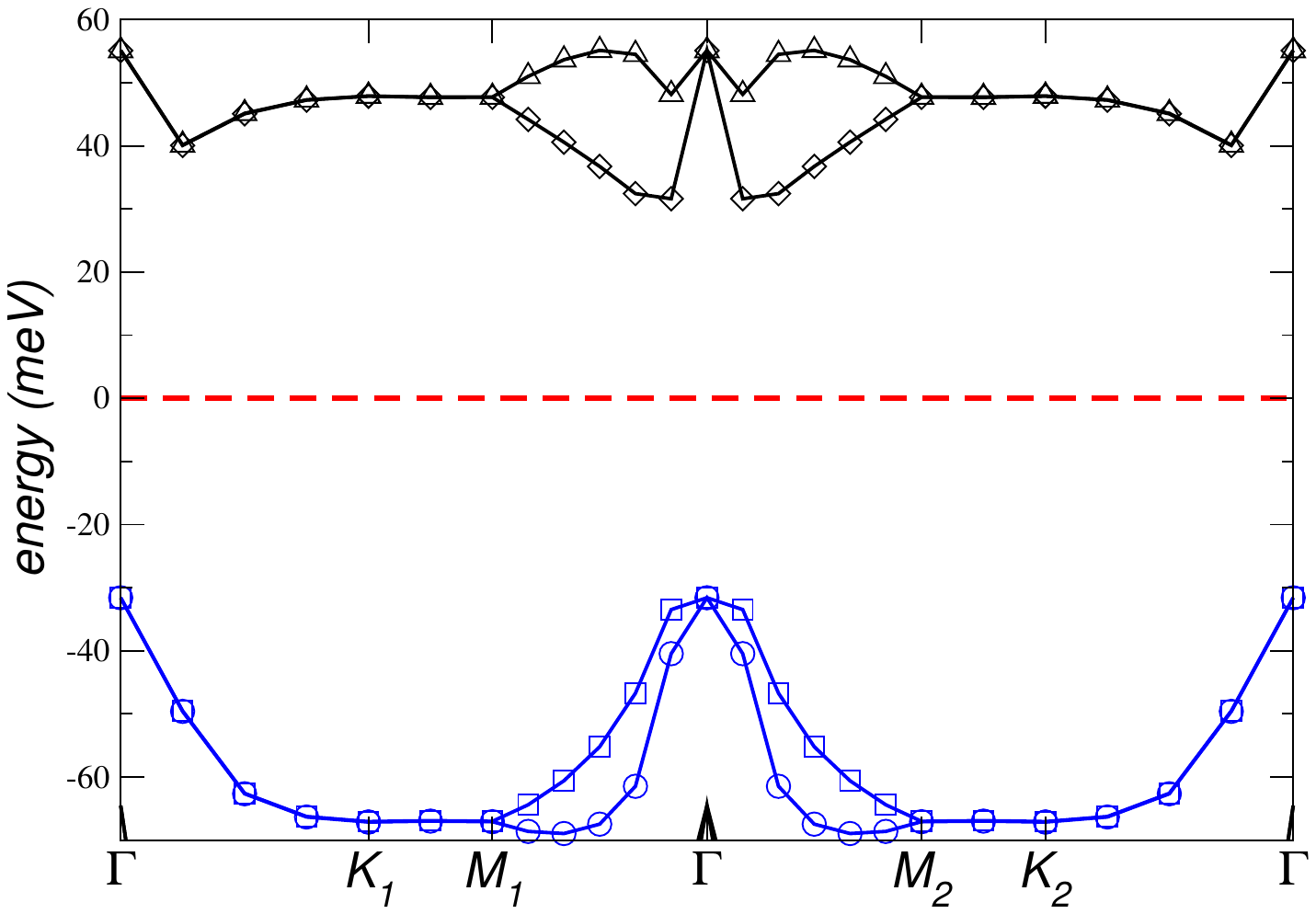}
    \caption{Hartree Fock bands for $\nu=0$ and $U=0.5$eV at $\epsilon=60$ (left) and $\epsilon=12$ (right). The chemical potential is set to zero and indicated by the red dashed line.}
    \label{BandsU0.5n0}
\end{figure}
\begin{figure}[h]
    \includegraphics[scale=0.2]{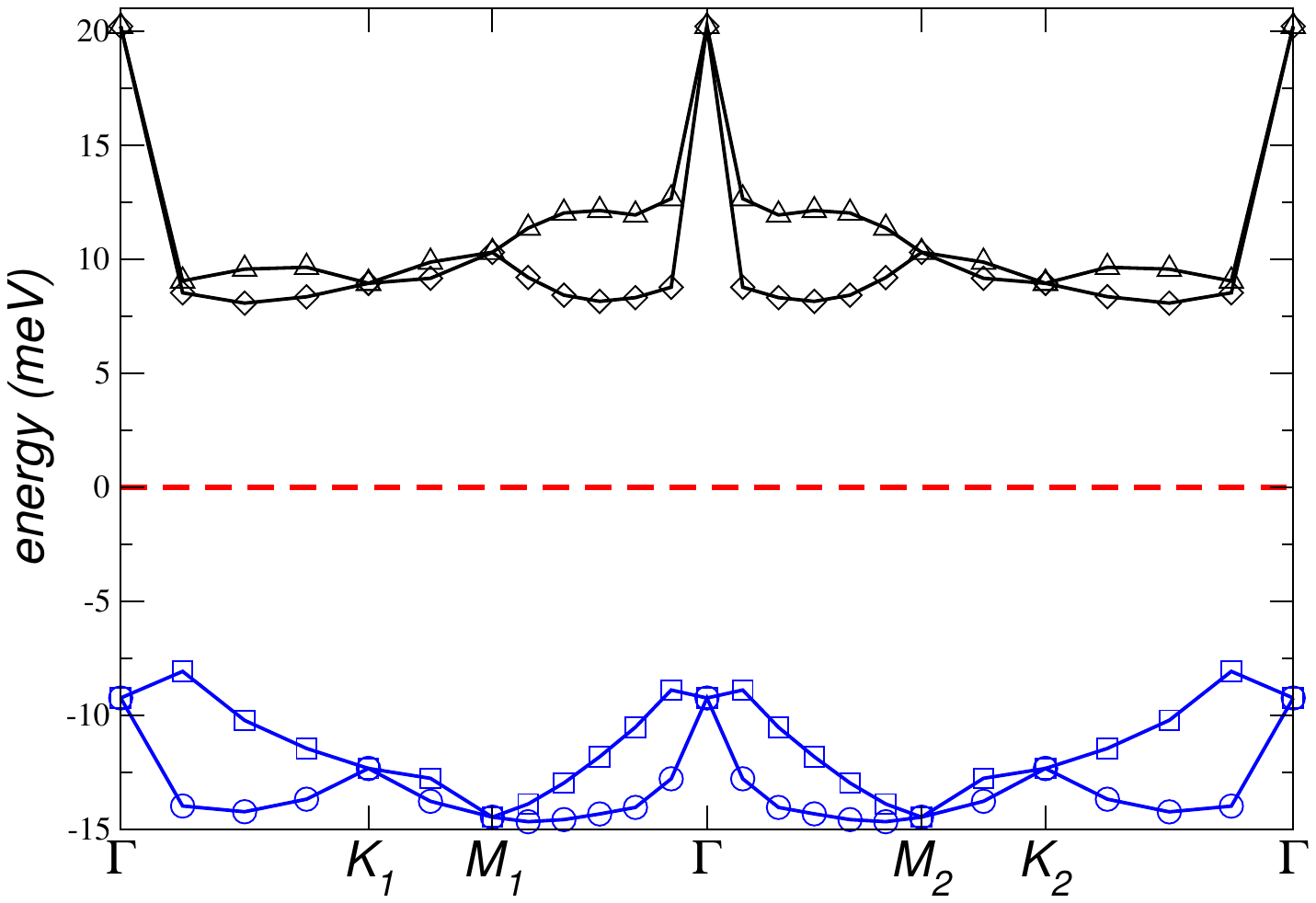}
    \includegraphics[scale=0.2]{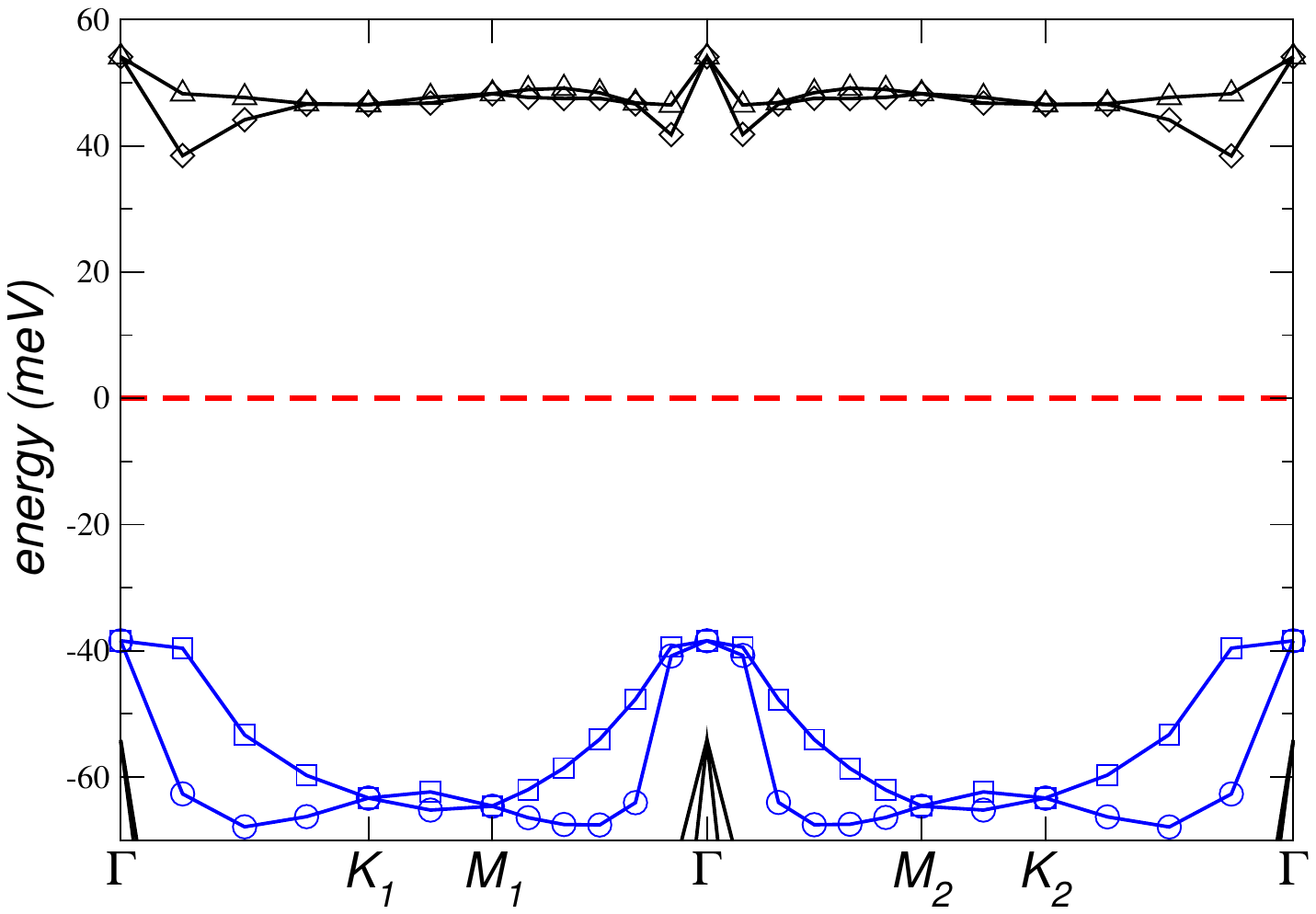}
    \caption{Hartree Fock bands for $\nu=0$ and $U=4.0$eV at $\epsilon=60$ (left) and $\epsilon=12$ (right). The chemical potential is set to zero and indicated by the red dashed line.}
    \label{BandsU4.0n0}
\end{figure}

\section{Renormalized contour plots}
Here, we present some typical contour plots for the Hartree-Fock bands. In Fig. \ref{ContourU0.5} and \ref{ContourU4.0}, the bands are shown for filling factor $\nu=-2$. The relative orientation is arbitrary. 
\begin{figure}[h]
    \includegraphics[scale=0.4]{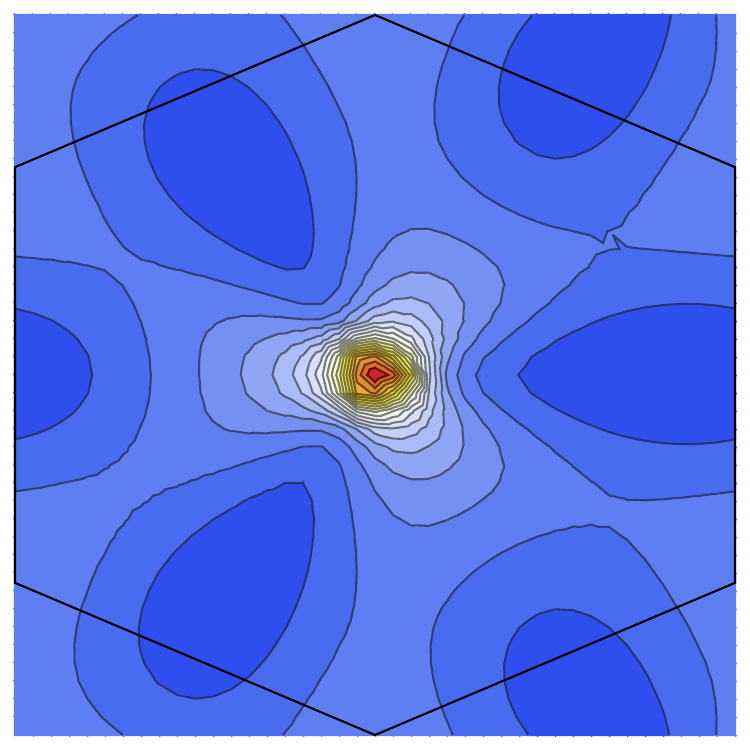}
    \includegraphics[scale=0.4]{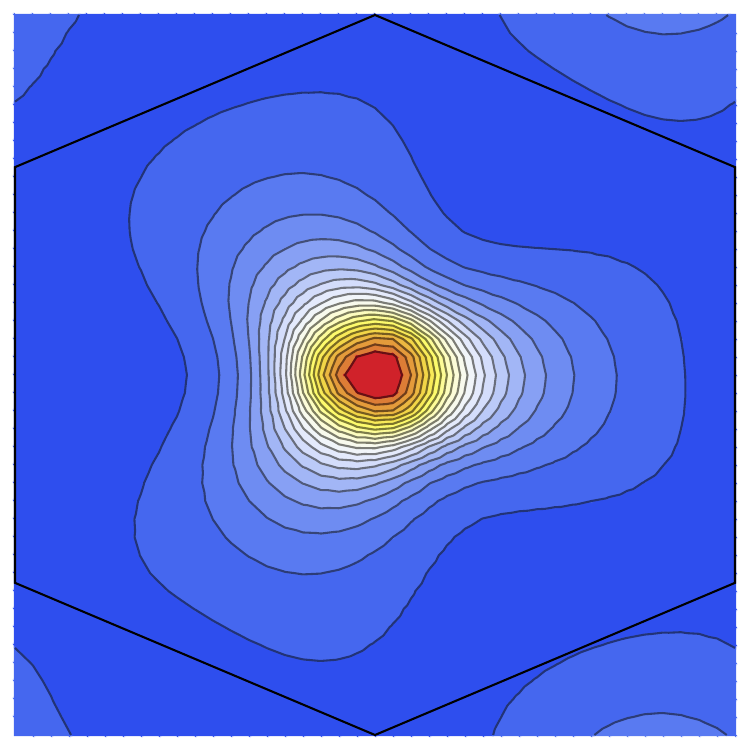}
    \caption{Energy contour plot of the Hartree Fock band for $\nu=-2$ and $U=0.5$eV at $\epsilon=60$ (left) and $\epsilon=12$ (right).}
    \label{ContourU0.5}
\end{figure}
\begin{figure}[h]
    \includegraphics[scale=0.4]{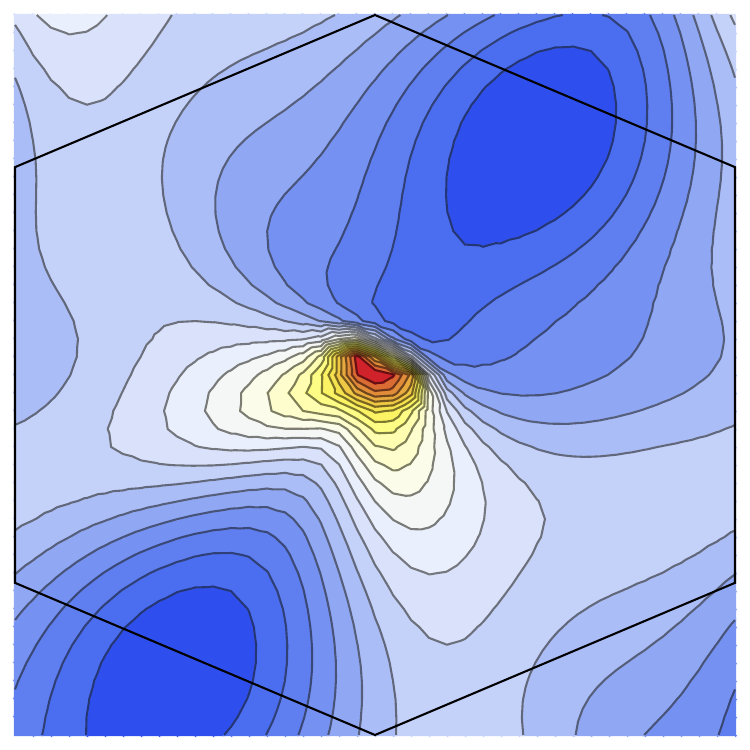}
    \includegraphics[scale=0.4]{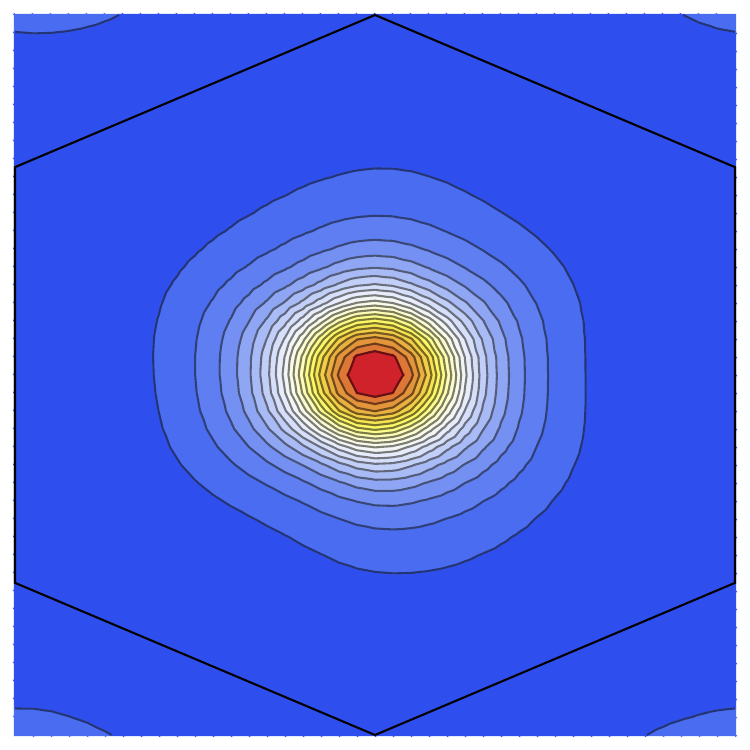}
    \caption{Energy contour plot of the Hartree Fock band for $\nu=-2$ and $U=4.0$eV at $\epsilon=60$ (left) and $\epsilon=12$ (right).}
    \label{ContourU4.0}
\end{figure}

In Fig. \ref{ContourU0.5n2} and \ref{ContourU4.0n2}, the bands are shown for filling factor $\nu=2$. Also here, the relative orientation is arbitrary. 
\begin{figure}[h]
    \includegraphics[scale=0.4]{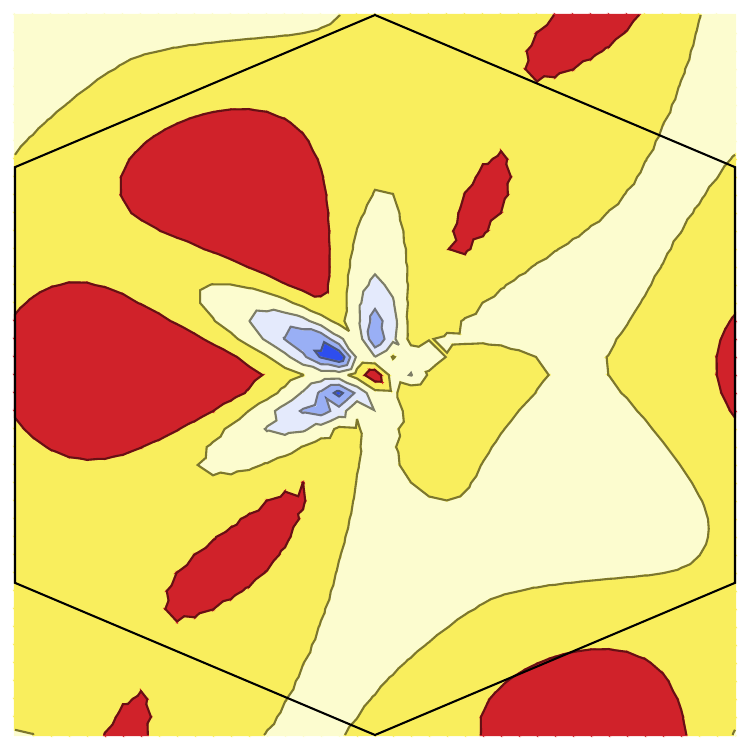}
    \includegraphics[scale=0.4]{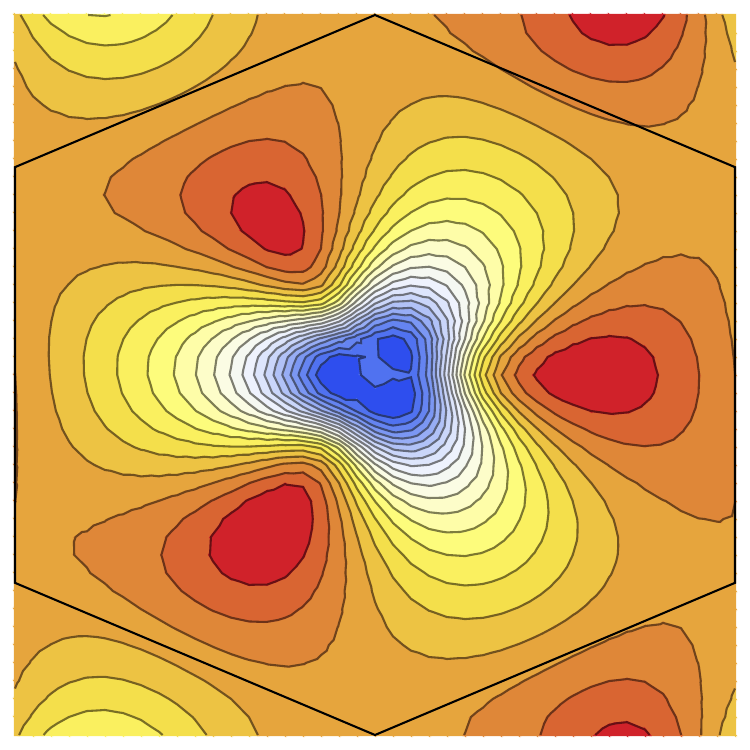}
    \caption{Energy contour plot of the Hartree Fock band for $\nu=2$ and $U=0.5$eV at $\epsilon=60$ (left) and $\epsilon=12$ (right). The chemical potential is set to zero and indicated by the red dashed line.}
    \label{ContourU0.5n2}
\end{figure}
\begin{figure}[h]
    \includegraphics[scale=0.4]{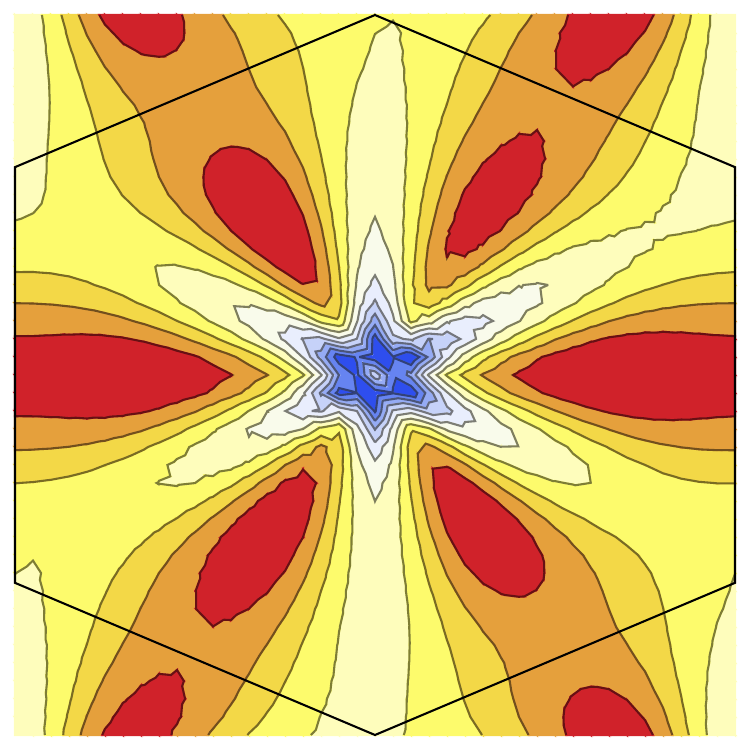}
    \includegraphics[scale=0.4]{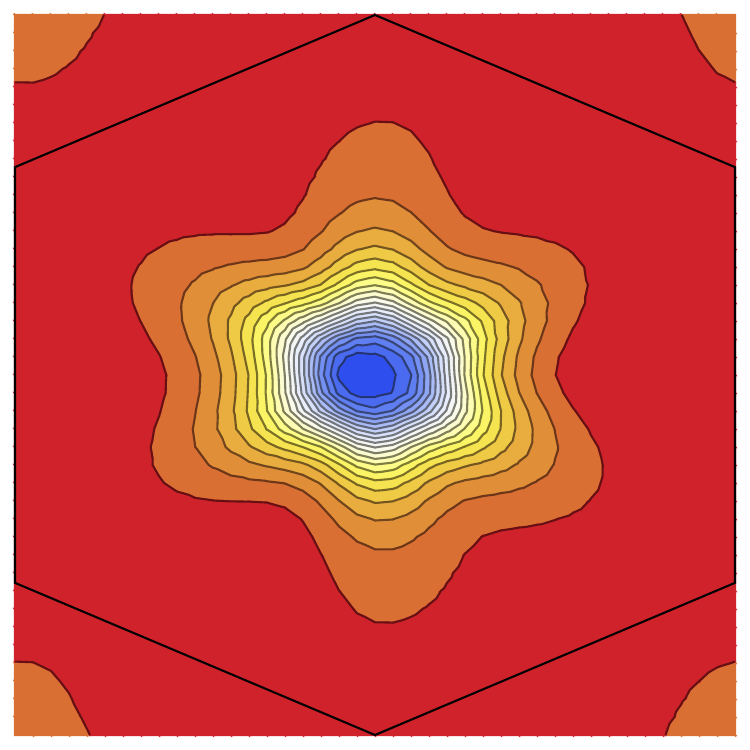}
    \caption{Energy contour plot of the Hartree Fock band for $\nu=2$ and $U=4.0$eV at $\epsilon=60$ (left) and $\epsilon=12$ (right). The chemical potential is set to zero and indicated by the red dashed line.}
    \label{ContourU4.0n2}
\end{figure}

Remarkably, there is a breaking of the $C_3$-symmetry for $n=-2$, $U=4$eV and $n=2$, $U=0.5$eV. In both cases, this symmetry is recovered if we set the onsite energy $U=0$. Also if we set $\alpha=0$ and $U$ finite, there is no symmetry breaking. This nematic state is thus due to an interplay of long-ranged and short-ranged interaction. Interestingly, this state generates an order parameter $\sigma_{x,y}\tau_{e,z}$ which is different for the two layers which can be interpreted as interlayer vortex.

\section{Appendix C: Space-dependent order parameter}
The order parameter presented in the phase diagram of the main text is the sum of local order parameters over the moir\'e unit cell. Here, we present the order parameter for valley polarization $\langle\tau_z\rangle$for some parameters, discriminating layer and sublattice.

We can also denote this quantity as additive flux as we use the formula
\begin{align}
\Im\Delta_{++}(\rr_i)=\Im\sum_{n=0,1,2}\rho_{\rr_i\pm\d_{n},\rr_i\pm\d_{n+1}}\;.
\end{align}
Alternatively, we can also define a different quantity which we label multiplicative flux as we use the formula
\begin{align}
\Im\Delta_{++}(\rr_i)=\Im\left(\prod_{n=0,1,2}\rho_{\rr_i\pm\d_{n},\rr_i\pm\d_{n+1}}\right)^{1/3}\;.
\end{align}
Even though both quantities yield similar order parameters, the spacial distribution is quite different. As the formulas using the sum have a clear interpretation also beyond the intra-sublattice, intravalley channel, we will only discuss the order parameters using this definition.

\begin{figure}[h]
    \includegraphics[scale=0.4]{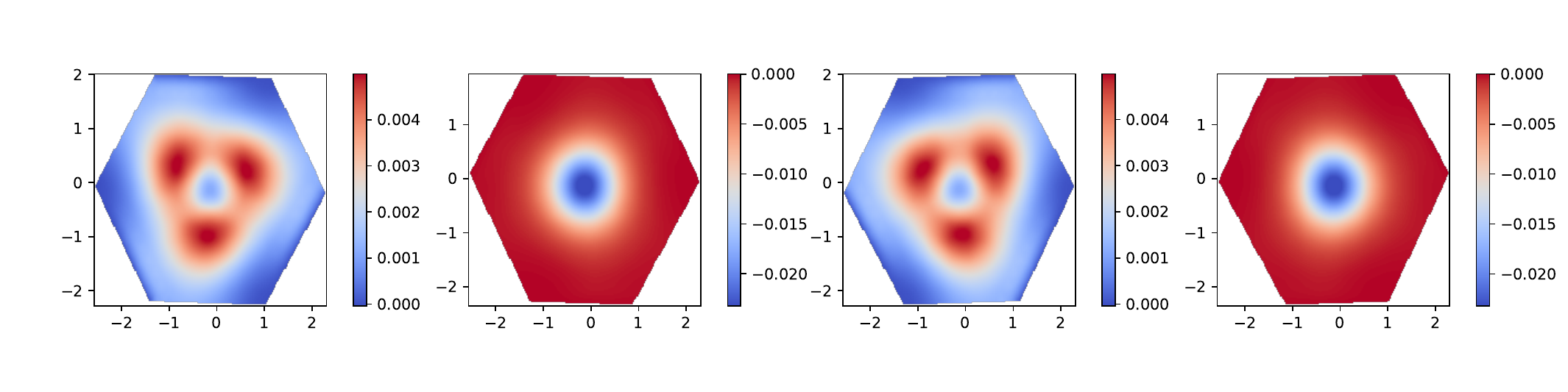}
    \includegraphics[scale=0.4]{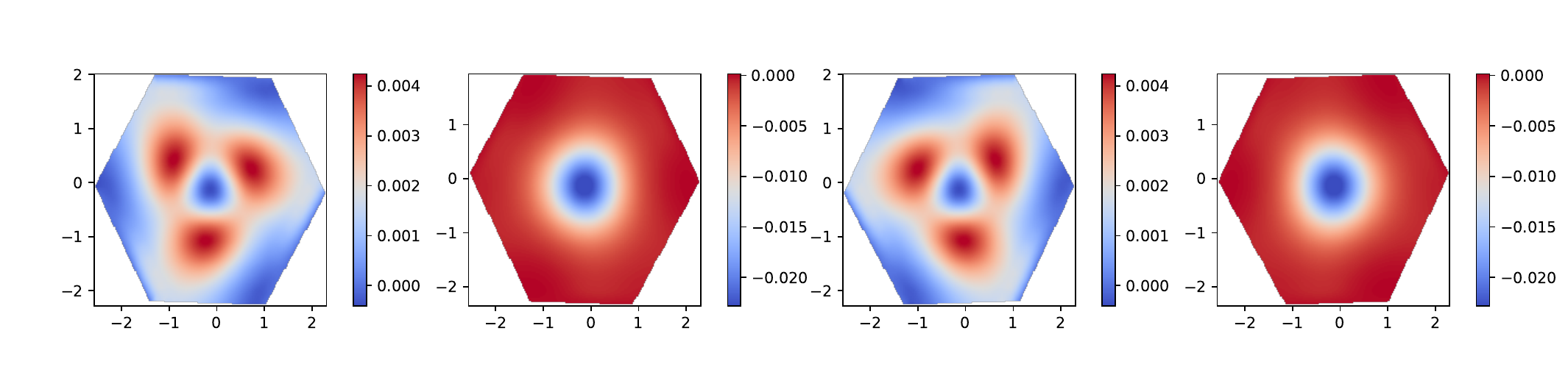}
    \caption{Additive flux of the Hartree Fock Hamiltonian for the sublattice A and layer 1 (left), B1 (center left), A2 (center right), B2 (right) with $U=0.5$eV for $\nu=-2$ and $\alpha=\frac{e^2}{4\pi\epsilon_0\epsilon}=0.1eV\times a$ (left) and $\alpha=\frac{e^2}{4\pi\epsilon_0\epsilon}=0.5eV\times a$ (right).}
    \label{ChernU0.5}
\end{figure}

\begin{figure}[h]
    \includegraphics[scale=0.4]{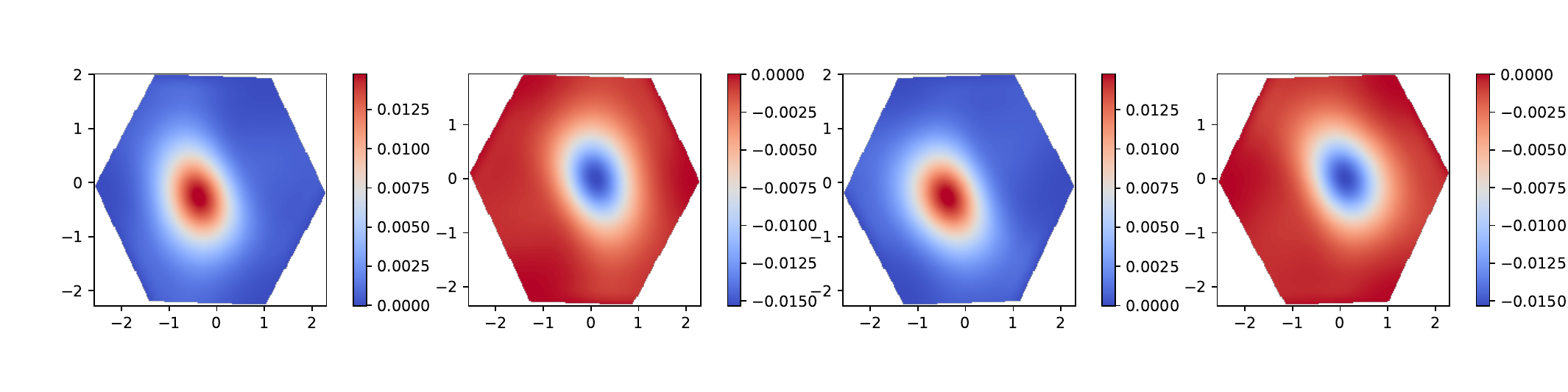}
    \includegraphics[scale=0.4]{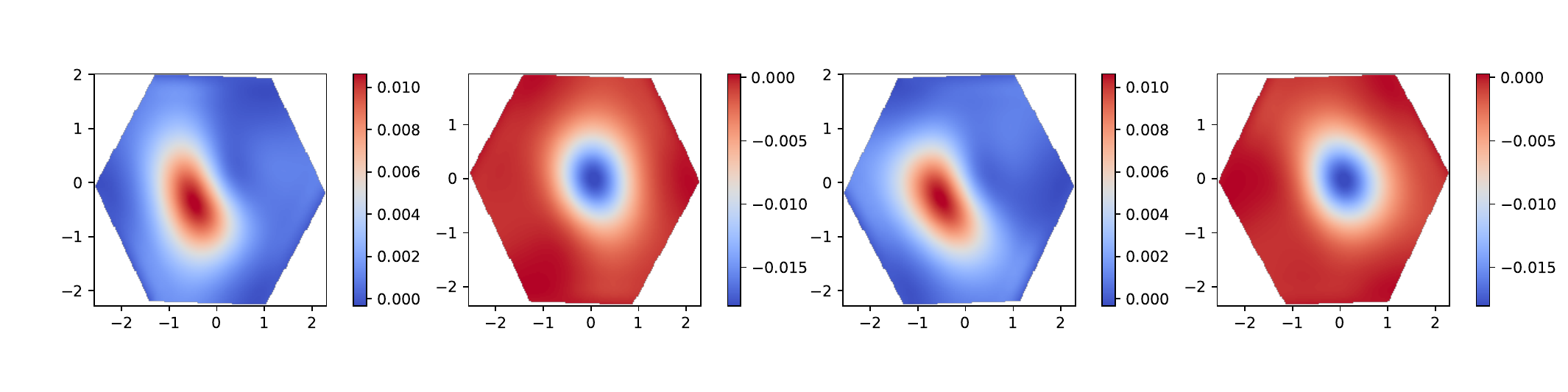}
    \caption{Additive flux of the Hartree Fock Hamiltonian for the sublattice A and layer 1 (left), B1 (center left), A2 (center right), B2 (right) with $U=4$eV for $n=-2$ and $\alpha=\frac{e^2}{4\pi\epsilon_0\epsilon}=0.1eV\times a$ (left) and $\alpha=\frac{e^2}{4\pi\epsilon_0\epsilon}=0.5eV\times a$ (right) for the valley polarized phase.}
    \label{BandsU4VP}
\end{figure}

\begin{figure}[h]
    \includegraphics[scale=0.4]{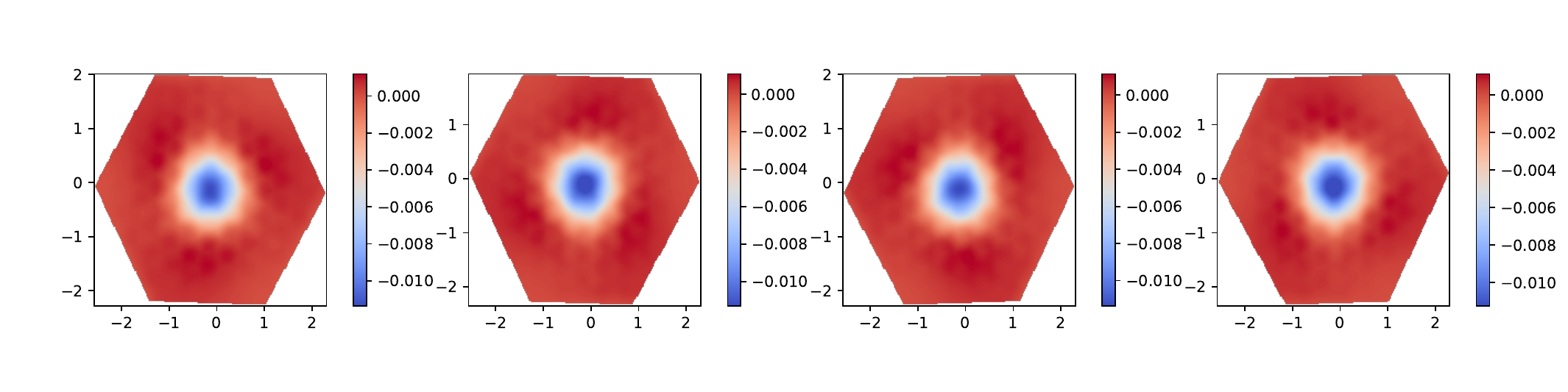}
    \includegraphics[scale=0.4]{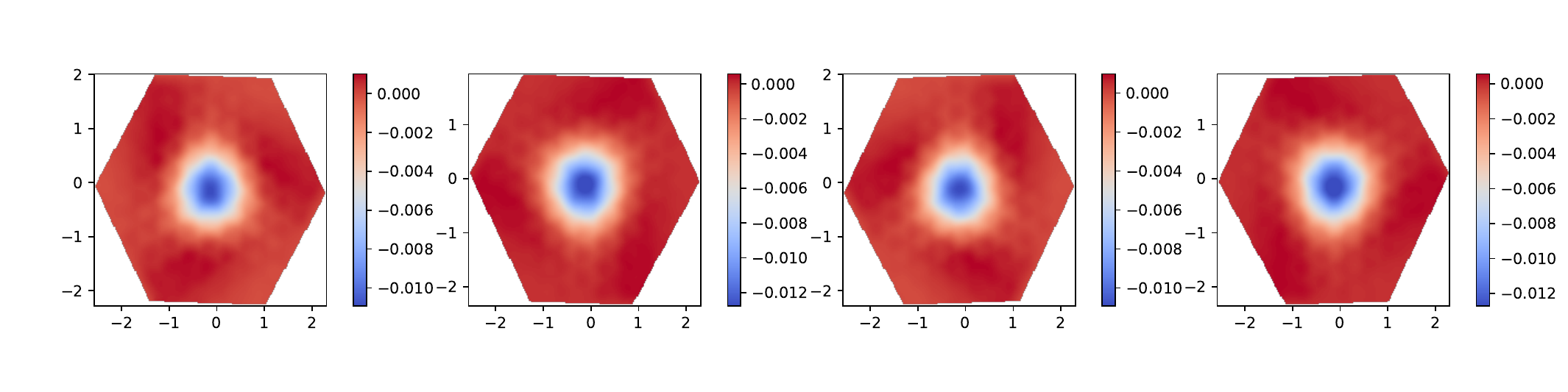}
    \caption{Additive flux of the Hartree Fock Hamiltonian for the sublattice A and layer 1 (left), B1 (center left), A2 (center right), B2 (right) with $U=4$eV for $n=-2$ and $\alpha=\frac{e^2}{4\pi\epsilon_0\epsilon}=0.1eV\times a$ (left) and $\alpha=\frac{e^2}{4\pi\epsilon_0\epsilon}=0.5eV\times a$ (right) for the valley coherent phase.}
    \label{BandsU4VC}
\end{figure}

\begin{figure}[h]
    \includegraphics[scale=0.4]{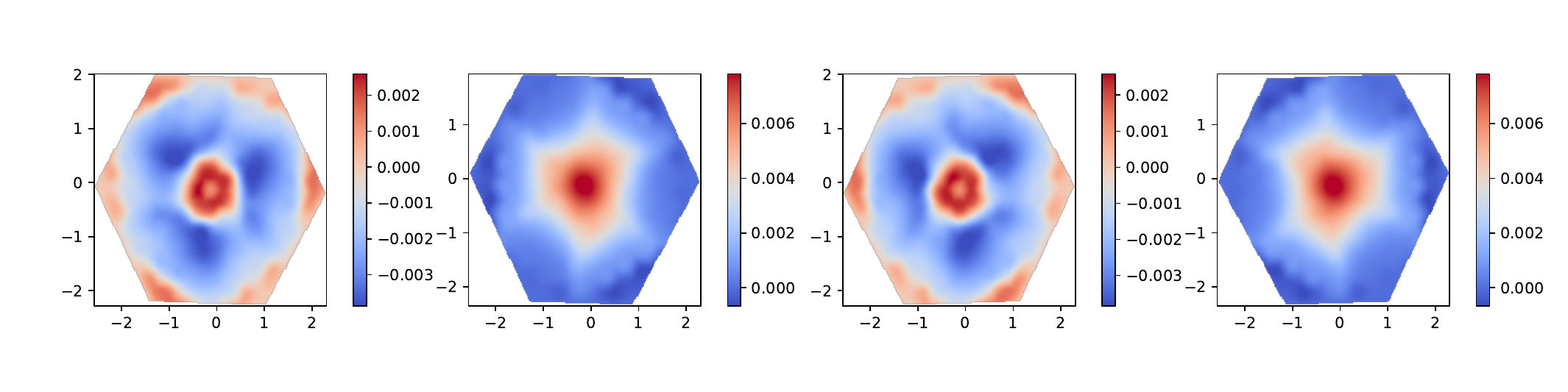}
    \includegraphics[scale=0.4]{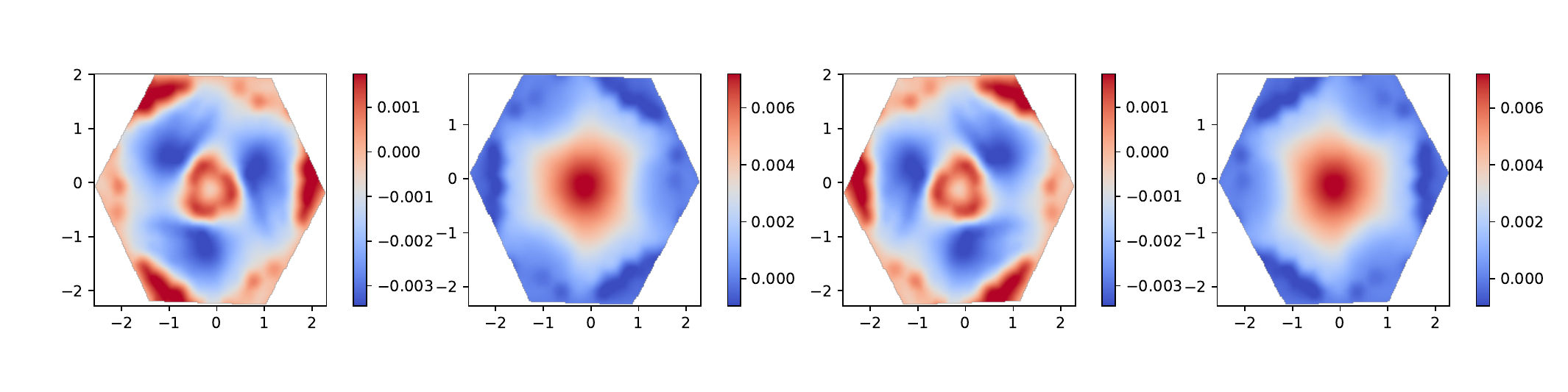}
    \caption{Multiplicable flux of the Hartree Fock Hamiltonian for the sublattice A and layer 1 (left), B1 (center left), A2 (center right), B2 (right) with $U=0.5$eV for $n=-2$ and $\alpha=\frac{e^2}{4\pi\epsilon_0\epsilon}=0.1eV\times a$ (left) and $\alpha=\frac{e^2}{4\pi\epsilon_0\epsilon}=0.5eV\times a$ (right).}
    \label{ChernU0.5}
\end{figure}

\begin{figure}[h]
    \includegraphics[scale=0.4]{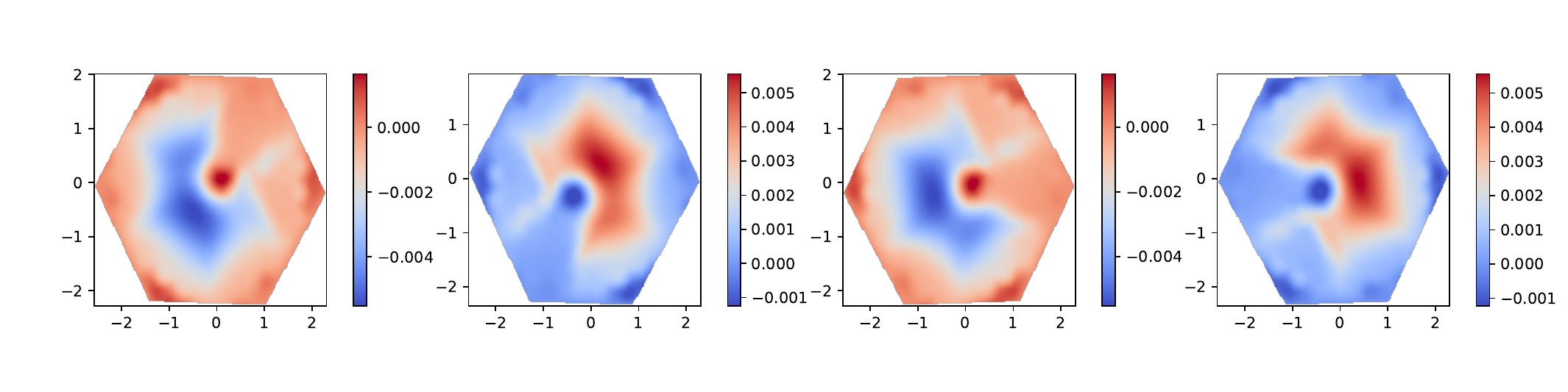}
    \includegraphics[scale=0.4]{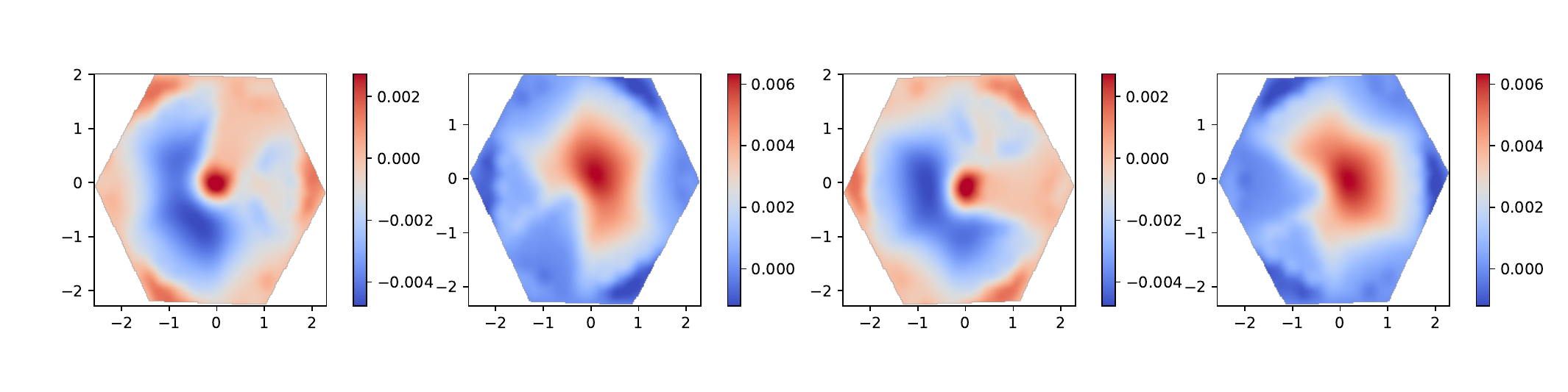}
    \caption{Multiplicable flux of the Hartree Fock Hamiltonian for the sublattice A and layer 1 (left), B1 (center left), A2 (center right), B2 (right) with $U=0.5$eV for $n=-2$ and $\alpha=\frac{e^2}{4\pi\epsilon_0\epsilon}=0.1eV\times a$ (left) and $\alpha=\frac{e^2}{4\pi\epsilon_0\epsilon}=0.5eV\times a$ (right) for the valley polarized phase.}
    \label{BandsU0.5}
\end{figure}

\begin{figure}[h]
    \includegraphics[scale=0.4]{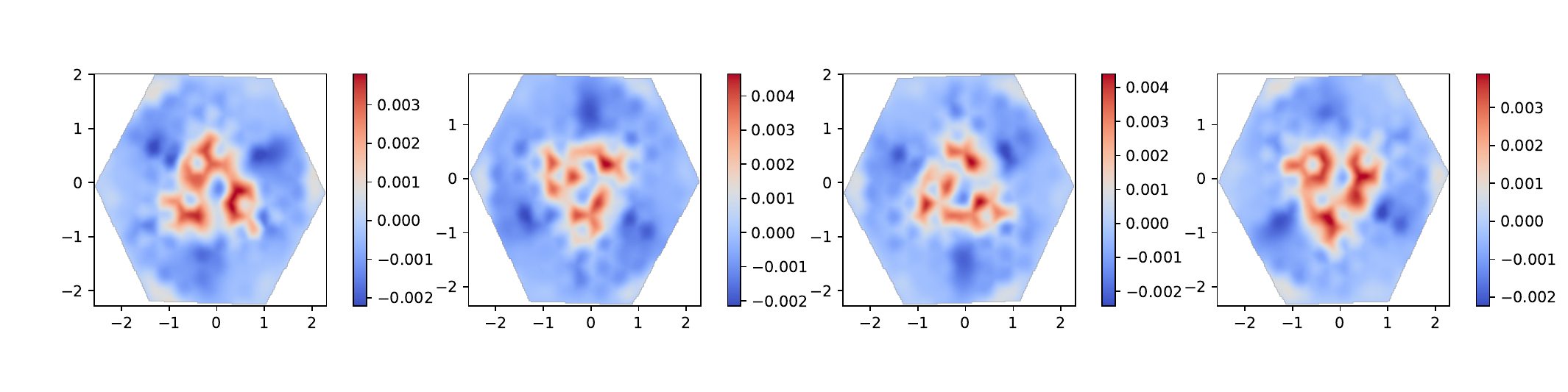}
    \includegraphics[scale=0.4]{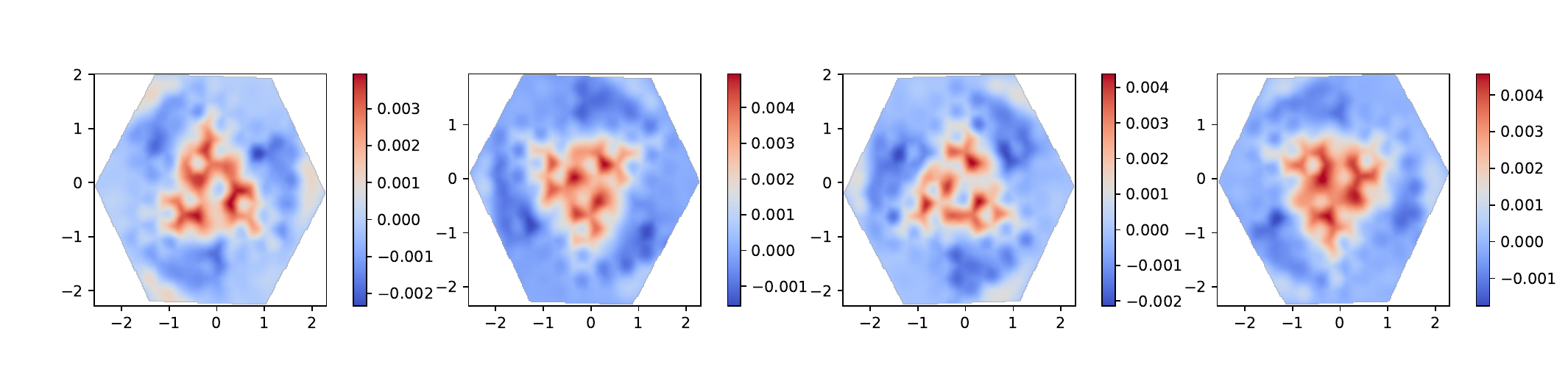}
    \caption{Multiplicable flux of the Hartree Fock Hamiltonian for the sublattice A and layer 1 (left), B1 (center left), A2 (center right), B2 (right) with $U=0.5$eV for $n=-2$ and $\alpha=\frac{e^2}{4\pi\epsilon_0\epsilon}=0.1eV\times a$ (left) and $\alpha=\frac{e^2}{4\pi\epsilon_0\epsilon}=0.5eV\times a$ (right) for the valley coherent phase.}
    \label{BandsU0.5}
\end{figure}

\end{widetext}


\begin{thebibliography}{63}%
\makeatletter
\providecommand \@ifxundefined [1]{%
 \@ifx{#1\undefined}
}%
\providecommand \@ifnum [1]{%
 \ifnum #1\expandafter \@firstoftwo
 \else \expandafter \@secondoftwo
 \fi
}%
\providecommand \@ifx [1]{%
 \ifx #1\expandafter \@firstoftwo
 \else \expandafter \@secondoftwo
 \fi
}%
\providecommand \natexlab [1]{#1}%
\providecommand \enquote  [1]{``#1''}%
\providecommand \bibnamefont  [1]{#1}%
\providecommand \bibfnamefont [1]{#1}%
\providecommand \citenamefont [1]{#1}%
\providecommand \href@noop [0]{\@secondoftwo}%
\providecommand \href [0]{\begingroup \@sanitize@url \@href}%
\providecommand \@href[1]{\@@startlink{#1}\@@href}%
\providecommand \@@href[1]{\endgroup#1\@@endlink}%
\providecommand \@sanitize@url [0]{\catcode `\\12\catcode `\$12\catcode
  `\&12\catcode `\#12\catcode `\^12\catcode `\_12\catcode `\%12\relax}%
\providecommand \@@startlink[1]{}%
\providecommand \@@endlink[0]{}%
\providecommand \url  [0]{\begingroup\@sanitize@url \@url }%
\providecommand \@url [1]{\endgroup\@href {#1}{\urlprefix }}%
\providecommand \urlprefix  [0]{URL }%
\providecommand \Eprint [0]{\href }%
\providecommand \doibase [0]{http://dx.doi.org/}%
\providecommand \selectlanguage [0]{\@gobble}%
\providecommand \bibinfo  [0]{\@secondoftwo}%
\providecommand \bibfield  [0]{\@secondoftwo}%
\providecommand \translation [1]{[#1]}%
\providecommand \BibitemOpen [0]{}%
\providecommand \bibitemStop [0]{}%
\providecommand \bibitemNoStop [0]{.\EOS\space}%
\providecommand \EOS [0]{\spacefactor3000\relax}%
\providecommand \BibitemShut  [1]{\csname bibitem#1\endcsname}%
\let\auto@bib@innerbib\@empty
%</preamble>
\bibitem [{\citenamefont {dos Santos}\ \emph {et~al.}(2007)\citenamefont {dos
  Santos}, \citenamefont {Peres},\ and\ \citenamefont
  {Neto}}]{Lopes_dos_Santos_2007}%
  \BibitemOpen
  \bibfield  {author} {\bibinfo {author} {\bibfnamefont {J.~M. B.~L.}\
  \bibnamefont {dos Santos}}, \bibinfo {author} {\bibfnamefont {N.~M.~R.}\
  \bibnamefont {Peres}}, \ and\ \bibinfo {author} {\bibfnamefont {A.~H.~C.}\
  \bibnamefont {Neto}},\ }\href
  {https://doi.org/10.1103%2Fphysrevlett.99.256802} {\bibfield  {journal}
  {\bibinfo  {journal} {Physical Review Letters}\ }\textbf {\bibinfo {volume}
  {99}} (\bibinfo {year} {2007})}\BibitemShut {NoStop}%
\bibitem [{\citenamefont {Li}\ \emph {et~al.}(2010)\citenamefont {Li},
  \citenamefont {Luican}, \citenamefont {Lopes~dos Santos}, \citenamefont
  {Castro~Neto}, \citenamefont {Reina}, \citenamefont {Kong},\ and\
  \citenamefont {Andrei}}]{Li10}%
  \BibitemOpen
  \bibfield  {author} {\bibinfo {author} {\bibfnamefont {G.}~\bibnamefont
  {Li}}, \bibinfo {author} {\bibfnamefont {A.}~\bibnamefont {Luican}}, \bibinfo
  {author} {\bibfnamefont {J.~M.~B.}\ \bibnamefont {Lopes~dos Santos}},
  \bibinfo {author} {\bibfnamefont {A.~H.}\ \bibnamefont {Castro~Neto}},
  \bibinfo {author} {\bibfnamefont {A.}~\bibnamefont {Reina}}, \bibinfo
  {author} {\bibfnamefont {J.}~\bibnamefont {Kong}}, \ and\ \bibinfo {author}
  {\bibfnamefont {E.~Y.}\ \bibnamefont {Andrei}},\ }\href {\doibase
  10.1038/nphys1463} {\bibfield  {journal} {\bibinfo  {journal} {Nature
  Physics}\ }\textbf {\bibinfo {volume} {6}},\ \bibinfo {pages} {109} (\bibinfo
  {year} {2010})}\BibitemShut {NoStop}%
\bibitem [{\citenamefont {Su\'arez~Morell}\ \emph {et~al.}(2010)\citenamefont
  {Su\'arez~Morell}, \citenamefont {Correa}, \citenamefont {Vargas},
  \citenamefont {Pacheco},\ and\ \citenamefont {Barticevic}}]{suarezmorell}%
  \BibitemOpen
  \bibfield  {author} {\bibinfo {author} {\bibfnamefont {E.}~\bibnamefont
  {Su\'arez~Morell}}, \bibinfo {author} {\bibfnamefont {J.~D.}\ \bibnamefont
  {Correa}}, \bibinfo {author} {\bibfnamefont {P.}~\bibnamefont {Vargas}},
  \bibinfo {author} {\bibfnamefont {M.}~\bibnamefont {Pacheco}}, \ and\
  \bibinfo {author} {\bibfnamefont {Z.}~\bibnamefont {Barticevic}},\ }\href
  {\doibase 10.1103/PhysRevB.82.121407} {\bibfield  {journal} {\bibinfo
  {journal} {Phys. Rev. B}\ }\textbf {\bibinfo {volume} {82}},\ \bibinfo
  {pages} {121407} (\bibinfo {year} {2010})}\BibitemShut {NoStop}%
\bibitem [{\citenamefont {Bistritzer}\ and\ \citenamefont
  {MacDonald}(2011)}]{bistritzer}%
  \BibitemOpen
  \bibfield  {author} {\bibinfo {author} {\bibfnamefont {R.}~\bibnamefont
  {Bistritzer}}\ and\ \bibinfo {author} {\bibfnamefont {A.~H.}\ \bibnamefont
  {MacDonald}},\ }\href {\doibase 10.1073/pnas.1108174108} {\bibfield
  {journal} {\bibinfo  {journal} {Proceedings of the National Academy of
  Sciences}\ }\textbf {\bibinfo {volume} {108}},\ \bibinfo {pages} {12233}
  (\bibinfo {year} {2011})}\BibitemShut {NoStop}%
\bibitem [{\citenamefont {Cao}\ \emph {et~al.}(2018{\natexlab{a}})\citenamefont
  {Cao}, \citenamefont {Fatemi}, \citenamefont {Demir}, \citenamefont {Fang},
  \citenamefont {Tomarken}, \citenamefont {Luo}, \citenamefont
  {Sanchez-Yamagishi}, \citenamefont {Watanabe}, \citenamefont {Taniguchi},
  \citenamefont {Kaxiras}, \citenamefont {Ashoori},\ and\ \citenamefont
  {Jarillo-Herrero}}]{Cao_2018}%
  \BibitemOpen
  \bibfield  {author} {\bibinfo {author} {\bibfnamefont {Y.}~\bibnamefont
  {Cao}}, \bibinfo {author} {\bibfnamefont {V.}~\bibnamefont {Fatemi}},
  \bibinfo {author} {\bibfnamefont {A.}~\bibnamefont {Demir}}, \bibinfo
  {author} {\bibfnamefont {S.}~\bibnamefont {Fang}}, \bibinfo {author}
  {\bibfnamefont {S.~L.}\ \bibnamefont {Tomarken}}, \bibinfo {author}
  {\bibfnamefont {J.~Y.}\ \bibnamefont {Luo}}, \bibinfo {author} {\bibfnamefont
  {J.~D.}\ \bibnamefont {Sanchez-Yamagishi}}, \bibinfo {author} {\bibfnamefont
  {K.}~\bibnamefont {Watanabe}}, \bibinfo {author} {\bibfnamefont
  {T.}~\bibnamefont {Taniguchi}}, \bibinfo {author} {\bibfnamefont
  {E.}~\bibnamefont {Kaxiras}}, \bibinfo {author} {\bibfnamefont {R.~C.}\
  \bibnamefont {Ashoori}}, \ and\ \bibinfo {author} {\bibfnamefont
  {P.}~\bibnamefont {Jarillo-Herrero}},\ }\href {\doibase 10.1038/nature26154}
  {\bibfield  {journal} {\bibinfo  {journal} {Nature}\ }\textbf {\bibinfo
  {volume} {556}},\ \bibinfo {pages} {80} (\bibinfo {year}
  {2018}{\natexlab{a}})}\BibitemShut {NoStop}%
\bibitem [{\citenamefont {Cao}\ \emph {et~al.}(2018{\natexlab{b}})\citenamefont
  {Cao}, \citenamefont {Fatemi}, \citenamefont {Fang}, \citenamefont
  {Watanabe}, \citenamefont {Taniguchi}, \citenamefont {Kaxiras},\ and\
  \citenamefont {Jarillo-Herrero}}]{Cao_2018unconv}%
  \BibitemOpen
  \bibfield  {author} {\bibinfo {author} {\bibfnamefont {Y.}~\bibnamefont
  {Cao}}, \bibinfo {author} {\bibfnamefont {V.}~\bibnamefont {Fatemi}},
  \bibinfo {author} {\bibfnamefont {S.}~\bibnamefont {Fang}}, \bibinfo {author}
  {\bibfnamefont {K.}~\bibnamefont {Watanabe}}, \bibinfo {author}
  {\bibfnamefont {T.}~\bibnamefont {Taniguchi}}, \bibinfo {author}
  {\bibfnamefont {E.}~\bibnamefont {Kaxiras}}, \ and\ \bibinfo {author}
  {\bibfnamefont {P.}~\bibnamefont {Jarillo-Herrero}},\ }\href {\doibase
  10.1038/nature26160} {\bibfield  {journal} {\bibinfo  {journal} {Nature}\
  }\textbf {\bibinfo {volume} {556}},\ \bibinfo {pages} {43} (\bibinfo {year}
  {2018}{\natexlab{b}})}\BibitemShut {NoStop}%
\bibitem [{\citenamefont {Lee}\ \emph {et~al.}(2006)\citenamefont {Lee},
  \citenamefont {Nagaosa},\ and\ \citenamefont {Wen}}]{cuprates}%
  \BibitemOpen
  \bibfield  {author} {\bibinfo {author} {\bibfnamefont {P.~A.}\ \bibnamefont
  {Lee}}, \bibinfo {author} {\bibfnamefont {N.}~\bibnamefont {Nagaosa}}, \ and\
  \bibinfo {author} {\bibfnamefont {X.-G.}\ \bibnamefont {Wen}},\ }\href
  {\doibase 10.1103/RevModPhys.78.17} {\bibfield  {journal} {\bibinfo
  {journal} {Rev. Mod. Phys.}\ }\textbf {\bibinfo {volume} {78}},\ \bibinfo
  {pages} {17} (\bibinfo {year} {2006})}\BibitemShut {NoStop}%
\bibitem [{\citenamefont {Sharpe}\ \emph {et~al.}(2019)\citenamefont {Sharpe},
  \citenamefont {Fox}, \citenamefont {Barnard}, \citenamefont {Finney},
  \citenamefont {Watanabe}, \citenamefont {Taniguchi}, \citenamefont
  {Kastner},\ and\ \citenamefont {Goldhaber-Gordon}}]{Sharpe19}%
  \BibitemOpen
  \bibfield  {author} {\bibinfo {author} {\bibfnamefont {A.~L.}\ \bibnamefont
  {Sharpe}}, \bibinfo {author} {\bibfnamefont {E.~J.}\ \bibnamefont {Fox}},
  \bibinfo {author} {\bibfnamefont {A.~W.}\ \bibnamefont {Barnard}}, \bibinfo
  {author} {\bibfnamefont {J.}~\bibnamefont {Finney}}, \bibinfo {author}
  {\bibfnamefont {K.}~\bibnamefont {Watanabe}}, \bibinfo {author}
  {\bibfnamefont {T.}~\bibnamefont {Taniguchi}}, \bibinfo {author}
  {\bibfnamefont {M.~A.}\ \bibnamefont {Kastner}}, \ and\ \bibinfo {author}
  {\bibfnamefont {D.}~\bibnamefont {Goldhaber-Gordon}},\ }\href {\doibase
  10.1126/science.aaw3780} {\bibfield  {journal} {\bibinfo  {journal}
  {Science}\ }\textbf {\bibinfo {volume} {365}},\ \bibinfo {pages} {605}
  (\bibinfo {year} {2019})}\BibitemShut {NoStop}%
\bibitem [{\citenamefont {Sharpe}\ \emph {et~al.}(2021)\citenamefont {Sharpe},
  \citenamefont {Fox}, \citenamefont {Barnard}, \citenamefont {Finney},
  \citenamefont {Watanabe}, \citenamefont {Taniguchi}, \citenamefont
  {Kastner},\ and\ \citenamefont {Goldhaber-Gordon}}]{Sharpe21}%
  \BibitemOpen
  \bibfield  {author} {\bibinfo {author} {\bibfnamefont {A.~L.}\ \bibnamefont
  {Sharpe}}, \bibinfo {author} {\bibfnamefont {E.~J.}\ \bibnamefont {Fox}},
  \bibinfo {author} {\bibfnamefont {A.~W.}\ \bibnamefont {Barnard}}, \bibinfo
  {author} {\bibfnamefont {J.}~\bibnamefont {Finney}}, \bibinfo {author}
  {\bibfnamefont {K.}~\bibnamefont {Watanabe}}, \bibinfo {author}
  {\bibfnamefont {T.}~\bibnamefont {Taniguchi}}, \bibinfo {author}
  {\bibfnamefont {M.~A.}\ \bibnamefont {Kastner}}, \ and\ \bibinfo {author}
  {\bibfnamefont {D.}~\bibnamefont {Goldhaber-Gordon}},\ }\href {\doibase
  10.1021/acs.nanolett.1c00696} {\bibfield  {journal} {\bibinfo  {journal}
  {Nano Letters}\ }\textbf {\bibinfo {volume} {21}},\ \bibinfo {pages} {4299}
  (\bibinfo {year} {2021})}\BibitemShut {NoStop}%
\bibitem [{\citenamefont {Moon}\ and\ \citenamefont
  {Koshino}(2012)}]{koshinohall}%
  \BibitemOpen
  \bibfield  {author} {\bibinfo {author} {\bibfnamefont {P.}~\bibnamefont
  {Moon}}\ and\ \bibinfo {author} {\bibfnamefont {M.}~\bibnamefont {Koshino}},\
  }\href {\doibase 10.1103/PhysRevB.85.195458} {\bibfield  {journal} {\bibinfo
  {journal} {Phys. Rev. B}\ }\textbf {\bibinfo {volume} {85}},\ \bibinfo
  {pages} {195458} (\bibinfo {year} {2012})}\BibitemShut {NoStop}%
\bibitem [{\citenamefont {Po}\ \emph {et~al.}(2018)\citenamefont {Po},
  \citenamefont {Zou}, \citenamefont {Vishwanath},\ and\ \citenamefont
  {Senthil}}]{PhysRevX.8.031089}%
  \BibitemOpen
  \bibfield  {author} {\bibinfo {author} {\bibfnamefont {H.~C.}\ \bibnamefont
  {Po}}, \bibinfo {author} {\bibfnamefont {L.}~\bibnamefont {Zou}}, \bibinfo
  {author} {\bibfnamefont {A.}~\bibnamefont {Vishwanath}}, \ and\ \bibinfo
  {author} {\bibfnamefont {T.}~\bibnamefont {Senthil}},\ }\href {\doibase
  10.1103/PhysRevX.8.031089} {\bibfield  {journal} {\bibinfo  {journal} {Phys.
  Rev. X}\ }\textbf {\bibinfo {volume} {8}},\ \bibinfo {pages} {031089}
  (\bibinfo {year} {2018})}\BibitemShut {NoStop}%
\bibitem [{\citenamefont {Ledwith}\ \emph {et~al.}(2020)\citenamefont
  {Ledwith}, \citenamefont {Tarnopolsky}, \citenamefont {Khalaf},\ and\
  \citenamefont {Vishwanath}}]{ledwith20}%
  \BibitemOpen
  \bibfield  {author} {\bibinfo {author} {\bibfnamefont {P.~J.}\ \bibnamefont
  {Ledwith}}, \bibinfo {author} {\bibfnamefont {G.}~\bibnamefont
  {Tarnopolsky}}, \bibinfo {author} {\bibfnamefont {E.}~\bibnamefont {Khalaf}},
  \ and\ \bibinfo {author} {\bibfnamefont {A.}~\bibnamefont {Vishwanath}},\
  }\href {\doibase 10.1103/PhysRevResearch.2.023237} {\bibfield  {journal}
  {\bibinfo  {journal} {Phys. Rev. Res.}\ }\textbf {\bibinfo {volume} {2}},\
  \bibinfo {pages} {023237} (\bibinfo {year} {2020})}\BibitemShut {NoStop}%
\bibitem [{\citenamefont {Xie}\ \emph {et~al.}(2021)\citenamefont {Xie},
  \citenamefont {Pierce}, \citenamefont {Park}, \citenamefont {Parker},
  \citenamefont {Khalaf}, \citenamefont {Ledwith}, \citenamefont {Cao},
  \citenamefont {Lee}, \citenamefont {Chen}, \citenamefont {Forrester},
  \citenamefont {Watanabe}, \citenamefont {Taniguchi}, \citenamefont
  {Vishwanath}, \citenamefont {Jarillo-Herrero},\ and\ \citenamefont
  {Yacoby}}]{Xie2021}%
  \BibitemOpen
  \bibfield  {author} {\bibinfo {author} {\bibfnamefont {Y.}~\bibnamefont
  {Xie}}, \bibinfo {author} {\bibfnamefont {A.~T.}\ \bibnamefont {Pierce}},
  \bibinfo {author} {\bibfnamefont {J.~M.}\ \bibnamefont {Park}}, \bibinfo
  {author} {\bibfnamefont {D.~E.}\ \bibnamefont {Parker}}, \bibinfo {author}
  {\bibfnamefont {E.}~\bibnamefont {Khalaf}}, \bibinfo {author} {\bibfnamefont
  {P.}~\bibnamefont {Ledwith}}, \bibinfo {author} {\bibfnamefont
  {Y.}~\bibnamefont {Cao}}, \bibinfo {author} {\bibfnamefont {S.~H.}\
  \bibnamefont {Lee}}, \bibinfo {author} {\bibfnamefont {S.}~\bibnamefont
  {Chen}}, \bibinfo {author} {\bibfnamefont {P.~R.}\ \bibnamefont {Forrester}},
  \bibinfo {author} {\bibfnamefont {K.}~\bibnamefont {Watanabe}}, \bibinfo
  {author} {\bibfnamefont {T.}~\bibnamefont {Taniguchi}}, \bibinfo {author}
  {\bibfnamefont {A.}~\bibnamefont {Vishwanath}}, \bibinfo {author}
  {\bibfnamefont {P.}~\bibnamefont {Jarillo-Herrero}}, \ and\ \bibinfo {author}
  {\bibfnamefont {A.}~\bibnamefont {Yacoby}},\ }\href {\doibase
  10.1038/s41586-021-04002-3} {\bibfield  {journal} {\bibinfo  {journal}
  {Nature}\ }\textbf {\bibinfo {volume} {600}},\ \bibinfo {pages} {439}
  (\bibinfo {year} {2021})}\BibitemShut {NoStop}%
\bibitem [{\citenamefont {Pierce}\ \emph {et~al.}(2021)\citenamefont {Pierce},
  \citenamefont {Xie}, \citenamefont {Park}, \citenamefont {Khalaf},
  \citenamefont {Lee}, \citenamefont {Cao}, \citenamefont {Parker},
  \citenamefont {Forrester}, \citenamefont {Chen}, \citenamefont {Watanabe},
  \citenamefont {Taniguchi}, \citenamefont {Vishwanath}, \citenamefont
  {Jarillo-Herrero},\ and\ \citenamefont {Yacoby}}]{Pierce2021}%
  \BibitemOpen
  \bibfield  {author} {\bibinfo {author} {\bibfnamefont {A.~T.}\ \bibnamefont
  {Pierce}}, \bibinfo {author} {\bibfnamefont {Y.}~\bibnamefont {Xie}},
  \bibinfo {author} {\bibfnamefont {J.~M.}\ \bibnamefont {Park}}, \bibinfo
  {author} {\bibfnamefont {E.}~\bibnamefont {Khalaf}}, \bibinfo {author}
  {\bibfnamefont {S.~H.}\ \bibnamefont {Lee}}, \bibinfo {author} {\bibfnamefont
  {Y.}~\bibnamefont {Cao}}, \bibinfo {author} {\bibfnamefont {D.~E.}\
  \bibnamefont {Parker}}, \bibinfo {author} {\bibfnamefont {P.~R.}\
  \bibnamefont {Forrester}}, \bibinfo {author} {\bibfnamefont {S.}~\bibnamefont
  {Chen}}, \bibinfo {author} {\bibfnamefont {K.}~\bibnamefont {Watanabe}},
  \bibinfo {author} {\bibfnamefont {T.}~\bibnamefont {Taniguchi}}, \bibinfo
  {author} {\bibfnamefont {A.}~\bibnamefont {Vishwanath}}, \bibinfo {author}
  {\bibfnamefont {P.}~\bibnamefont {Jarillo-Herrero}}, \ and\ \bibinfo {author}
  {\bibfnamefont {A.}~\bibnamefont {Yacoby}},\ }\href {\doibase
  10.1038/s41567-021-01347-4} {\bibfield  {journal} {\bibinfo  {journal}
  {Nature Physics}\ }\textbf {\bibinfo {volume} {17}},\ \bibinfo {pages} {1210}
  (\bibinfo {year} {2021})}\BibitemShut {NoStop}%
\bibitem [{\citenamefont {Oh}\ \emph {et~al.}(2021)\citenamefont {Oh},
  \citenamefont {Nuckolls}, \citenamefont {Wong}, \citenamefont {Lee},
  \citenamefont {Liu}, \citenamefont {Watanabe}, \citenamefont {Taniguchi},\
  and\ \citenamefont {Yazdani}}]{Oh_2021}%
  \BibitemOpen
  \bibfield  {author} {\bibinfo {author} {\bibfnamefont {M.}~\bibnamefont
  {Oh}}, \bibinfo {author} {\bibfnamefont {K.~P.}\ \bibnamefont {Nuckolls}},
  \bibinfo {author} {\bibfnamefont {D.}~\bibnamefont {Wong}}, \bibinfo {author}
  {\bibfnamefont {R.~L.}\ \bibnamefont {Lee}}, \bibinfo {author} {\bibfnamefont
  {X.}~\bibnamefont {Liu}}, \bibinfo {author} {\bibfnamefont {K.}~\bibnamefont
  {Watanabe}}, \bibinfo {author} {\bibfnamefont {T.}~\bibnamefont {Taniguchi}},
  \ and\ \bibinfo {author} {\bibfnamefont {A.}~\bibnamefont {Yazdani}},\ }\href
  {\doibase 10.1038/s41586-021-04121-x} {\bibfield  {journal} {\bibinfo
  {journal} {Nature}\ }\textbf {\bibinfo {volume} {600}},\ \bibinfo {pages}
  {240} (\bibinfo {year} {2021})}\BibitemShut {NoStop}%
\bibitem [{\citenamefont {Roy}\ and\ \citenamefont {Juri\ifmmode \check{c}\else
  \v{c}\fi{}i\ifmmode~\acute{c}\else \'{c}\fi{}}(2019)}]{bitanroy}%
  \BibitemOpen
  \bibfield  {author} {\bibinfo {author} {\bibfnamefont {B.}~\bibnamefont
  {Roy}}\ and\ \bibinfo {author} {\bibfnamefont {V.}~\bibnamefont {Juri\ifmmode
  \check{c}\else \v{c}\fi{}i\ifmmode~\acute{c}\else \'{c}\fi{}}},\ }\href
  {\doibase 10.1103/PhysRevB.99.121407} {\bibfield  {journal} {\bibinfo
  {journal} {Phys. Rev. B}\ }\textbf {\bibinfo {volume} {99}},\ \bibinfo
  {pages} {121407} (\bibinfo {year} {2019})}\BibitemShut {NoStop}%
\bibitem [{\citenamefont {Goodwin}\ \emph {et~al.}(2019)\citenamefont
  {Goodwin}, \citenamefont {Corsetti}, \citenamefont {Mostofi},\ and\
  \citenamefont {Lischner}}]{goodwin}%
  \BibitemOpen
  \bibfield  {author} {\bibinfo {author} {\bibfnamefont {Z.~A.~H.}\
  \bibnamefont {Goodwin}}, \bibinfo {author} {\bibfnamefont {F.}~\bibnamefont
  {Corsetti}}, \bibinfo {author} {\bibfnamefont {A.~A.}\ \bibnamefont
  {Mostofi}}, \ and\ \bibinfo {author} {\bibfnamefont {J.}~\bibnamefont
  {Lischner}},\ }\href {\doibase 10.1103/PhysRevB.100.235424} {\bibfield
  {journal} {\bibinfo  {journal} {Phys. Rev. B}\ }\textbf {\bibinfo {volume}
  {100}},\ \bibinfo {pages} {235424} (\bibinfo {year} {2019})}\BibitemShut
  {NoStop}%
\bibitem [{\citenamefont {Lake}\ \emph {et~al.}(2022)\citenamefont {Lake},
  \citenamefont {Patri},\ and\ \citenamefont {Senthil}}]{Lake22}%
  \BibitemOpen
  \bibfield  {author} {\bibinfo {author} {\bibfnamefont {E.}~\bibnamefont
  {Lake}}, \bibinfo {author} {\bibfnamefont {A.~S.}\ \bibnamefont {Patri}}, \
  and\ \bibinfo {author} {\bibfnamefont {T.}~\bibnamefont {Senthil}},\ }\href
  {\doibase 10.1103/PhysRevB.106.104506} {\bibfield  {journal} {\bibinfo
  {journal} {Phys. Rev. B}\ }\textbf {\bibinfo {volume} {106}},\ \bibinfo
  {pages} {104506} (\bibinfo {year} {2022})}\BibitemShut {NoStop}%
\bibitem [{\citenamefont {Cao}\ \emph {et~al.}(2020)\citenamefont {Cao},
  \citenamefont {Chowdhury}, \citenamefont {Rodan-Legrain}, \citenamefont
  {Rubies-Bigorda}, \citenamefont {Watanabe}, \citenamefont {Taniguchi},
  \citenamefont {Senthil},\ and\ \citenamefont {Jarillo-Herrero}}]{Cao20}%
  \BibitemOpen
  \bibfield  {author} {\bibinfo {author} {\bibfnamefont {Y.}~\bibnamefont
  {Cao}}, \bibinfo {author} {\bibfnamefont {D.}~\bibnamefont {Chowdhury}},
  \bibinfo {author} {\bibfnamefont {D.}~\bibnamefont {Rodan-Legrain}}, \bibinfo
  {author} {\bibfnamefont {O.}~\bibnamefont {Rubies-Bigorda}}, \bibinfo
  {author} {\bibfnamefont {K.}~\bibnamefont {Watanabe}}, \bibinfo {author}
  {\bibfnamefont {T.}~\bibnamefont {Taniguchi}}, \bibinfo {author}
  {\bibfnamefont {T.}~\bibnamefont {Senthil}}, \ and\ \bibinfo {author}
  {\bibfnamefont {P.}~\bibnamefont {Jarillo-Herrero}},\ }\href {\doibase
  10.1103/PhysRevLett.124.076801} {\bibfield  {journal} {\bibinfo  {journal}
  {Phys. Rev. Lett.}\ }\textbf {\bibinfo {volume} {124}},\ \bibinfo {pages}
  {076801} (\bibinfo {year} {2020})}\BibitemShut {NoStop}%
\bibitem [{\citenamefont {Gonz\'alez}\ and\ \citenamefont
  {Stauber}(2020)}]{Gonzalez20}%
  \BibitemOpen
  \bibfield  {author} {\bibinfo {author} {\bibfnamefont {J.}~\bibnamefont
  {Gonz\'alez}}\ and\ \bibinfo {author} {\bibfnamefont {T.}~\bibnamefont
  {Stauber}},\ }\href {\doibase 10.1103/PhysRevLett.124.186801} {\bibfield
  {journal} {\bibinfo  {journal} {Phys. Rev. Lett.}\ }\textbf {\bibinfo
  {volume} {124}},\ \bibinfo {pages} {186801} (\bibinfo {year}
  {2020})}\BibitemShut {NoStop}%
\bibitem [{\citenamefont {Jaoui}\ \emph {et~al.}(2022)\citenamefont {Jaoui},
  \citenamefont {Das}, \citenamefont {Di~Battista}, \citenamefont
  {D{\'\i}ez-M{\'e}rida}, \citenamefont {Lu}, \citenamefont {Watanabe},
  \citenamefont {Taniguchi}, \citenamefont {Ishizuka}, \citenamefont
  {Levitov},\ and\ \citenamefont {Efetov}}]{Jaoui22}%
  \BibitemOpen
  \bibfield  {author} {\bibinfo {author} {\bibfnamefont {A.}~\bibnamefont
  {Jaoui}}, \bibinfo {author} {\bibfnamefont {I.}~\bibnamefont {Das}}, \bibinfo
  {author} {\bibfnamefont {G.}~\bibnamefont {Di~Battista}}, \bibinfo {author}
  {\bibfnamefont {J.}~\bibnamefont {D{\'\i}ez-M{\'e}rida}}, \bibinfo {author}
  {\bibfnamefont {X.}~\bibnamefont {Lu}}, \bibinfo {author} {\bibfnamefont
  {K.}~\bibnamefont {Watanabe}}, \bibinfo {author} {\bibfnamefont
  {T.}~\bibnamefont {Taniguchi}}, \bibinfo {author} {\bibfnamefont
  {H.}~\bibnamefont {Ishizuka}}, \bibinfo {author} {\bibfnamefont
  {L.}~\bibnamefont {Levitov}}, \ and\ \bibinfo {author} {\bibfnamefont
  {D.~K.}\ \bibnamefont {Efetov}},\ }\href {\doibase
  10.1038/s41567-022-01556-5} {\bibfield  {journal} {\bibinfo  {journal}
  {Nature Physics}\ }\textbf {\bibinfo {volume} {18}},\ \bibinfo {pages} {633}
  (\bibinfo {year} {2022})}\BibitemShut {NoStop}%
\bibitem [{\citenamefont {Stauber}\ and\ \citenamefont
  {Gonz{\'a}lez}(2022)}]{Stauber22}%
  \BibitemOpen
  \bibfield  {author} {\bibinfo {author} {\bibfnamefont {T.}~\bibnamefont
  {Stauber}}\ and\ \bibinfo {author} {\bibfnamefont {J.}~\bibnamefont
  {Gonz{\'a}lez}},\ }\href {\doibase 10.1038/s41567-022-01573-4} {\bibfield
  {journal} {\bibinfo  {journal} {Nature Physics}\ }\textbf {\bibinfo {volume}
  {18}},\ \bibinfo {pages} {619} (\bibinfo {year} {2022})}\BibitemShut
  {NoStop}%
\bibitem [{\citenamefont {Lian}\ \emph {et~al.}(2019)\citenamefont {Lian},
  \citenamefont {Wang},\ and\ \citenamefont {Bernevig}}]{Lian_2019}%
  \BibitemOpen
  \bibfield  {author} {\bibinfo {author} {\bibfnamefont {B.}~\bibnamefont
  {Lian}}, \bibinfo {author} {\bibfnamefont {Z.}~\bibnamefont {Wang}}, \ and\
  \bibinfo {author} {\bibfnamefont {B.~A.}\ \bibnamefont {Bernevig}},\
  }\href@noop {} {\bibfield  {journal} {\bibinfo  {journal} {Physical Review
  Letters}\ }\textbf {\bibinfo {volume} {122}} (\bibinfo {year}
  {2019})}\BibitemShut {NoStop}%
\bibitem [{\citenamefont {Wu}\ \emph {et~al.}(2018)\citenamefont {Wu},
  \citenamefont {MacDonald},\ and\ \citenamefont {Martin}}]{mcdonaldphonon}%
  \BibitemOpen
  \bibfield  {author} {\bibinfo {author} {\bibfnamefont {F.}~\bibnamefont
  {Wu}}, \bibinfo {author} {\bibfnamefont {A.~H.}\ \bibnamefont {MacDonald}}, \
  and\ \bibinfo {author} {\bibfnamefont {I.}~\bibnamefont {Martin}},\ }\href
  {\doibase 10.1103/PhysRevLett.121.257001} {\bibfield  {journal} {\bibinfo
  {journal} {Phys. Rev. Lett.}\ }\textbf {\bibinfo {volume} {121}},\ \bibinfo
  {pages} {257001} (\bibinfo {year} {2018})}\BibitemShut {NoStop}%
\bibitem [{\citenamefont {Park}\ \emph {et~al.}(2022)\citenamefont {Park},
  \citenamefont {Cao}, \citenamefont {Xia}, \citenamefont {Sun}, \citenamefont
  {Watanabe}, \citenamefont {Taniguchi},\ and\ \citenamefont
  {Jarillo-Herrero}}]{Park22}%
  \BibitemOpen
  \bibfield  {author} {\bibinfo {author} {\bibfnamefont {J.~M.}\ \bibnamefont
  {Park}}, \bibinfo {author} {\bibfnamefont {Y.}~\bibnamefont {Cao}}, \bibinfo
  {author} {\bibfnamefont {L.-Q.}\ \bibnamefont {Xia}}, \bibinfo {author}
  {\bibfnamefont {S.}~\bibnamefont {Sun}}, \bibinfo {author} {\bibfnamefont
  {K.}~\bibnamefont {Watanabe}}, \bibinfo {author} {\bibfnamefont
  {T.}~\bibnamefont {Taniguchi}}, \ and\ \bibinfo {author} {\bibfnamefont
  {P.}~\bibnamefont {Jarillo-Herrero}},\ }\href {\doibase
  10.1038/s41563-022-01287-1} {\bibfield  {journal} {\bibinfo  {journal}
  {Nature Materials}\ }\textbf {\bibinfo {volume} {21}},\ \bibinfo {pages}
  {877} (\bibinfo {year} {2022})}\BibitemShut {NoStop}%
\bibitem [{\citenamefont {Park}\ \emph {et~al.}(2021)\citenamefont {Park},
  \citenamefont {Cao}, \citenamefont {Watanabe}, \citenamefont {Taniguchi},\
  and\ \citenamefont {Jarillo-Herrero}}]{Park21}%
  \BibitemOpen
  \bibfield  {author} {\bibinfo {author} {\bibfnamefont {J.~M.}\ \bibnamefont
  {Park}}, \bibinfo {author} {\bibfnamefont {Y.}~\bibnamefont {Cao}}, \bibinfo
  {author} {\bibfnamefont {K.}~\bibnamefont {Watanabe}}, \bibinfo {author}
  {\bibfnamefont {T.}~\bibnamefont {Taniguchi}}, \ and\ \bibinfo {author}
  {\bibfnamefont {P.}~\bibnamefont {Jarillo-Herrero}},\ }\href {\doibase
  10.1038/s41586-021-03192-0} {\bibfield  {journal} {\bibinfo  {journal}
  {Nature}\ }\textbf {\bibinfo {volume} {590}},\ \bibinfo {pages} {249}
  (\bibinfo {year} {2021})}\BibitemShut {NoStop}%
\bibitem [{\citenamefont {Hao}\ \emph {et~al.}(2021)\citenamefont {Hao},
  \citenamefont {Zimmerman}, \citenamefont {Ledwith}, \citenamefont {Khalaf},
  \citenamefont {Najafabadi}, \citenamefont {Watanabe}, \citenamefont
  {Taniguchi}, \citenamefont {Vishwanath},\ and\ \citenamefont {Kim}}]{Zeyu21}%
  \BibitemOpen
  \bibfield  {author} {\bibinfo {author} {\bibfnamefont {Z.}~\bibnamefont
  {Hao}}, \bibinfo {author} {\bibfnamefont {A.~M.}\ \bibnamefont {Zimmerman}},
  \bibinfo {author} {\bibfnamefont {P.}~\bibnamefont {Ledwith}}, \bibinfo
  {author} {\bibfnamefont {E.}~\bibnamefont {Khalaf}}, \bibinfo {author}
  {\bibfnamefont {D.~H.}\ \bibnamefont {Najafabadi}}, \bibinfo {author}
  {\bibfnamefont {K.}~\bibnamefont {Watanabe}}, \bibinfo {author}
  {\bibfnamefont {T.}~\bibnamefont {Taniguchi}}, \bibinfo {author}
  {\bibfnamefont {A.}~\bibnamefont {Vishwanath}}, \ and\ \bibinfo {author}
  {\bibfnamefont {P.}~\bibnamefont {Kim}},\ }\href@noop {} {\bibfield
  {journal} {\bibinfo  {journal} {Science}\ }\textbf {\bibinfo {volume}
  {371}},\ \bibinfo {pages} {1133} (\bibinfo {year} {2021})}\BibitemShut
  {NoStop}%
\bibitem [{\citenamefont {Cao}\ \emph {et~al.}(2021{\natexlab{a}})\citenamefont
  {Cao}, \citenamefont {Park}, \citenamefont {Watanabe}, \citenamefont
  {Taniguchi},\ and\ \citenamefont {Jarillo-Herrero}}]{Cao21}%
  \BibitemOpen
  \bibfield  {author} {\bibinfo {author} {\bibfnamefont {Y.}~\bibnamefont
  {Cao}}, \bibinfo {author} {\bibfnamefont {J.~M.}\ \bibnamefont {Park}},
  \bibinfo {author} {\bibfnamefont {K.}~\bibnamefont {Watanabe}}, \bibinfo
  {author} {\bibfnamefont {T.}~\bibnamefont {Taniguchi}}, \ and\ \bibinfo
  {author} {\bibfnamefont {P.}~\bibnamefont {Jarillo-Herrero}},\ }\href
  {\doibase 10.1038/s41586-021-03685-y} {\bibfield  {journal} {\bibinfo
  {journal} {Nature}\ }\textbf {\bibinfo {volume} {595}},\ \bibinfo {pages}
  {526} (\bibinfo {year} {2021}{\natexlab{a}})}\BibitemShut {NoStop}%
\bibitem [{\citenamefont {Lake}\ and\ \citenamefont {Senthil}(2021)}]{Lake21}%
  \BibitemOpen
  \bibfield  {author} {\bibinfo {author} {\bibfnamefont {E.}~\bibnamefont
  {Lake}}\ and\ \bibinfo {author} {\bibfnamefont {T.}~\bibnamefont {Senthil}},\
  }\href {\doibase 10.1103/PhysRevB.104.174505} {\bibfield  {journal} {\bibinfo
   {journal} {Phys. Rev. B}\ }\textbf {\bibinfo {volume} {104}},\ \bibinfo
  {pages} {174505} (\bibinfo {year} {2021})}\BibitemShut {NoStop}%
\bibitem [{\citenamefont {Christos}\ \emph {et~al.}(2022)\citenamefont
  {Christos}, \citenamefont {Sachdev},\ and\ \citenamefont
  {Scheurer}}]{Christos22}%
  \BibitemOpen
  \bibfield  {author} {\bibinfo {author} {\bibfnamefont {M.}~\bibnamefont
  {Christos}}, \bibinfo {author} {\bibfnamefont {S.}~\bibnamefont {Sachdev}}, \
  and\ \bibinfo {author} {\bibfnamefont {M.~S.}\ \bibnamefont {Scheurer}},\
  }\href {\doibase 10.1103/PhysRevX.12.021018} {\bibfield  {journal} {\bibinfo
  {journal} {Phys. Rev. X}\ }\textbf {\bibinfo {volume} {12}},\ \bibinfo
  {pages} {021018} (\bibinfo {year} {2022})}\BibitemShut {NoStop}%
\bibitem [{\citenamefont {Christos}\ \emph {et~al.}(2023)\citenamefont
  {Christos}, \citenamefont {Sachdev},\ and\ \citenamefont
  {Scheurer}}]{Christos23}%
  \BibitemOpen
  \bibfield  {author} {\bibinfo {author} {\bibfnamefont {M.}~\bibnamefont
  {Christos}}, \bibinfo {author} {\bibfnamefont {S.}~\bibnamefont {Sachdev}}, \
  and\ \bibinfo {author} {\bibfnamefont {M.~S.}\ \bibnamefont {Scheurer}},\
  }\href {\doibase 10.1038/s41467-023-42471-4} {\bibfield  {journal} {\bibinfo
  {journal} {Nature Communications}\ }\textbf {\bibinfo {volume} {14}},\
  \bibinfo {pages} {7134} (\bibinfo {year} {2023})}\BibitemShut {NoStop}%
\bibitem [{\citenamefont {Gonz{\'a}lez}\ and\ \citenamefont
  {Stauber}(2023)}]{Gonzalez23}%
  \BibitemOpen
  \bibfield  {author} {\bibinfo {author} {\bibfnamefont {J.}~\bibnamefont
  {Gonz{\'a}lez}}\ and\ \bibinfo {author} {\bibfnamefont {T.}~\bibnamefont
  {Stauber}},\ }\href {\doibase 10.1038/s41467-023-38250-w} {\bibfield
  {journal} {\bibinfo  {journal} {Nature Communications}\ }\textbf {\bibinfo
  {volume} {14}},\ \bibinfo {pages} {2746} (\bibinfo {year}
  {2023})}\BibitemShut {NoStop}%
\bibitem [{\citenamefont {Kang}\ and\ \citenamefont {Vafek}(2019)}]{Kang19}%
  \BibitemOpen
  \bibfield  {author} {\bibinfo {author} {\bibfnamefont {J.}~\bibnamefont
  {Kang}}\ and\ \bibinfo {author} {\bibfnamefont {O.}~\bibnamefont {Vafek}},\
  }\href {\doibase 10.1103/PhysRevLett.122.246401} {\bibfield  {journal}
  {\bibinfo  {journal} {Phys. Rev. Lett.}\ }\textbf {\bibinfo {volume} {122}},\
  \bibinfo {pages} {246401} (\bibinfo {year} {2019})}\BibitemShut {NoStop}%
\bibitem [{\citenamefont {Seo}\ \emph {et~al.}(2019)\citenamefont {Seo},
  \citenamefont {Kotov},\ and\ \citenamefont {Uchoa}}]{seo19}%
  \BibitemOpen
  \bibfield  {author} {\bibinfo {author} {\bibfnamefont {K.}~\bibnamefont
  {Seo}}, \bibinfo {author} {\bibfnamefont {V.~N.}\ \bibnamefont {Kotov}}, \
  and\ \bibinfo {author} {\bibfnamefont {B.}~\bibnamefont {Uchoa}},\ }\href
  {\doibase 10.1103/PhysRevLett.122.246402} {\bibfield  {journal} {\bibinfo
  {journal} {Phys. Rev. Lett.}\ }\textbf {\bibinfo {volume} {122}},\ \bibinfo
  {pages} {246402} (\bibinfo {year} {2019})}\BibitemShut {NoStop}%
\bibitem [{\citenamefont {Stepanov}\ \emph
  {et~al.}(2020{\natexlab{a}})\citenamefont {Stepanov}, \citenamefont {Das},
  \citenamefont {Lu}, \citenamefont {Fahimniya}, \citenamefont {Watanabe},
  \citenamefont {Taniguchi}, \citenamefont {Koppens}, \citenamefont {Lischner},
  \citenamefont {Levitov},\ and\ \citenamefont {Efetov}}]{Stepanov2020}%
  \BibitemOpen
  \bibfield  {author} {\bibinfo {author} {\bibfnamefont {P.}~\bibnamefont
  {Stepanov}}, \bibinfo {author} {\bibfnamefont {I.}~\bibnamefont {Das}},
  \bibinfo {author} {\bibfnamefont {X.}~\bibnamefont {Lu}}, \bibinfo {author}
  {\bibfnamefont {A.}~\bibnamefont {Fahimniya}}, \bibinfo {author}
  {\bibfnamefont {K.}~\bibnamefont {Watanabe}}, \bibinfo {author}
  {\bibfnamefont {T.}~\bibnamefont {Taniguchi}}, \bibinfo {author}
  {\bibfnamefont {F.~H.~L.}\ \bibnamefont {Koppens}}, \bibinfo {author}
  {\bibfnamefont {J.}~\bibnamefont {Lischner}}, \bibinfo {author}
  {\bibfnamefont {L.}~\bibnamefont {Levitov}}, \ and\ \bibinfo {author}
  {\bibfnamefont {D.~K.}\ \bibnamefont {Efetov}},\ }\href {\doibase
  10.1038/s41586-020-2459-6} {\bibfield  {journal} {\bibinfo  {journal}
  {Nature}\ }\textbf {\bibinfo {volume} {583}},\ \bibinfo {pages} {375}
  (\bibinfo {year} {2020}{\natexlab{a}})}\BibitemShut {NoStop}%
\bibitem [{\citenamefont {Bultinck}\ \emph {et~al.}(2020)\citenamefont
  {Bultinck}, \citenamefont {Khalaf}, \citenamefont {Liu}, \citenamefont
  {Chatterjee}, \citenamefont {Vishwanath},\ and\ \citenamefont
  {Zaletel}}]{Bultinck20}%
  \BibitemOpen
  \bibfield  {author} {\bibinfo {author} {\bibfnamefont {N.}~\bibnamefont
  {Bultinck}}, \bibinfo {author} {\bibfnamefont {E.}~\bibnamefont {Khalaf}},
  \bibinfo {author} {\bibfnamefont {S.}~\bibnamefont {Liu}}, \bibinfo {author}
  {\bibfnamefont {S.}~\bibnamefont {Chatterjee}}, \bibinfo {author}
  {\bibfnamefont {A.}~\bibnamefont {Vishwanath}}, \ and\ \bibinfo {author}
  {\bibfnamefont {M.~P.}\ \bibnamefont {Zaletel}},\ }\href {\doibase
  10.1103/PhysRevX.10.031034} {\bibfield  {journal} {\bibinfo  {journal} {Phys.
  Rev. X}\ }\textbf {\bibinfo {volume} {10}},\ \bibinfo {pages} {031034}
  (\bibinfo {year} {2020})}\BibitemShut {NoStop}%
\bibitem [{\citenamefont {Song}\ \emph {et~al.}(2021)\citenamefont {Song},
  \citenamefont {Lian}, \citenamefont {Regnault},\ and\ \citenamefont
  {Bernevig}}]{bernevig221}%
  \BibitemOpen
  \bibfield  {author} {\bibinfo {author} {\bibfnamefont {Z.-D.}\ \bibnamefont
  {Song}}, \bibinfo {author} {\bibfnamefont {B.}~\bibnamefont {Lian}}, \bibinfo
  {author} {\bibfnamefont {N.}~\bibnamefont {Regnault}}, \ and\ \bibinfo
  {author} {\bibfnamefont {B.~A.}\ \bibnamefont {Bernevig}},\ }\href {\doibase
  10.1103/PhysRevB.103.205412} {\bibfield  {journal} {\bibinfo  {journal}
  {Phys. Rev. B}\ }\textbf {\bibinfo {volume} {103}},\ \bibinfo {pages}
  {205412} (\bibinfo {year} {2021})}\BibitemShut {NoStop}%
\bibitem [{\citenamefont {Bernevig}\ \emph {et~al.}(2021)\citenamefont
  {Bernevig}, \citenamefont {Song}, \citenamefont {Regnault},\ and\
  \citenamefont {Lian}}]{bernevig321}%
  \BibitemOpen
  \bibfield  {author} {\bibinfo {author} {\bibfnamefont {B.~A.}\ \bibnamefont
  {Bernevig}}, \bibinfo {author} {\bibfnamefont {Z.-D.}\ \bibnamefont {Song}},
  \bibinfo {author} {\bibfnamefont {N.}~\bibnamefont {Regnault}}, \ and\
  \bibinfo {author} {\bibfnamefont {B.}~\bibnamefont {Lian}},\ }\href {\doibase
  10.1103/PhysRevB.103.205413} {\bibfield  {journal} {\bibinfo  {journal}
  {Phys. Rev. B}\ }\textbf {\bibinfo {volume} {103}},\ \bibinfo {pages}
  {205413} (\bibinfo {year} {2021})}\BibitemShut {NoStop}%
\bibitem [{\citenamefont {Lian}\ \emph {et~al.}(2021)\citenamefont {Lian},
  \citenamefont {Song}, \citenamefont {Regnault}, \citenamefont {Efetov},
  \citenamefont {Yazdani},\ and\ \citenamefont {Bernevig}}]{bernevig421}%
  \BibitemOpen
  \bibfield  {author} {\bibinfo {author} {\bibfnamefont {B.}~\bibnamefont
  {Lian}}, \bibinfo {author} {\bibfnamefont {Z.-D.}\ \bibnamefont {Song}},
  \bibinfo {author} {\bibfnamefont {N.}~\bibnamefont {Regnault}}, \bibinfo
  {author} {\bibfnamefont {D.~K.}\ \bibnamefont {Efetov}}, \bibinfo {author}
  {\bibfnamefont {A.}~\bibnamefont {Yazdani}}, \ and\ \bibinfo {author}
  {\bibfnamefont {B.~A.}\ \bibnamefont {Bernevig}},\ }\href {\doibase
  10.1103/PhysRevB.103.205414} {\bibfield  {journal} {\bibinfo  {journal}
  {Phys. Rev. B}\ }\textbf {\bibinfo {volume} {103}},\ \bibinfo {pages}
  {205414} (\bibinfo {year} {2021})}\BibitemShut {NoStop}%
\bibitem [{\citenamefont {C\ifmmode \u{a}\else \u{a}\fi{}lug\ifmmode~\u{a}\else
  \u{a}\fi{}ru}\ \emph {et~al.}(2022)\citenamefont {C\ifmmode \u{a}\else
  \u{a}\fi{}lug\ifmmode~\u{a}\else \u{a}\fi{}ru}, \citenamefont {Regnault},
  \citenamefont {Oh}, \citenamefont {Nuckolls}, \citenamefont {Wong},
  \citenamefont {Lee}, \citenamefont {Yazdani}, \citenamefont {Vafek},\ and\
  \citenamefont {Bernevig}}]{dimitru22}%
  \BibitemOpen
  \bibfield  {author} {\bibinfo {author} {\bibfnamefont {D.}~\bibnamefont
  {C\ifmmode \u{a}\else \u{a}\fi{}lug\ifmmode~\u{a}\else \u{a}\fi{}ru}},
  \bibinfo {author} {\bibfnamefont {N.}~\bibnamefont {Regnault}}, \bibinfo
  {author} {\bibfnamefont {M.}~\bibnamefont {Oh}}, \bibinfo {author}
  {\bibfnamefont {K.~P.}\ \bibnamefont {Nuckolls}}, \bibinfo {author}
  {\bibfnamefont {D.}~\bibnamefont {Wong}}, \bibinfo {author} {\bibfnamefont
  {R.~L.}\ \bibnamefont {Lee}}, \bibinfo {author} {\bibfnamefont
  {A.}~\bibnamefont {Yazdani}}, \bibinfo {author} {\bibfnamefont
  {O.}~\bibnamefont {Vafek}}, \ and\ \bibinfo {author} {\bibfnamefont {B.~A.}\
  \bibnamefont {Bernevig}},\ }\href {\doibase 10.1103/PhysRevLett.129.117602}
  {\bibfield  {journal} {\bibinfo  {journal} {Phys. Rev. Lett.}\ }\textbf
  {\bibinfo {volume} {129}},\ \bibinfo {pages} {117602} (\bibinfo {year}
  {2022})}\BibitemShut {NoStop}%
\bibitem [{\citenamefont {Wagner}\ \emph {et~al.}(2022)\citenamefont {Wagner},
  \citenamefont {Kwan}, \citenamefont {Bultinck}, \citenamefont {Simon},\ and\
  \citenamefont {Parameswaran}}]{Wagner22}%
  \BibitemOpen
  \bibfield  {author} {\bibinfo {author} {\bibfnamefont {G.}~\bibnamefont
  {Wagner}}, \bibinfo {author} {\bibfnamefont {Y.~H.}\ \bibnamefont {Kwan}},
  \bibinfo {author} {\bibfnamefont {N.}~\bibnamefont {Bultinck}}, \bibinfo
  {author} {\bibfnamefont {S.~H.}\ \bibnamefont {Simon}}, \ and\ \bibinfo
  {author} {\bibfnamefont {S.~A.}\ \bibnamefont {Parameswaran}},\ }\href
  {\doibase 10.1103/PhysRevLett.128.156401} {\bibfield  {journal} {\bibinfo
  {journal} {Phys. Rev. Lett.}\ }\textbf {\bibinfo {volume} {128}},\ \bibinfo
  {pages} {156401} (\bibinfo {year} {2022})}\BibitemShut {NoStop}%
\bibitem [{\citenamefont {Liu}\ \emph {et~al.}(2021)\citenamefont {Liu},
  \citenamefont {Wang}, \citenamefont {Watanabe}, \citenamefont {Taniguchi},
  \citenamefont {Vafek},\ and\ \citenamefont {Li}}]{Liu21}%
  \BibitemOpen
  \bibfield  {author} {\bibinfo {author} {\bibfnamefont {X.}~\bibnamefont
  {Liu}}, \bibinfo {author} {\bibfnamefont {Z.}~\bibnamefont {Wang}}, \bibinfo
  {author} {\bibfnamefont {K.}~\bibnamefont {Watanabe}}, \bibinfo {author}
  {\bibfnamefont {T.}~\bibnamefont {Taniguchi}}, \bibinfo {author}
  {\bibfnamefont {O.}~\bibnamefont {Vafek}}, \ and\ \bibinfo {author}
  {\bibfnamefont {J.~I.~A.}\ \bibnamefont {Li}},\ }\href {\doibase
  10.1126/science.abb8754} {\bibfield  {journal} {\bibinfo  {journal}
  {Science}\ }\textbf {\bibinfo {volume} {371}},\ \bibinfo {pages} {1261}
  (\bibinfo {year} {2021})},\ \Eprint
  {http://arxiv.org/abs/https://www.science.org/doi/pdf/10.1126/science.abb8754}
  {https://www.science.org/doi/pdf/10.1126/science.abb8754} \BibitemShut
  {NoStop}%
\bibitem [{\citenamefont {Nuckolls}\ \emph {et~al.}(2023)\citenamefont
  {Nuckolls}, \citenamefont {Lee}, \citenamefont {Oh}, \citenamefont {Wong},
  \citenamefont {Soejima}, \citenamefont {Hong}, \citenamefont {C{\u a}lug{\u
  a}ru}, \citenamefont {Herzog-Arbeitman}, \citenamefont {Bernevig},
  \citenamefont {Watanabe}, \citenamefont {Taniguchi}, \citenamefont
  {Regnault}, \citenamefont {Zaletel},\ and\ \citenamefont
  {Yazdani}}]{Nuckolls23}%
  \BibitemOpen
  \bibfield  {author} {\bibinfo {author} {\bibfnamefont {K.~P.}\ \bibnamefont
  {Nuckolls}}, \bibinfo {author} {\bibfnamefont {R.~L.}\ \bibnamefont {Lee}},
  \bibinfo {author} {\bibfnamefont {M.}~\bibnamefont {Oh}}, \bibinfo {author}
  {\bibfnamefont {D.}~\bibnamefont {Wong}}, \bibinfo {author} {\bibfnamefont
  {T.}~\bibnamefont {Soejima}}, \bibinfo {author} {\bibfnamefont {J.~P.}\
  \bibnamefont {Hong}}, \bibinfo {author} {\bibfnamefont {D.}~\bibnamefont
  {C{\u a}lug{\u a}ru}}, \bibinfo {author} {\bibfnamefont {J.}~\bibnamefont
  {Herzog-Arbeitman}}, \bibinfo {author} {\bibfnamefont {B.~A.}\ \bibnamefont
  {Bernevig}}, \bibinfo {author} {\bibfnamefont {K.}~\bibnamefont {Watanabe}},
  \bibinfo {author} {\bibfnamefont {T.}~\bibnamefont {Taniguchi}}, \bibinfo
  {author} {\bibfnamefont {N.}~\bibnamefont {Regnault}}, \bibinfo {author}
  {\bibfnamefont {M.~P.}\ \bibnamefont {Zaletel}}, \ and\ \bibinfo {author}
  {\bibfnamefont {A.}~\bibnamefont {Yazdani}},\ }\href {\doibase
  10.1038/s41586-023-06226-x} {\bibfield  {journal} {\bibinfo  {journal}
  {Nature}\ }\textbf {\bibinfo {volume} {620}},\ \bibinfo {pages} {525}
  (\bibinfo {year} {2023})}\BibitemShut {NoStop}%
\bibitem [{\citenamefont {Kim}\ \emph {et~al.}(2023)\citenamefont {Kim},
  \citenamefont {Choi}, \citenamefont {Lantagne-Hurtubise}, \citenamefont
  {Lewandowski}, \citenamefont {Thomson}, \citenamefont {Kong}, \citenamefont
  {Zhou}, \citenamefont {Baum}, \citenamefont {Zhang}, \citenamefont {Holleis},
  \citenamefont {Watanabe}, \citenamefont {Taniguchi}, \citenamefont {Young},
  \citenamefont {Alicea},\ and\ \citenamefont {Nadj-Perge}}]{Kim23}%
  \BibitemOpen
  \bibfield  {author} {\bibinfo {author} {\bibfnamefont {H.}~\bibnamefont
  {Kim}}, \bibinfo {author} {\bibfnamefont {Y.}~\bibnamefont {Choi}}, \bibinfo
  {author} {\bibfnamefont {{\'E}.}~\bibnamefont {Lantagne-Hurtubise}}, \bibinfo
  {author} {\bibfnamefont {C.}~\bibnamefont {Lewandowski}}, \bibinfo {author}
  {\bibfnamefont {A.}~\bibnamefont {Thomson}}, \bibinfo {author} {\bibfnamefont
  {L.}~\bibnamefont {Kong}}, \bibinfo {author} {\bibfnamefont {H.}~\bibnamefont
  {Zhou}}, \bibinfo {author} {\bibfnamefont {E.}~\bibnamefont {Baum}}, \bibinfo
  {author} {\bibfnamefont {Y.}~\bibnamefont {Zhang}}, \bibinfo {author}
  {\bibfnamefont {L.}~\bibnamefont {Holleis}}, \bibinfo {author} {\bibfnamefont
  {K.}~\bibnamefont {Watanabe}}, \bibinfo {author} {\bibfnamefont
  {T.}~\bibnamefont {Taniguchi}}, \bibinfo {author} {\bibfnamefont {A.~F.}\
  \bibnamefont {Young}}, \bibinfo {author} {\bibfnamefont {J.}~\bibnamefont
  {Alicea}}, \ and\ \bibinfo {author} {\bibfnamefont {S.}~\bibnamefont
  {Nadj-Perge}},\ }\href {\doibase 10.1038/s41586-023-06663-8} {\bibfield
  {journal} {\bibinfo  {journal} {Nature}\ }\textbf {\bibinfo {volume} {623}},\
  \bibinfo {pages} {942} (\bibinfo {year} {2023})}\BibitemShut {NoStop}%
\bibitem [{\citenamefont {Cao}\ \emph {et~al.}(2021{\natexlab{b}})\citenamefont
  {Cao}, \citenamefont {Rodan-Legrain}, \citenamefont {Park}, \citenamefont
  {Yuan}, \citenamefont {Watanabe}, \citenamefont {Taniguchi}, \citenamefont
  {Fernandes}, \citenamefont {Fu},\ and\ \citenamefont
  {Jarillo-Herrero}}]{Cao21Nematic}%
  \BibitemOpen
  \bibfield  {author} {\bibinfo {author} {\bibfnamefont {Y.}~\bibnamefont
  {Cao}}, \bibinfo {author} {\bibfnamefont {D.}~\bibnamefont {Rodan-Legrain}},
  \bibinfo {author} {\bibfnamefont {J.~M.}\ \bibnamefont {Park}}, \bibinfo
  {author} {\bibfnamefont {N.~F.~Q.}\ \bibnamefont {Yuan}}, \bibinfo {author}
  {\bibfnamefont {K.}~\bibnamefont {Watanabe}}, \bibinfo {author}
  {\bibfnamefont {T.}~\bibnamefont {Taniguchi}}, \bibinfo {author}
  {\bibfnamefont {R.~M.}\ \bibnamefont {Fernandes}}, \bibinfo {author}
  {\bibfnamefont {L.}~\bibnamefont {Fu}}, \ and\ \bibinfo {author}
  {\bibfnamefont {P.}~\bibnamefont {Jarillo-Herrero}},\ }\href {\doibase
  10.1126/science.abc2836} {\bibfield  {journal} {\bibinfo  {journal}
  {Science}\ }\textbf {\bibinfo {volume} {372}},\ \bibinfo {pages} {264}
  (\bibinfo {year} {2021}{\natexlab{b}})},\ \Eprint
  {http://arxiv.org/abs/https://www.science.org/doi/pdf/10.1126/science.abc2836}
  {https://www.science.org/doi/pdf/10.1126/science.abc2836} \BibitemShut
  {NoStop}%
\bibitem [{\citenamefont {Carr}\ \emph {et~al.}(2018)\citenamefont {Carr},
  \citenamefont {Fang}, \citenamefont {Jarillo-Herrero},\ and\ \citenamefont
  {Kaxiras}}]{jarillopressure}%
  \BibitemOpen
  \bibfield  {author} {\bibinfo {author} {\bibfnamefont {S.}~\bibnamefont
  {Carr}}, \bibinfo {author} {\bibfnamefont {S.}~\bibnamefont {Fang}}, \bibinfo
  {author} {\bibfnamefont {P.}~\bibnamefont {Jarillo-Herrero}}, \ and\ \bibinfo
  {author} {\bibfnamefont {E.}~\bibnamefont {Kaxiras}},\ }\href {\doibase
  10.1103/PhysRevB.98.085144} {\bibfield  {journal} {\bibinfo  {journal} {Phys.
  Rev. B}\ }\textbf {\bibinfo {volume} {98}},\ \bibinfo {pages} {085144}
  (\bibinfo {year} {2018})}\BibitemShut {NoStop}%
\bibitem [{\citenamefont {Yankowitz}\ \emph {et~al.}(2019)\citenamefont
  {Yankowitz}, \citenamefont {Chen}, \citenamefont {Polshyn}, \citenamefont
  {Zhang}, \citenamefont {Watanabe}, \citenamefont {Taniguchi}, \citenamefont
  {Graf}, \citenamefont {Young},\ and\ \citenamefont {Dean}}]{Yankowitz_2019}%
  \BibitemOpen
  \bibfield  {author} {\bibinfo {author} {\bibfnamefont {M.}~\bibnamefont
  {Yankowitz}}, \bibinfo {author} {\bibfnamefont {S.}~\bibnamefont {Chen}},
  \bibinfo {author} {\bibfnamefont {H.}~\bibnamefont {Polshyn}}, \bibinfo
  {author} {\bibfnamefont {Y.}~\bibnamefont {Zhang}}, \bibinfo {author}
  {\bibfnamefont {K.}~\bibnamefont {Watanabe}}, \bibinfo {author}
  {\bibfnamefont {T.}~\bibnamefont {Taniguchi}}, \bibinfo {author}
  {\bibfnamefont {D.}~\bibnamefont {Graf}}, \bibinfo {author} {\bibfnamefont
  {A.~F.}\ \bibnamefont {Young}}, \ and\ \bibinfo {author} {\bibfnamefont
  {C.~R.}\ \bibnamefont {Dean}},\ }\href {\doibase 10.1126/science.aav1910}
  {\bibfield  {journal} {\bibinfo  {journal} {Science}\ }\textbf {\bibinfo
  {volume} {363}},\ \bibinfo {pages} {1059} (\bibinfo {year}
  {2019})}\BibitemShut {NoStop}%
\bibitem [{\citenamefont {Kwan}\ \emph {et~al.}(2023)\citenamefont {Kwan},
  \citenamefont {Wagner}, \citenamefont {Bultinck}, \citenamefont {Simon},
  \citenamefont {Berg},\ and\ \citenamefont {Parameswaran}}]{Kwan23}%
  \BibitemOpen
  \bibfield  {author} {\bibinfo {author} {\bibfnamefont {Y.~H.}\ \bibnamefont
  {Kwan}}, \bibinfo {author} {\bibfnamefont {G.}~\bibnamefont {Wagner}},
  \bibinfo {author} {\bibfnamefont {N.}~\bibnamefont {Bultinck}}, \bibinfo
  {author} {\bibfnamefont {S.~H.}\ \bibnamefont {Simon}}, \bibinfo {author}
  {\bibfnamefont {E.}~\bibnamefont {Berg}}, \ and\ \bibinfo {author}
  {\bibfnamefont {S.~A.}\ \bibnamefont {Parameswaran}},\ }\href@noop {}
  {\enquote {\bibinfo {title} {Electron-phonon coupling and competing kekul\'e
  orders in twisted bilayer graphene},}\ } (\bibinfo {year} {2023}),\ \Eprint
  {http://arxiv.org/abs/2303.13602} {arXiv:2303.13602 [cond-mat.str-el]}
  \BibitemShut {NoStop}%
\bibitem [{SM()}]{SM}%
  \BibitemOpen
  \href@noop {} {}\bibinfo {note} {See Supplemental Material for details on the
  tight-binding approach and Hartree-Fock theory, the algorithm to obtain the
  reduced density matrix, discussion of the phase diagram as well as additional
  plots of the band structure.}\BibitemShut {Stop}%
\bibitem [{\citenamefont {Moon}\ and\ \citenamefont {Koshino}(2013)}]{Moon13}%
  \BibitemOpen
  \bibfield  {author} {\bibinfo {author} {\bibfnamefont {P.}~\bibnamefont
  {Moon}}\ and\ \bibinfo {author} {\bibfnamefont {M.}~\bibnamefont {Koshino}},\
  }\href {\doibase 10.1103/PhysRevB.87.205404} {\bibfield  {journal} {\bibinfo
  {journal} {Phys. Rev. B}\ }\textbf {\bibinfo {volume} {87}},\ \bibinfo
  {pages} {205404} (\bibinfo {year} {2013})}\BibitemShut {NoStop}%
\bibitem [{\citenamefont {Throckmorton}\ and\ \citenamefont
  {Vafek}(2012)}]{Throckmorton12}%
  \BibitemOpen
  \bibfield  {author} {\bibinfo {author} {\bibfnamefont {R.~E.}\ \bibnamefont
  {Throckmorton}}\ and\ \bibinfo {author} {\bibfnamefont {O.}~\bibnamefont
  {Vafek}},\ }\href@noop {} {\bibfield  {journal} {\bibinfo  {journal} {Phys.
  Rev. B}\ }\textbf {\bibinfo {volume} {86}},\ \bibinfo {pages} {115447}
  (\bibinfo {year} {2012})}\BibitemShut {NoStop}%
\bibitem [{\citenamefont {Saito}\ \emph {et~al.}(2020)\citenamefont {Saito},
  \citenamefont {Ge}, \citenamefont {Watanabe}, \citenamefont {Taniguchi},\
  and\ \citenamefont {Young}}]{Saito_2020}%
  \BibitemOpen
  \bibfield  {author} {\bibinfo {author} {\bibfnamefont {Y.}~\bibnamefont
  {Saito}}, \bibinfo {author} {\bibfnamefont {J.}~\bibnamefont {Ge}}, \bibinfo
  {author} {\bibfnamefont {K.}~\bibnamefont {Watanabe}}, \bibinfo {author}
  {\bibfnamefont {T.}~\bibnamefont {Taniguchi}}, \ and\ \bibinfo {author}
  {\bibfnamefont {A.~F.}\ \bibnamefont {Young}},\ }\href {\doibase
  10.1038/s41567-020-0928-3} {\bibfield  {journal} {\bibinfo  {journal} {Nature
  Physics}\ }\textbf {\bibinfo {volume} {16}},\ \bibinfo {pages} {926}
  (\bibinfo {year} {2020})}\BibitemShut {NoStop}%
\bibitem [{\citenamefont {Stepanov}\ \emph
  {et~al.}(2020{\natexlab{b}})\citenamefont {Stepanov}, \citenamefont {Das},
  \citenamefont {Lu}, \citenamefont {Fahimniya}, \citenamefont {Watanabe},
  \citenamefont {Taniguchi}, \citenamefont {Koppens}, \citenamefont {Lischner},
  \citenamefont {Levitov},\ and\ \citenamefont {Efetov}}]{Stepanov_2020}%
  \BibitemOpen
  \bibfield  {author} {\bibinfo {author} {\bibfnamefont {P.}~\bibnamefont
  {Stepanov}}, \bibinfo {author} {\bibfnamefont {I.}~\bibnamefont {Das}},
  \bibinfo {author} {\bibfnamefont {X.}~\bibnamefont {Lu}}, \bibinfo {author}
  {\bibfnamefont {A.}~\bibnamefont {Fahimniya}}, \bibinfo {author}
  {\bibfnamefont {K.}~\bibnamefont {Watanabe}}, \bibinfo {author}
  {\bibfnamefont {T.}~\bibnamefont {Taniguchi}}, \bibinfo {author}
  {\bibfnamefont {F.~H.~L.}\ \bibnamefont {Koppens}}, \bibinfo {author}
  {\bibfnamefont {J.}~\bibnamefont {Lischner}}, \bibinfo {author}
  {\bibfnamefont {L.}~\bibnamefont {Levitov}}, \ and\ \bibinfo {author}
  {\bibfnamefont {D.~K.}\ \bibnamefont {Efetov}},\ }\href {\doibase
  10.1038/s41586-020-2459-6} {\bibfield  {journal} {\bibinfo  {journal}
  {Nature}\ }\textbf {\bibinfo {volume} {583}},\ \bibinfo {pages} {375}
  (\bibinfo {year} {2020}{\natexlab{b}})}\BibitemShut {NoStop}%
\bibitem [{\citenamefont {Kohn}(1964)}]{Kohn64}%
  \BibitemOpen
  \bibfield  {author} {\bibinfo {author} {\bibfnamefont {W.}~\bibnamefont
  {Kohn}},\ }\href {\doibase 10.1103/PhysRev.133.A171} {\bibfield  {journal}
  {\bibinfo  {journal} {Phys. Rev.}\ }\textbf {\bibinfo {volume} {133}},\
  \bibinfo {pages} {A171} (\bibinfo {year} {1964})}\BibitemShut {NoStop}%
\bibitem [{\citenamefont {Kohn}(1996)}]{Kohn96}%
  \BibitemOpen
  \bibfield  {author} {\bibinfo {author} {\bibfnamefont {W.}~\bibnamefont
  {Kohn}},\ }\href {\doibase 10.1103/PhysRevLett.76.3168} {\bibfield  {journal}
  {\bibinfo  {journal} {Phys. Rev. Lett.}\ }\textbf {\bibinfo {volume} {76}},\
  \bibinfo {pages} {3168} (\bibinfo {year} {1996})}\BibitemShut {NoStop}%
\bibitem [{\citenamefont {Song}\ and\ \citenamefont {Bernevig}(2022)}]{Song22}%
  \BibitemOpen
  \bibfield  {author} {\bibinfo {author} {\bibfnamefont {Z.-D.}\ \bibnamefont
  {Song}}\ and\ \bibinfo {author} {\bibfnamefont {B.~A.}\ \bibnamefont
  {Bernevig}},\ }\href {\doibase 10.1103/PhysRevLett.129.047601} {\bibfield
  {journal} {\bibinfo  {journal} {Phys. Rev. Lett.}\ }\textbf {\bibinfo
  {volume} {129}},\ \bibinfo {pages} {047601} (\bibinfo {year}
  {2022})}\BibitemShut {NoStop}%
\bibitem [{\citenamefont {S\'anchez}\ and\ \citenamefont
  {Stauber}()}]{Sanchez24}%
  \BibitemOpen
  \bibfield  {author} {\bibinfo {author} {\bibfnamefont {M.~S.}\ \bibnamefont
  {S\'anchez}}\ and\ \bibinfo {author} {\bibfnamefont {T.}~\bibnamefont
  {Stauber}},\ }\href@noop {} {\bibinfo  {journal} {arXiv:2302.00884}\
  }\BibitemShut {NoStop}%
\bibitem [{\citenamefont {Gonzalez}\ and\ \citenamefont
  {Stauber}(2021)}]{Gonzalez21}%
  \BibitemOpen
\bibfield  {journal} {  }\bibfield  {author} {\bibinfo {author} {\bibfnamefont
  {J.}~\bibnamefont {Gonzalez}}\ and\ \bibinfo {author} {\bibfnamefont
  {T.}~\bibnamefont {Stauber}},\ }\href
  {https://doi.org/10.1103%2Fphysrevb.104.115110} {\bibfield  {journal}
  {\bibinfo  {journal} {Physical Review B}\ }\textbf {\bibinfo {volume} {104}}
  (\bibinfo {year} {2021})}\BibitemShut {NoStop}%
\bibitem [{\citenamefont {Kohn}\ and\ \citenamefont
  {Luttinger}(1965)}]{Kohn65}%
  \BibitemOpen
  \bibfield  {author} {\bibinfo {author} {\bibfnamefont {W.}~\bibnamefont
  {Kohn}}\ and\ \bibinfo {author} {\bibfnamefont {J.~M.}\ \bibnamefont
  {Luttinger}},\ }\href {\doibase 10.1103/PhysRevLett.15.524} {\bibfield
  {journal} {\bibinfo  {journal} {Phys. Rev. Lett.}\ }\textbf {\bibinfo
  {volume} {15}},\ \bibinfo {pages} {524} (\bibinfo {year} {1965})}\BibitemShut
  {NoStop}%
\bibitem [{\citenamefont {Baranov}\ \emph {et~al.}(1992)\citenamefont
  {Baranov}, \citenamefont {Chubukov},\ and\ \citenamefont
  {Yu.~Kagan}}]{Baranov92}%
  \BibitemOpen
  \bibfield  {author} {\bibinfo {author} {\bibfnamefont {M.~A.}\ \bibnamefont
  {Baranov}}, \bibinfo {author} {\bibfnamefont {A.~V.}\ \bibnamefont
  {Chubukov}}, \ and\ \bibinfo {author} {\bibfnamefont {M.}~\bibnamefont
  {Yu.~Kagan}},\ }\href {\doibase 10.1142/S0217979292001249} {\bibfield
  {journal} {\bibinfo  {journal} {International Journal of Modern Physics B}\
  }\textbf {\bibinfo {volume} {06}},\ \bibinfo {pages} {2471} (\bibinfo {year}
  {1992})}\BibitemShut {NoStop}%
\bibitem [{\citenamefont {Scalapino}\ \emph {et~al.}(1987)\citenamefont
  {Scalapino}, \citenamefont {Loh},\ and\ \citenamefont
  {Hirsch}}]{Scalapino87}%
  \BibitemOpen
  \bibfield  {author} {\bibinfo {author} {\bibfnamefont {D.~J.}\ \bibnamefont
  {Scalapino}}, \bibinfo {author} {\bibfnamefont {E.}~\bibnamefont {Loh}}, \
  and\ \bibinfo {author} {\bibfnamefont {J.~E.}\ \bibnamefont {Hirsch}},\
  }\href {\doibase 10.1103/PhysRevB.35.6694} {\bibfield  {journal} {\bibinfo
  {journal} {Phys. Rev. B}\ }\textbf {\bibinfo {volume} {35}},\ \bibinfo
  {pages} {6694} (\bibinfo {year} {1987})}\BibitemShut {NoStop}%
\bibitem [{\citenamefont {Zhang}\ \emph {et~al.}(2023)\citenamefont {Zhang},
  \citenamefont {Polski}, \citenamefont {Thomson}, \citenamefont
  {Lantagne-Hurtubise}, \citenamefont {Lewandowski}, \citenamefont {Zhou},
  \citenamefont {Watanabe}, \citenamefont {Taniguchi}, \citenamefont {Alicea},\
  and\ \citenamefont {Nadj-Perge}}]{Zhang23}%
  \BibitemOpen
  \bibfield  {author} {\bibinfo {author} {\bibfnamefont {Y.}~\bibnamefont
  {Zhang}}, \bibinfo {author} {\bibfnamefont {R.}~\bibnamefont {Polski}},
  \bibinfo {author} {\bibfnamefont {A.}~\bibnamefont {Thomson}}, \bibinfo
  {author} {\bibfnamefont {{\'E}.}~\bibnamefont {Lantagne-Hurtubise}}, \bibinfo
  {author} {\bibfnamefont {C.}~\bibnamefont {Lewandowski}}, \bibinfo {author}
  {\bibfnamefont {H.}~\bibnamefont {Zhou}}, \bibinfo {author} {\bibfnamefont
  {K.}~\bibnamefont {Watanabe}}, \bibinfo {author} {\bibfnamefont
  {T.}~\bibnamefont {Taniguchi}}, \bibinfo {author} {\bibfnamefont
  {J.}~\bibnamefont {Alicea}}, \ and\ \bibinfo {author} {\bibfnamefont
  {S.}~\bibnamefont {Nadj-Perge}},\ }\href {\doibase
  10.1038/s41586-022-05446-x} {\bibfield  {journal} {\bibinfo  {journal}
  {Nature}\ }\textbf {\bibinfo {volume} {613}},\ \bibinfo {pages} {268}
  (\bibinfo {year} {2023})}\BibitemShut {NoStop}%
\bibitem [{\citenamefont {Holleis}\ \emph {et~al.}()\citenamefont {Holleis},
  \citenamefont {Patterson}, \citenamefont {Zhang}, \citenamefont {Yoo},
  \citenamefont {Zhou}, \citenamefont {Taniguchi}, \citenamefont {Watanabe},
  \citenamefont {Nadj-Perge},\ and\ \citenamefont {Young}}]{Holleis24}%
  \BibitemOpen
  \bibfield  {author} {\bibinfo {author} {\bibfnamefont {L.}~\bibnamefont
  {Holleis}}, \bibinfo {author} {\bibfnamefont {C.~L.}\ \bibnamefont
  {Patterson}}, \bibinfo {author} {\bibfnamefont {Y.}~\bibnamefont {Zhang}},
  \bibinfo {author} {\bibfnamefont {H.~M.}\ \bibnamefont {Yoo}}, \bibinfo
  {author} {\bibfnamefont {H.}~\bibnamefont {Zhou}}, \bibinfo {author}
  {\bibfnamefont {T.}~\bibnamefont {Taniguchi}}, \bibinfo {author}
  {\bibfnamefont {K.}~\bibnamefont {Watanabe}}, \bibinfo {author}
  {\bibfnamefont {S.}~\bibnamefont {Nadj-Perge}}, \ and\ \bibinfo {author}
  {\bibfnamefont {A.~F.}\ \bibnamefont {Young}},\ }\href@noop {} {\bibinfo
  {journal} {arXiv:2303.00742}\ }\BibitemShut {NoStop}%
\end{thebibliography}
\end{document}